\documentclass[journal]{IEEEtran}
\usepackage{multirow}
\usepackage{amsmath,graphicx}
\usepackage{algpseudocode}
\usepackage{multirow}
\usepackage{array}
\usepackage{colortbl}
\usepackage{booktabs}
\usepackage{rotating}
\usepackage{caption}
\usepackage{subcaption}
\usepackage[normalem]{ulem} 
\usepackage{booktabs}
\usepackage{tikz}
\usepackage{makecell}
\usepackage[margin=0.55in]{geometry}

\begin{document}

\title{Quality Prediction on Deep Generative Images}

\author{Hyunsuk Ko \textit{Member, IEEE} , Dae Yeol Lee, Seunghyun Cho, and Alan C. Bovik \textit{Fellow, IEEE} 
\thanks{This work was supported by Institute for Information and communications
Technology Promotion (IITP) grant funded by the Korea government (MSIT) 2017-0-00072,
Development of Audio/Video Coding and Light Field Media Fundamental Technologies 
for Ultra Realistic Tera-media.}%
\thanks{Hyunsuk Ko is with the Division of Electrical Engineering, Hanyang University ERICA,
Ansan, Rep. of Korea (\textit{Corresponding author}, e-mai: hyunsuk@hanyang.ac.kr).}
\thanks{Seunghyun Cho  is with the Department of Information and Communication Engineering,
Changwon, Rep. of Korea (e-mai: scho@kyungnam.ac.kr).}
\thanks{Dae Yeol Lee and A. C. Bovik are with the Department of Electrical and Computer 
Engineering at The University of Texas at Austin, Austin, TX, 78712, USA
(e-mail: daelee711@gmail.com, bovik@ece.utexas.edu).}
\thanks{Citation information: DOI 10.1109/TIP.2020.2987180, IEEE Transactions on Image Processing}
\thanks{Link to the abstract(Early Access) https://ieeexplore.ieee.org/document/9069418}
}

\maketitle

\makeatletter
\def\ps@IEEEtitlepagestyle{
  \def\@oddfoot{\mycopyrightnotice}
  \def\@evenfoot{}
}
\def\mycopyrightnotice{
  {\footnotesize
  \begin{minipage}{\textwidth}
  \centering
 \copyright~20XX IEEE.  Personal use of this material is permitted.  Permission from IEEE must be obtained for all other uses, in any current or future media, including reprinting/republishing this material for advertising or promotional purposes, creating new collective works, for resale or redistribution to servers or lists, or reuse of any copyrighted component of this work in other works.
  \end{minipage}
  }
}

\begin{abstract}
In recent years, deep neural networks have been utilized in a wide variety of applications including
image generation. In particular, generative adversarial networks (GANs) are able to produce
highly realistic pictures as part of tasks such as image compression. As with standard compression, 
it is desirable to be able to automatically assess the perceptual quality of generative images to monitor 
and control the encode process. However, existing image quality algorithms are ineffective on GAN 
generated content, especially on textured regions and at high compressions. Here we propose a new 
``naturalness''-based image quality predictor for generative images. Our new GAN picture quality 
predictor is built  using a multi-stage parallel boosting system based on structural similarity features 
and measurements of statistical similarity. To enable model development and testing, we also 
constructed a subjective GAN image quality database containing (distorted) GAN images and 
collected human opinions of them. Our experimental results indicate that our proposed GAN IQA
model delivers superior quality predictions on the generative image datasets, as well as on  
traditional image quality datasets. 
\end{abstract}


\section{Introduction}

\IEEEPARstart{D}{eep} neural networks (DNNs) have been applied to 
broad swathes of applications beyond traditional computer vision, including super-resolution
\cite{cit:Dong2016}, detection~\cite{cit:Shin2016} and classification 
\cite{cit:Krizhevsky2012} problems, and even for composing music~\cite{cit:Huang2016} 
and creating computer-generated art work~\cite{cit:Gatys2016}. Of particular interest are
generative adversarial networks (GANs)~\cite{cit:Goodfellow2014}, which can learn models 
of highly non-linear distributions in an unsupervised manner. For example, GANs have been 
shown to be able to compute highly realistic, naturalistic images by capturing both global semantic 
information and local textural descriptions of real-world image data~\cite{cit:Ledig2017, cit:Brock2019}.
 In this direction, GANs 
have recently been utilized for image/video compression~\cite{cit:Agustsson2018}. 
One approach is to create a hybrid codec combining a GAN with a legacy codec such as 
H.264 or HEVC (High Efficiency Video Coding)~\cite{cit:H264, cit:HEVC}. At the decoder side, 
uncompressed regions can be synthesized using a pre-trained GAN that processes transmitted 
texture parameters. Another approach is to encode the entire full-resolution image using a GAN, 
which could be particularly effective at very low bitrates~\cite{cit:Kim2018}. 
While research on GAN-based compression is in its infancy, expectations are high, because 
generative images are often visually pleasing and GAN-based compression has the potential
to supply competitive compression efficiency. \\
\begin{figure}
        \centering
        \begin{subfigure}[t]{0.22\textwidth}
                \centering
                \includegraphics[width=\textwidth]{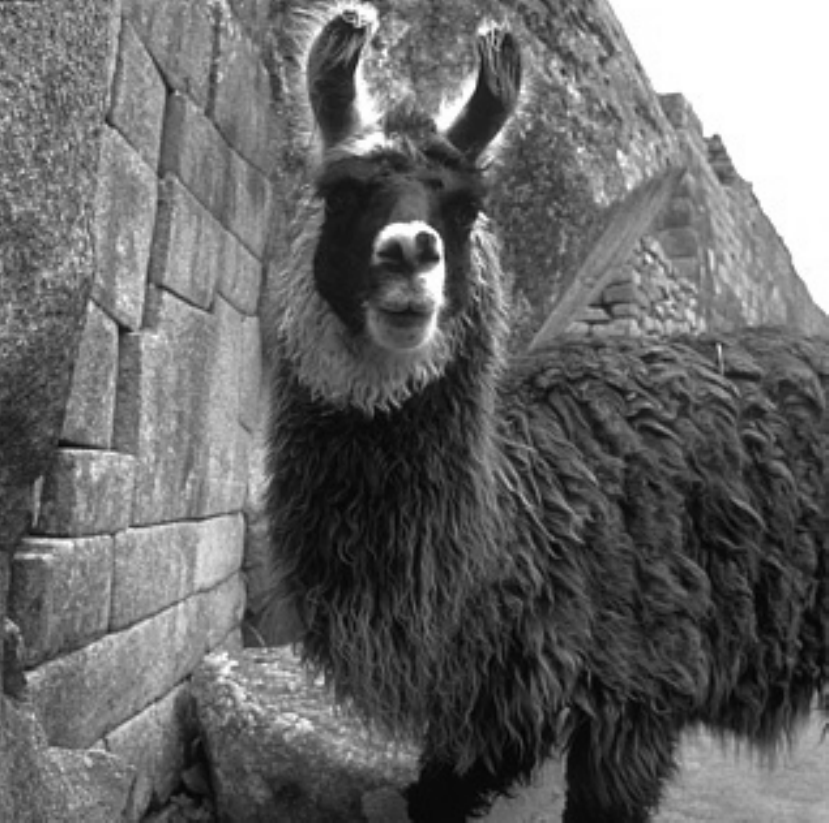}
                \caption{Original Image}
		\label{fig:exorig1}
        \end{subfigure}
		\quad 
        \begin{subfigure}[t]{0.22\textwidth}
                \centering
                \includegraphics[width=\textwidth]{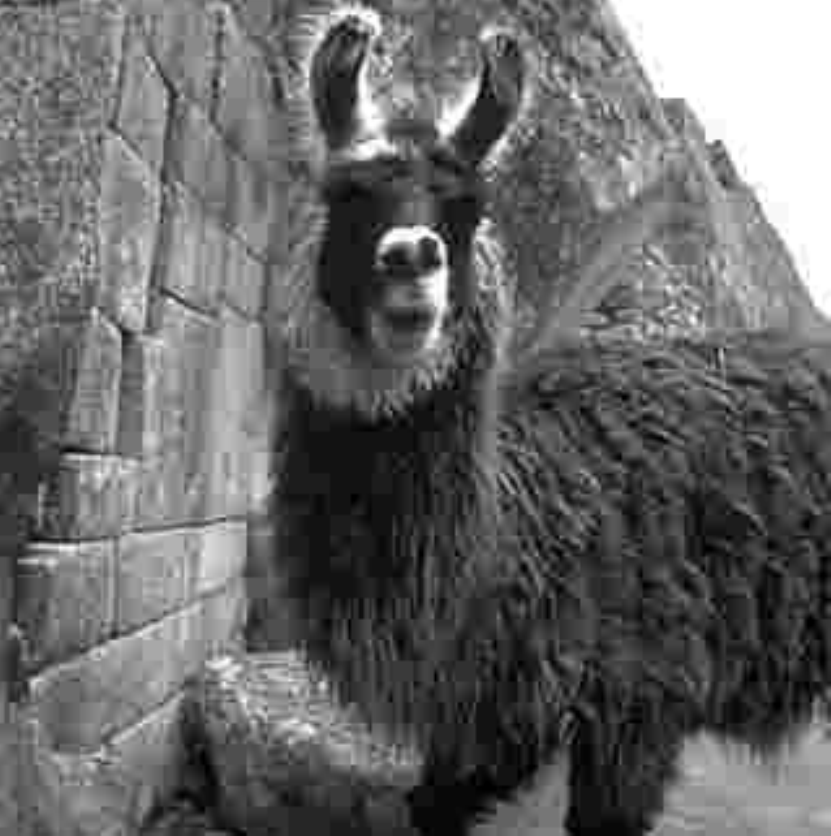}
                \caption{JPEG (PSNR: 25.3dB)}
		\label{fig:exjpeg}
        \end{subfigure}
\\
        \begin{subfigure}[t]{0.22\textwidth}
                \centering
                \includegraphics[width=\textwidth]{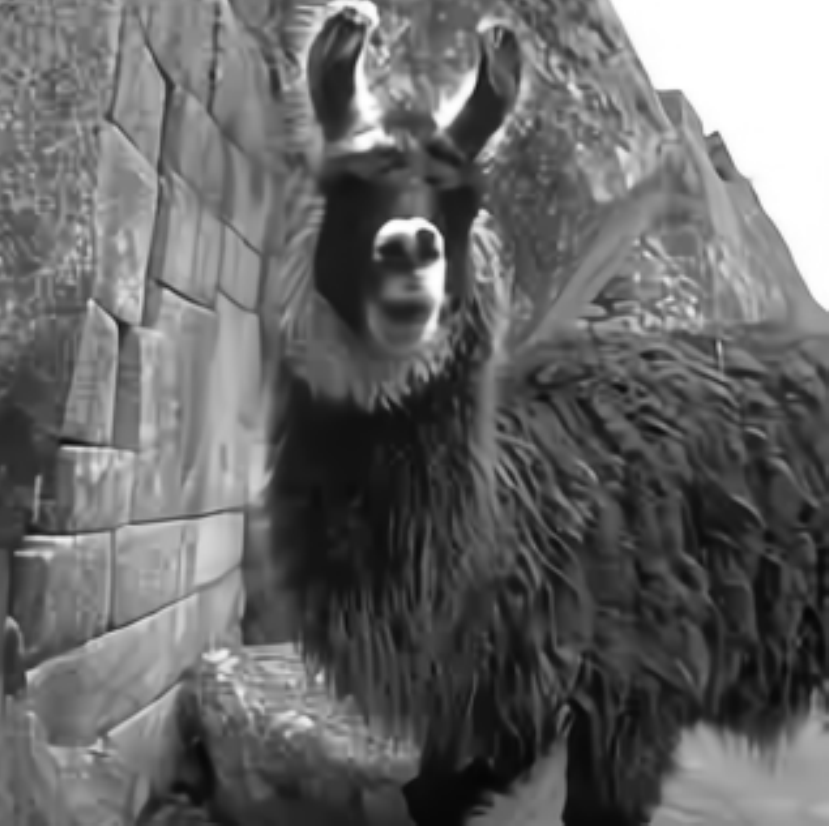}
                \caption{CNN (PSNR: 26.3dB)}
		\label{fig:exbase}
        \end{subfigure}
		\quad 
        \begin{subfigure}[t]{0.22\textwidth}
                \centering
                \includegraphics[width=\textwidth]{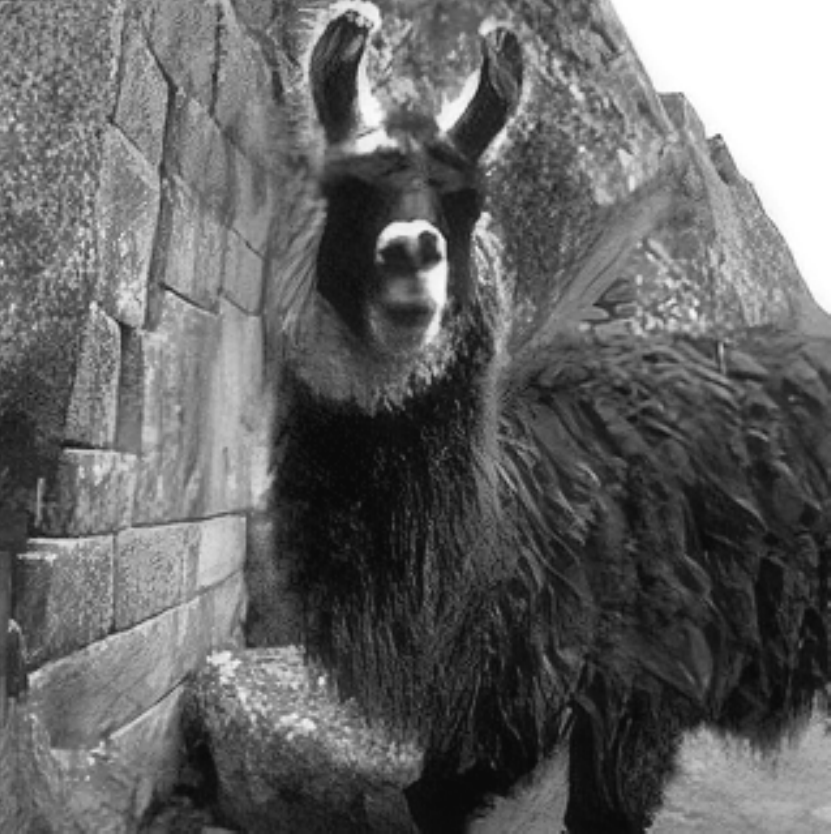}
                \caption{GAN (PSNR: 23.5dB)}
		\label{fig:exgan}
        \end{subfigure}
\caption{Examples of DNN based generative images: (a) original image,
(b) JPEG-coded image, (c) CNN-generated image, and (d) GAN image.}
\label{fig:ex1}
  \vspace{-5mm}
\end{figure}
An important aspect of the development of generative image compression systems 
is the ability to objectively measure the perceptual quality of the reconstructed 
(decoded) images. Since the characteristics of generative images are quite different from 
those of natural images, existing image quality assessment (IQA) models are inadequate 
for measuring the quality of generative images. The principle reason for this is that 
images generated by a GAN may appear quite realistic and similar to an original, 
yet may match it poorly based on pixel comparisons. 
For example, an original image and three versions of it created using different schemes
are shown in Fig. \ref{fig:ex1}. Here, the JPEG-coded image (Fig. \ref{fig:exjpeg}) is 
afflicted by clearly visible blocking artifacts, while an image that was generated
using a CNN (convolutional neural network) is blurred. 
By comparison, the GAN-generated image appears looks as natural and realistic as 
the original. However, the peak signal to noise ratio (PSNR) values indicate otherwise, 
indicating the inconsistency of pixel comparisons when evaluating GAN images.\\
\indent Towards addressing this problem, we propose a novel full-reference quality assessment 
model for analyzing generative images. The contributions that we make are summarized as follows: 
\begin{itemize}
\item Our proposed GAN IQA model utilizes measurements of both statistical similarity 
and structural similarity between a reference image and a possibly distorted version of it. 
Statistical similarity is expressed by multiple quality-related histogram 
distances computed between the reference and test images. These measures effectively
capture the textural characteristic of GAN-generated images and their perceptual similarity 
to the images they were generated from. The later is realized by 
deriving quality-sensitive spatial and spectral structure features based on the
singular value decomposition (SVD). The final predicted scores are generated using 
a multi-stage parallel boosting system based on support vector regression (SVR). 
\item We built a generative image database by using a GAN designed for artifact removal.
The new database comprises reference images that span a wide 
range of natural scene characteristics. These images were systematically distorted using 
JPEG compression as well as by generation by both CNNs and GANs (i.e. with 
an adversarial loss term). Unlike most other datasets, our 
database is divided into four data subsets, one containing full-frame images, and three 
subsets containing patches having different structural peculiarities, to enable a deeper 
analysis of both the global and local attributes of generative images. We also conducted 
a human subjective test utilizing the pairwise comparison method, yielding a substantial 
set of MOS (mean opinion scores). 
\item We conducted a comprehensive set of algorithm comparison experiments.
First, we analyzed the per-feature group efficacy of the statistical and structural features.
In addition, we also compared our proposed model with thirteen existing full-reference 
picture quality models as well as two recent deep learning system-based IQA modes. 
Furthermore, we tested our model on the three traditional image quality databases:
LIVE, TID2013 and CSIQ, where it is shown to have the capability to predict 
the perceptual quality of natural distorted images, as well as generative images.
Lastly, we performed a set of cross-database tests to evaluate the database
independence of our model.  
\end{itemize}

The rest of the paper is organized as follows. We review related work on image
quality assessment and GAN-based image/video compression in 
Section~\ref{sec:related}. The construction of our generative image database
is detailed in Section~\ref{sec:database}, while the proposed generative image 
quality model is introduced in Section \ref{sec:metric}. Experimental 
results and their discussions are provided in Section~\ref{sec:performance}. 
Finally, concluding remarks are given in Section~\ref{sec:conclusions}.

\section{Related Work}\label{sec:related}
\subsection{Image Quality Assessment}\label{sec:2.1}
IQA models are usually classified as full-reference (FR), reduced-reference (RR),
or no-reference (NR), depending on the availability of a reference image. A variety of 
picture quality engines based on natural scene statistic (NSS) features have been 
proposed, which do not rely on any strong distortion hypotheses, such as specific impairment types.
Instead, these models exploit certain statistical regularities that exist in natural scene data, 
which are disturbed or lost in the presence of image distortions. Moorthy {\it et al.} and Saad {\it et al.}
proposed ``quality-aware'' NSS features in the wavelet domain~\cite{cit:Moorthy2011} and in the 
discrete cosine transform domain~\cite{cit:Saad2012}, respectively, while Mittal {\it et al.} developed 
similar features in the spatial domain~\cite{cit:Mittal2013}. These and many other IQA models use 
machine learning to capture the highly non-linear relationship between handcrafted NSS 
features and human judgments of picture quality.
Other examples include Li {\it et al.}~\cite{cit:Li2011} who deployed a general regression neural 
network to learn a mapping between several features and picture quality, including phase 
congruency, entropy and the gradient. Other examples include the multi-metric fusion 
(MMF) model~\cite{cit:Liu2013}, the 3-D multi scorers fusion model~\cite{cit:Ko2017}, and a 
sparse representation of NSS features~\cite{cit:He2012}.\\
\indent More recently, IQA models developed using deep learning have been emerging. 
Li {\it et al.}~\cite{cit:Li2015} fed the NSS-related features into a stacked 
autoencoder to reinforce quality prediction accuracy while Ghadiyaram {\it et al.} 
\cite{cit:Ghadiyaram2017} deployed an a large variety of NSS features to train a deep 
belief network (DBN) to predict subjective picture quality. 
Rather than designing handcrafted features, the authors of~\cite{cit:Kim2017} 
automatically learned features when training a CNN to generate local quality maps and 
conduct NR IQA. In~\cite{cit:Jong2017}, Kim {\it et al.} proposed a CNN-based FR IQA model
that learns the underlying data distribution and uses it to optimize a set of visual weights,
without using any prior knowledge of the HVS. A survey of deep learning methods for IQA
is given in~\cite{cit:Zeng2017}. 

\subsection{GAN Based Image/Video Compression}\label{sec:2.2}
Methods of using DNNs for data compression have recently become an active 
area of research. Over the last few years,  the most popular DNN
architecture for image compression has been various forms of the auto-encoder
\cite{cit:Balle2017, cit:Li2018, cit:Mentzer2018, cit:Minnen2018}. An autoencoder based
image compressor generates latent vectors as bit-streams, then encodes them using entropy 
coding. They usually use a mean-squared error (MSE) or perceptual loss functions like MS-SSIM 
~\cite{cit:Wang2004, cit:Wang2003} to reduce perceived distortions between the original and the 
decompressed images. Another popular trend is to take the adversarial loss of GAN
into account, because it  is capable of maintaining both global structure and local 
texture even of very low bitrates. However, the aforementioned loss terms may
fail when reconstructing semantic information, because they favor the preservation
of pixel-wise fidelity. Rippel {\it et al.}~\cite{cit:Rippel2017} used an adversarial 
loss to train a deep compression system, while Santurkar {\it et al.} trained a GAN framework
to decode thumbnail images~\cite{cit:Santurkar2018}. Eirikur {\it et al.} \cite{cit:Agustsson2018} 
proposed a GAN-based compression system targeting bitrates below 0.1 bit per pixel. 
Their system realistically synthesizes image objects and textures like streets and trees.
They also increase the coding gain using a semantic label map.\\
\indent GANs can be also used to pre-/post-process compressed images. 
For example, Galteri {\it et al.} presented a feed-forward model trained by a
GAN to remove compression artifacts~\cite{cit:Galteri2017}. The authors in \cite{cit:Ledig2017} 
train a GAN to perform image super-resolution (SR) with photo-realism for upscaling factors
as large as 4x. While GAN-based video coding is still in early stages of development, 
a number of researchers are trying to replace the traditional hybrid codec framework 
(e.g., H.264 or HEVC) with end-to-end deep learning frameworks. In these efforts, the focus 
of the GAN is primary on prediction and reconstruction~\cite{cit:Oh2015, cit:Vondrick2016}, 
quantization~\cite{cit:Ranzato2014} and pixel motion estimation~\cite{cit:Finn2016, cit:Liu2017}.
Recently, Kim {\it et al.} proposed a soft edge-guided conditional GAN framework targeting 
streaming videos at very low bitrates~\cite{cit:Kim2018}. \\
\indent The notation of ``picture quality'' takes on a somewhat different flavor in the context of 
assessing generative images, which is why a new family of picture quality models is needed. 
This is particularly true in the context of FR IQA, as exemplified by the 
SSIM~\cite{cit:Wang2004, cit:Wang2003} and VIF~\cite{cit:Sheikh2006} models. 
FR models that make pixel-wise comparisons - even over neighborhoods or in a 
transformed space - are ultimately best characterized as perceptual fidelity measurements. 
NR IQA models such as BRISQUE~\cite{cit:Mittal2012} and NIQE~\cite{cit:Mittal2013}
supply a different way, where only the appearance of quality of an image is assessed,
 but these algorithms do not make use
of valuable reference information. Our approach shares elements of both; a reference 
picture is deployed in the evaluation, but the GAN IQA evaluator also assesses the test 
picture in regards to its intrinsic natural quality. This is important since, when an image is 
compressed, textured areas may be replaced by generative (synthesized) content instead of 
being encoded directly~\cite{cit:Agustsson2018}. For example, the furry parts of the llama
generated by the GAN in Fig.~\ref{fig:exgan} appear sharp and natural more so than 
in  Figs. \ref{fig:exjpeg} and \ref{fig:exbase}, giving a more visually pleasing appearance
although the content is not a good pixel-wise match to the original furry parts. 
Maintaining the `naturalness' of generative images or subimages, while still contributing
a good representation of the original can be important, since it may afford the possibility
of much higher compressions while ensuring that the generative images or subimages 
appear both highly similar to the original as well as naturalistic.

\section{Generative Image Database}\label{sec:database}

The new database contains four subsets; one of full-frames images and three of image patches.
The patch subsets were designed to exhibit different degrees of structural complexity.
Specifically, we grouped them as: 1) random structured patches, 2) regular structured (pattern) 
patches and 3) high-level structured (face) patches. We conducted a subjective human study 
on these images and patches to supply ground truth in our effort to learn a mapping between 
generative images and human opinions of them. 

\subsection{Database Generation}\label{sec:3.1}
\begin{figure}[t]
\centering
\includegraphics[width=0.44\textwidth]{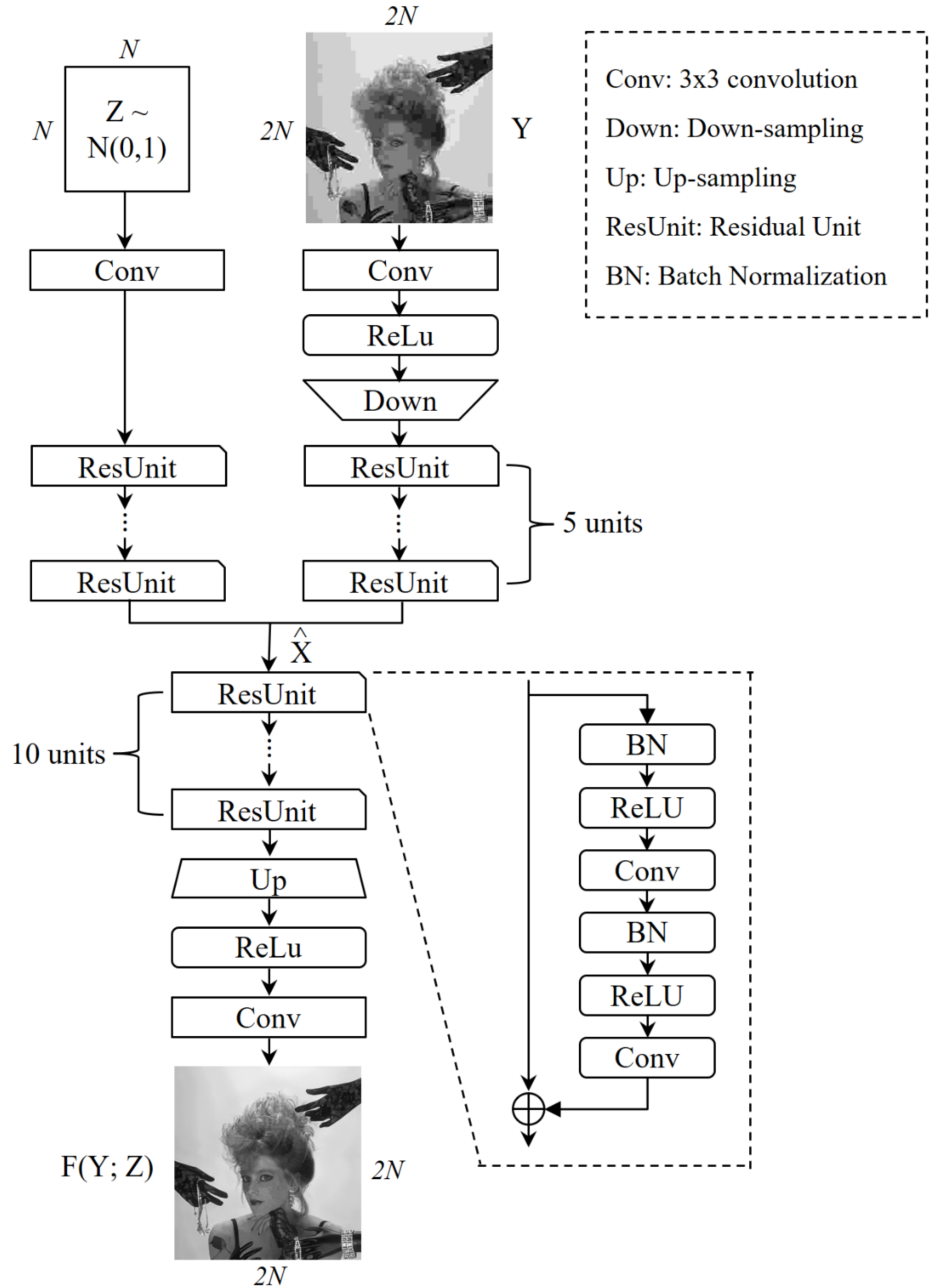}
\caption{The GAN architecture used to build the generative image database.}
\label{fig:onetomany}
  \vspace{-5mm}
\end{figure}

\begin{figure*}
        \centering
        \begin{subfigure}[t]{0.17\textwidth}
                \centering
                \includegraphics[width=\textwidth]{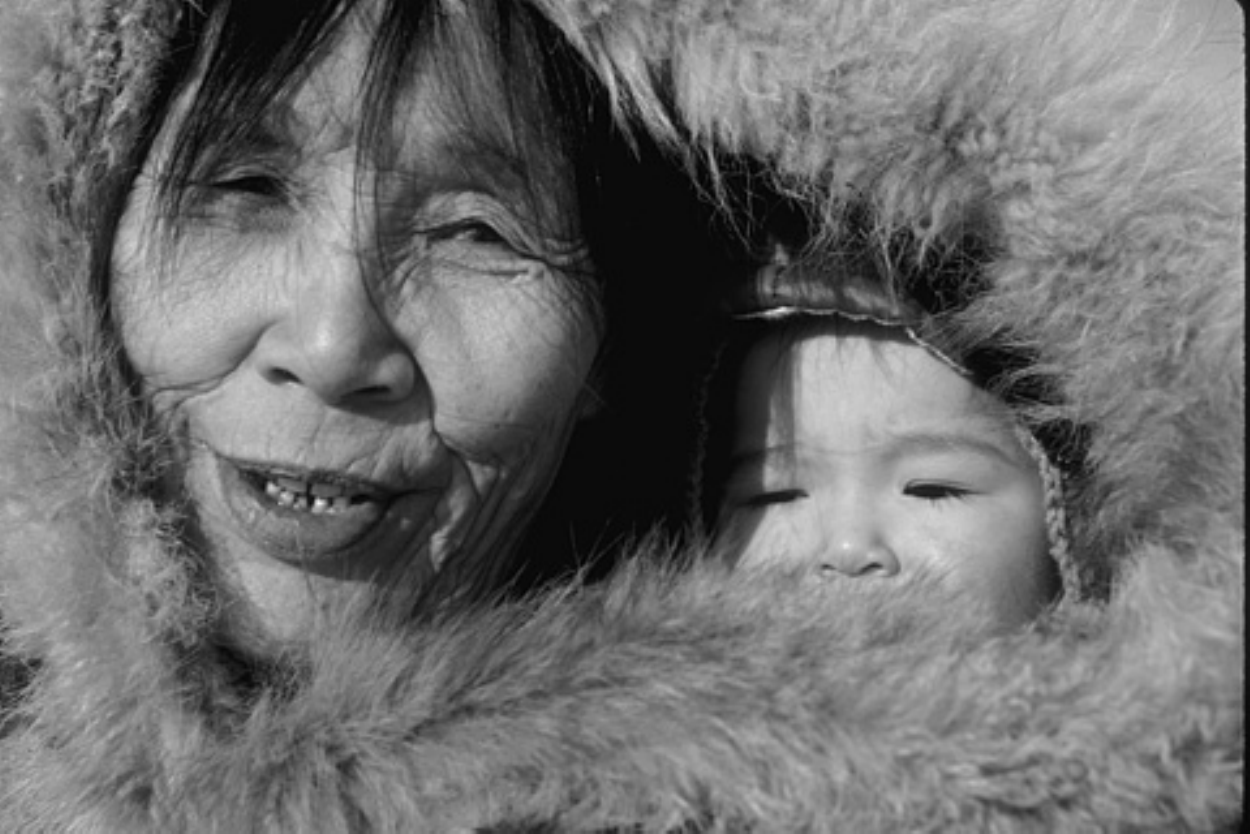}
                \caption{}
		\label{fig:frRef1}
        \end{subfigure}
		\quad
        \begin{subfigure}[t]{0.17\textwidth}
                \centering
                \includegraphics[width=\textwidth]{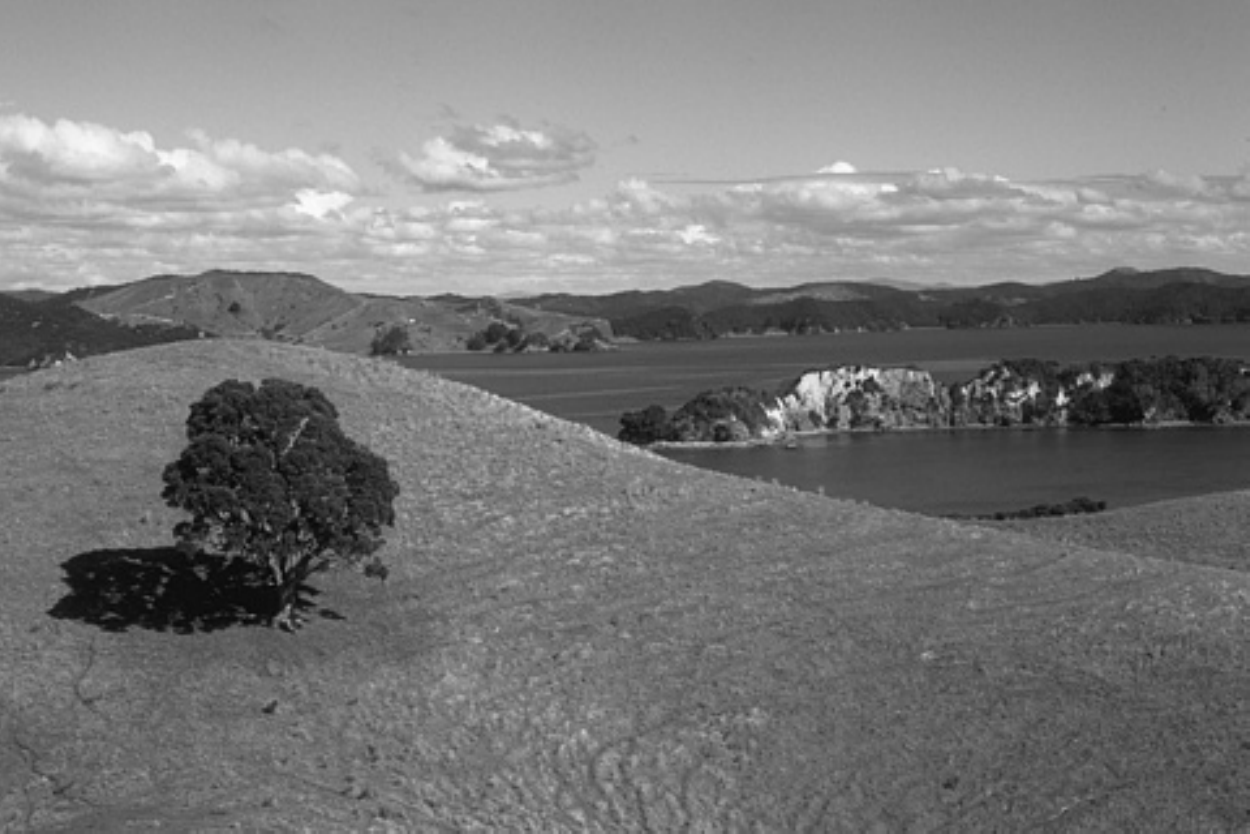}
                \caption{}
		\label{fig:frRef2}
        \end{subfigure}
		\quad
        \begin{subfigure}[t]{0.17\textwidth}
                \centering
                \includegraphics[width=\textwidth]{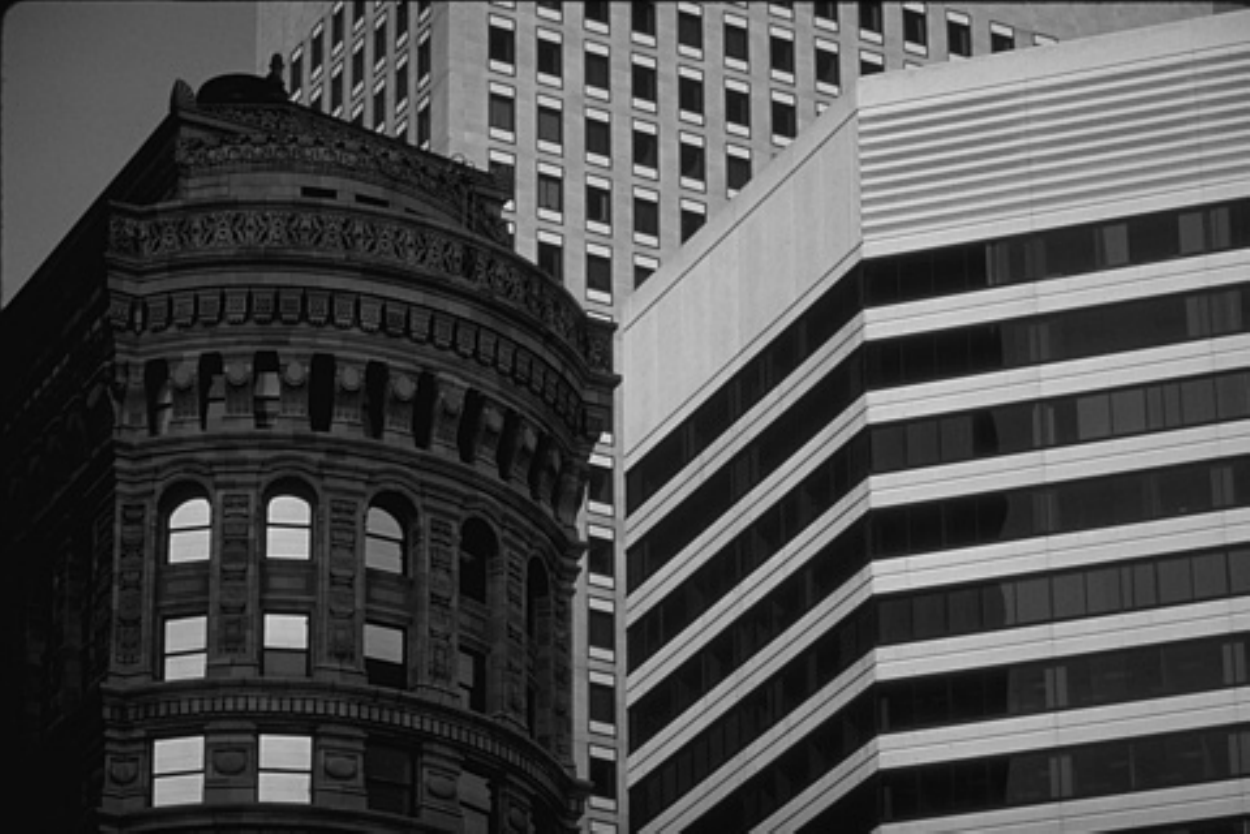}
                \caption{}
		\label{fig:frRef3}
        \end{subfigure}
		\quad
        \begin{subfigure}[t]{0.17\textwidth}
                \centering
                \includegraphics[width=\textwidth]{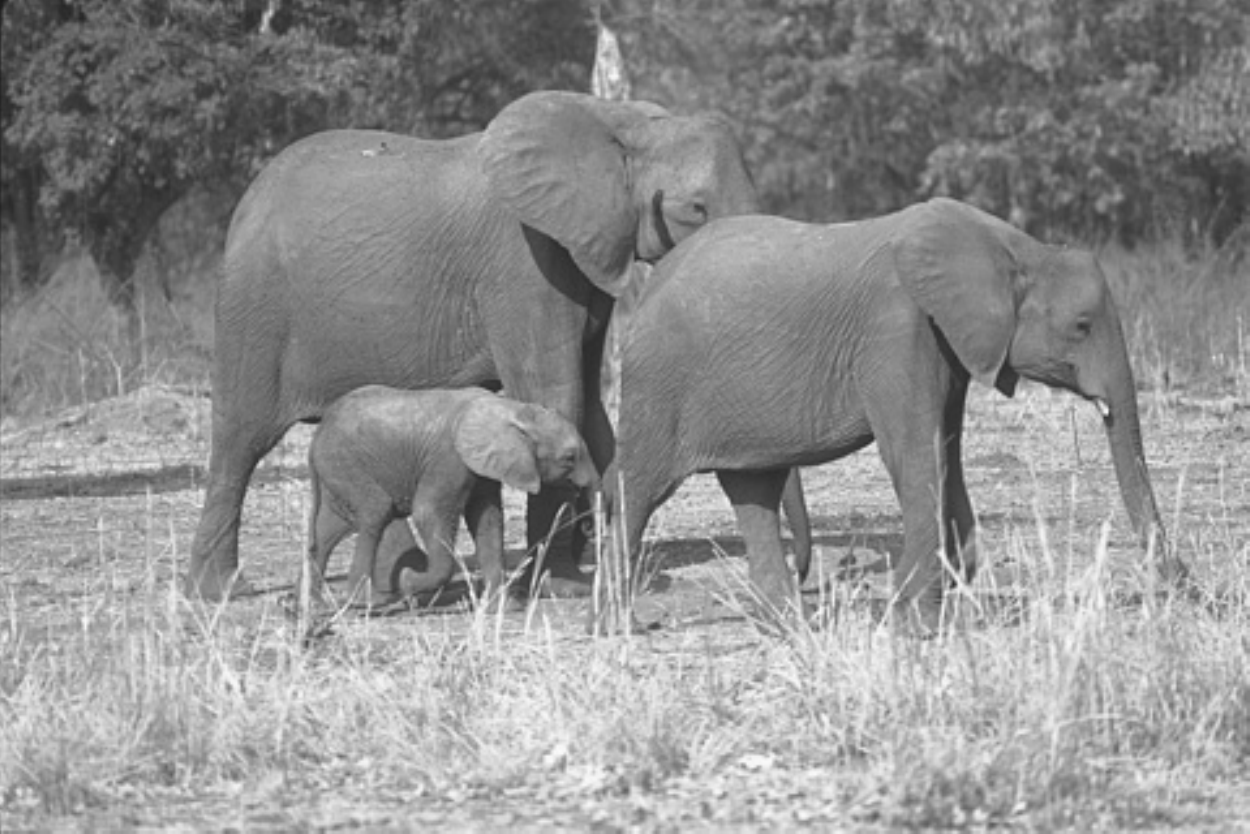}
                \caption{}
		\label{fig:frRef4}
        \end{subfigure}
		\quad
        \begin{subfigure}[t]{0.17\textwidth}
                \centering
                \includegraphics[width=\textwidth]{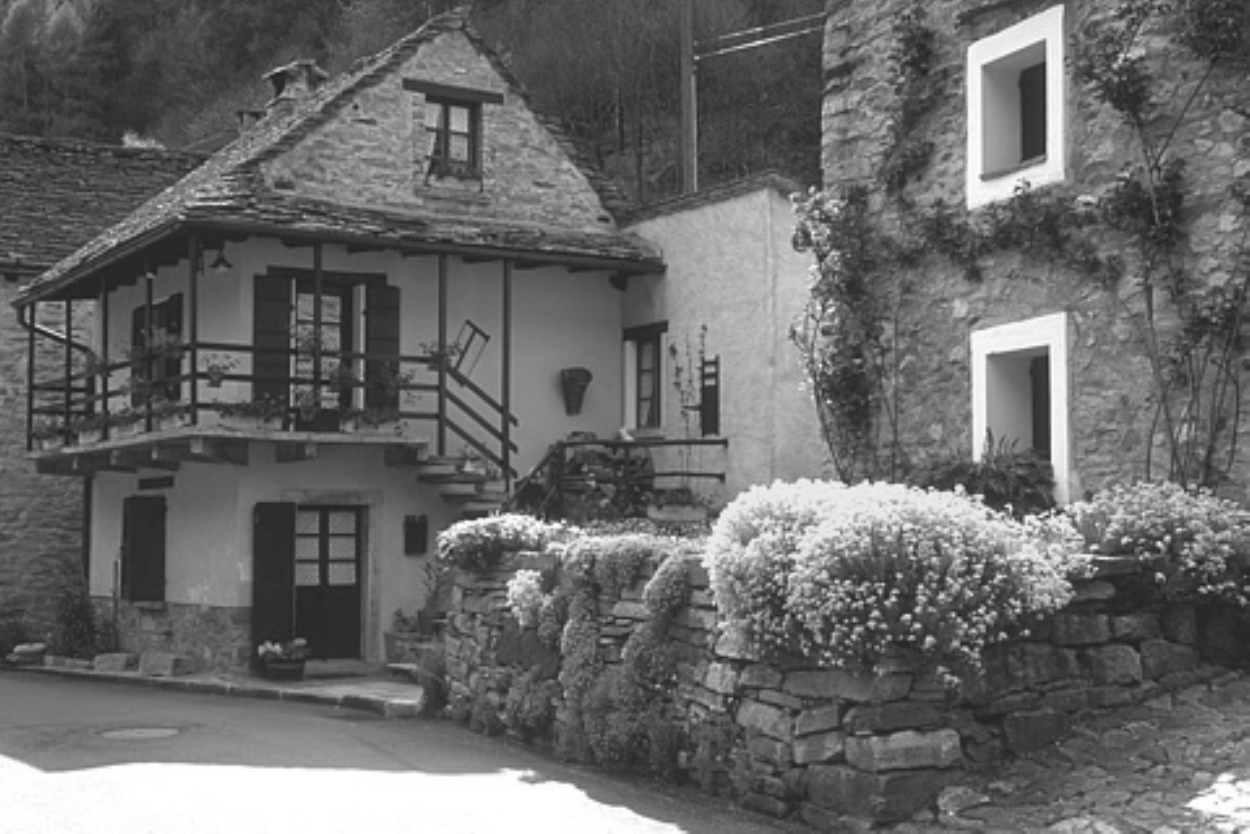}
                \caption{}
		\label{fig:frRef5}
        \end{subfigure}
\\
        \begin{subfigure}[t]{0.17\textwidth}
                \centering
                \includegraphics[width=\textwidth]{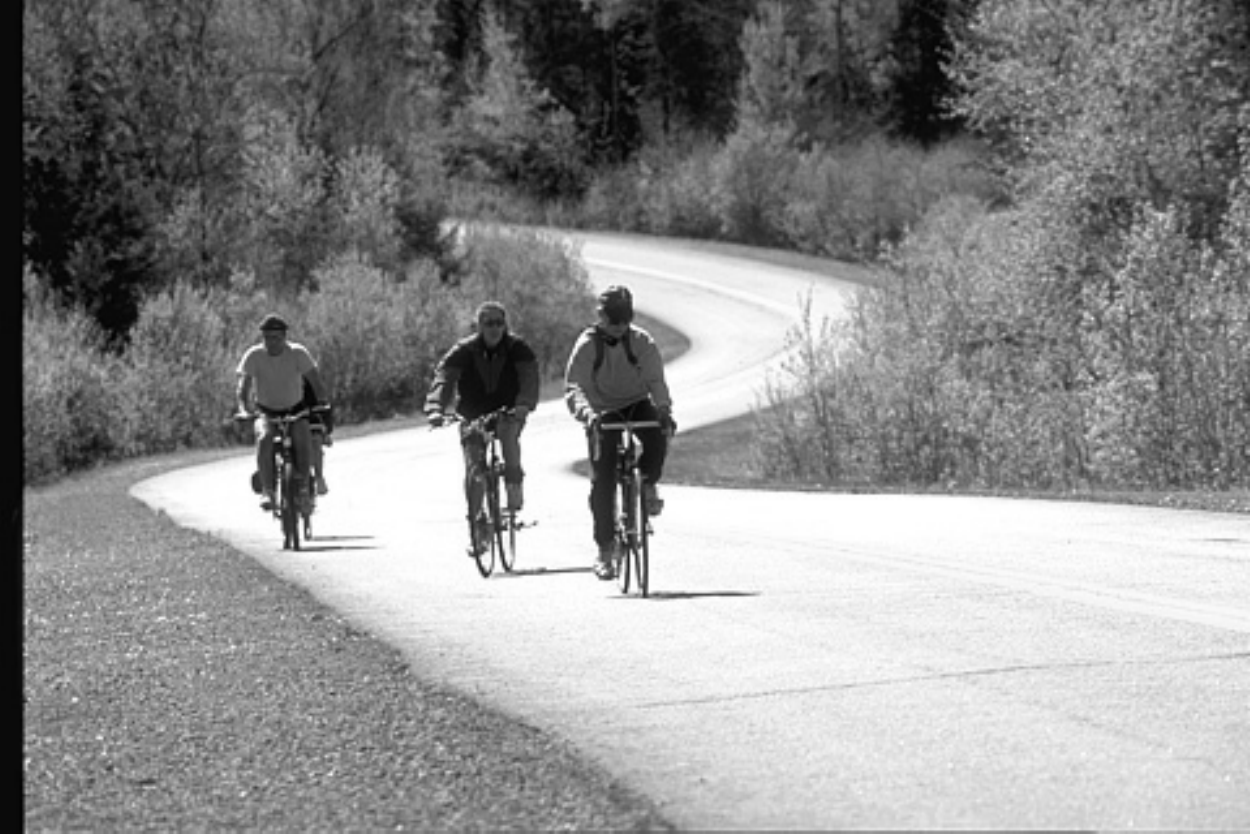}
                \caption{}
		\label{fig:frRef6}
        \end{subfigure}
		\quad
        \begin{subfigure}[t]{0.17\textwidth}
                \centering
                \includegraphics[width=\textwidth]{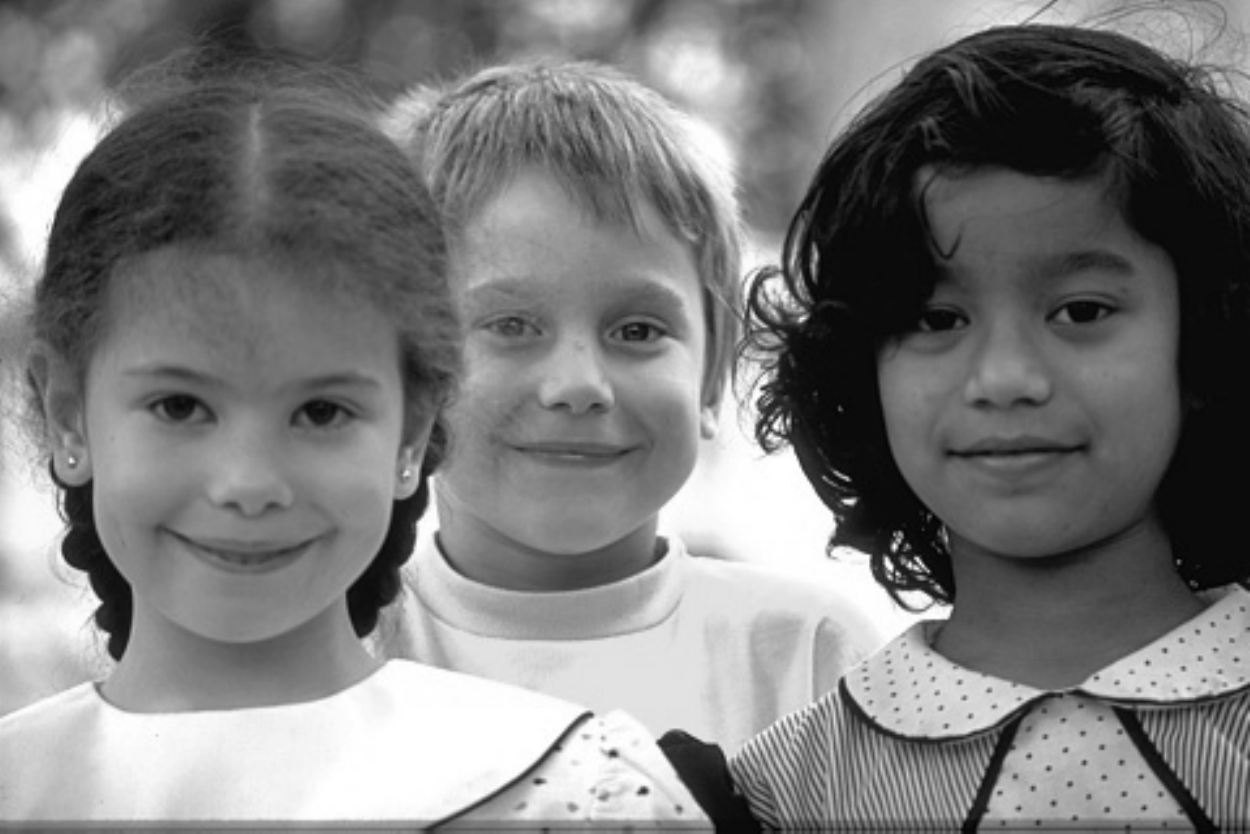}
                \caption{}
		\label{fig:frRef7}
        \end{subfigure}
		\quad
        \begin{subfigure}[t]{0.12\textwidth}
                \centering
                \includegraphics[width=\textwidth]{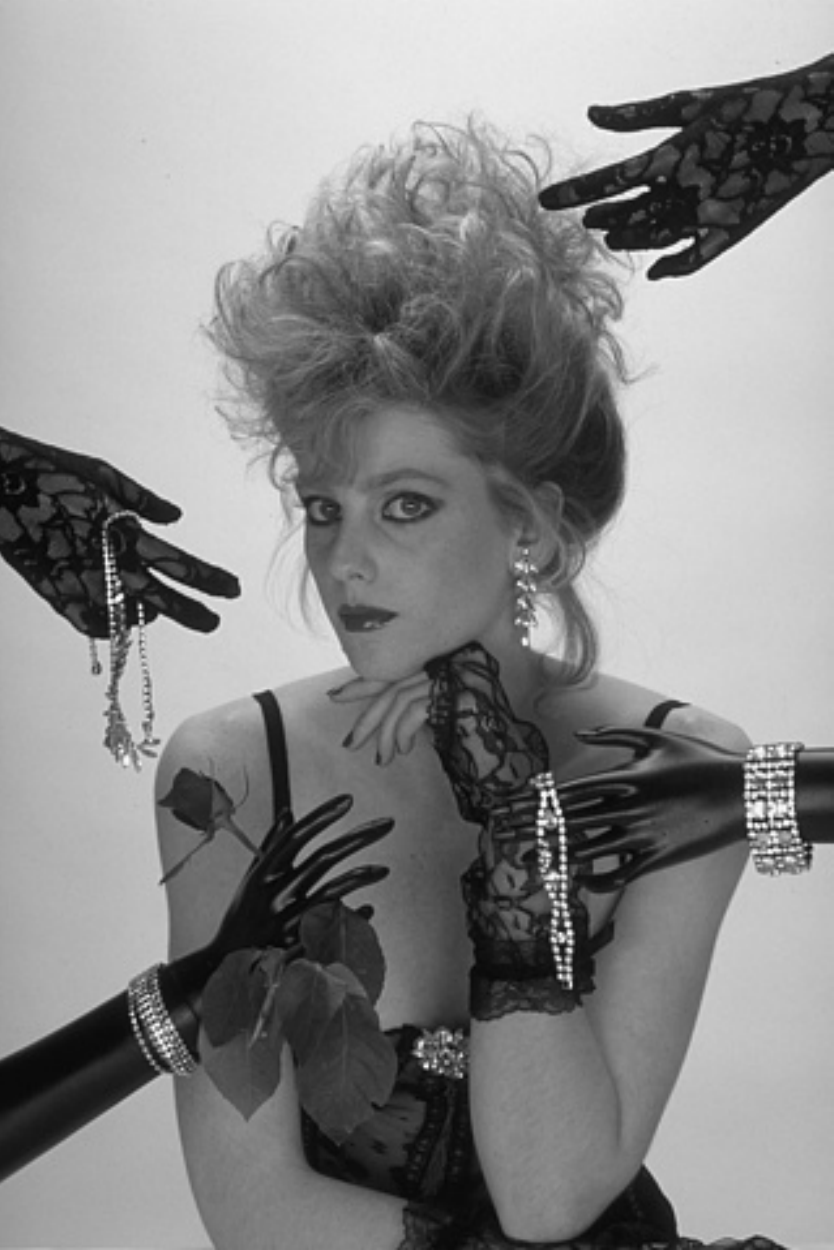}
                \caption{}
		\label{fig:frRef8}
        \end{subfigure}
		\quad
        \begin{subfigure}[t]{0.12\textwidth}
                \centering
                \includegraphics[width=\textwidth]{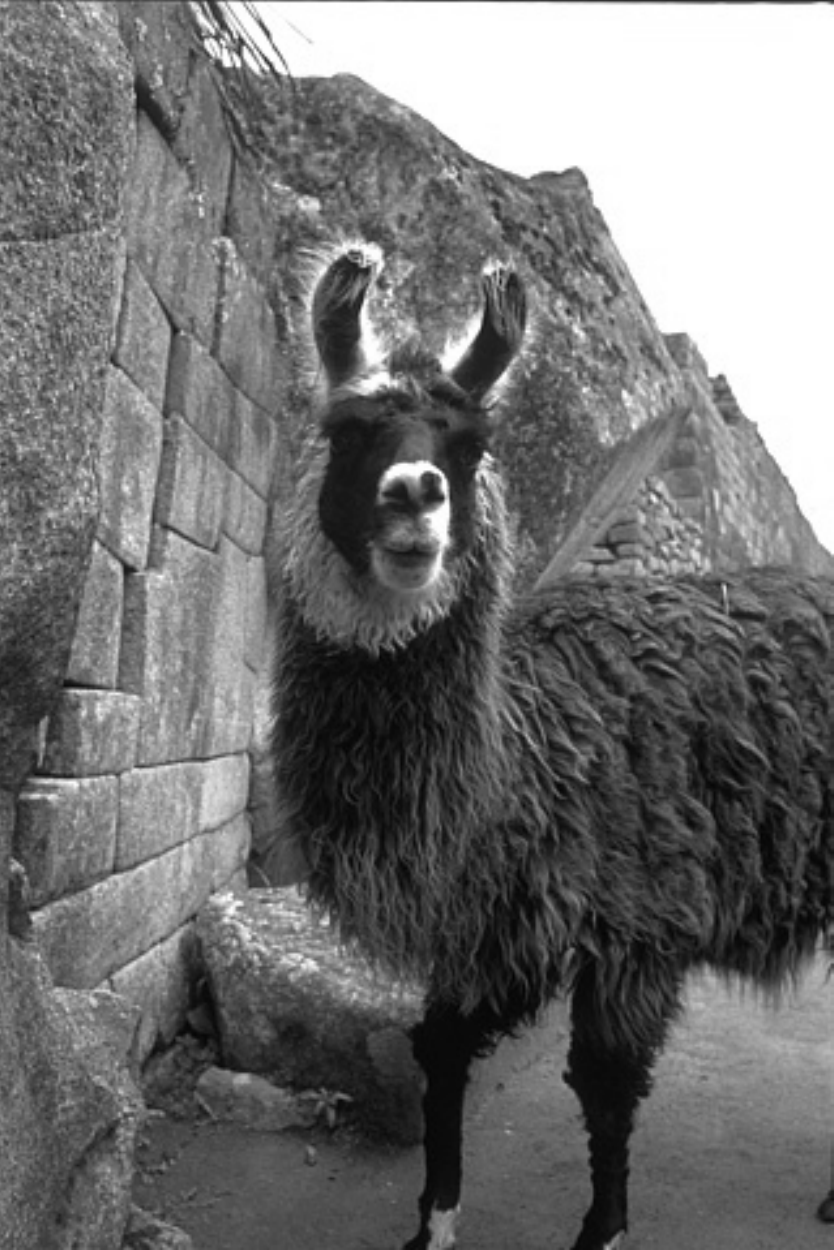}
                \caption{}
		\label{fig:frRef9}
        \end{subfigure}
\\
        \begin{subfigure}[t]{0.13\textwidth}
                \centering
                \includegraphics[width=\textwidth]{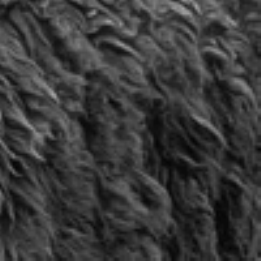}
                \caption{}
		\label{fig:randomRef1}
        \end{subfigure}
		\quad
        \begin{subfigure}[t]{0.13\textwidth}
                \centering
                \includegraphics[width=\textwidth]{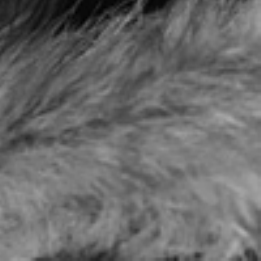}
                \caption{}
		\label{fig:randomRef2}
        \end{subfigure}
		\quad
        \begin{subfigure}[t]{0.13\textwidth}
                \centering
                \includegraphics[width=\textwidth]{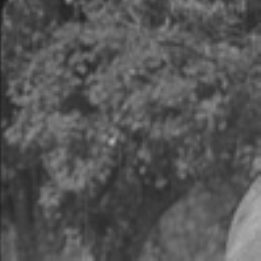}
                \caption{}
		\label{fig:randomRef3}
        \end{subfigure}
		\quad
        \begin{subfigure}[t]{0.13\textwidth}
                \centering
                \includegraphics[width=\textwidth]{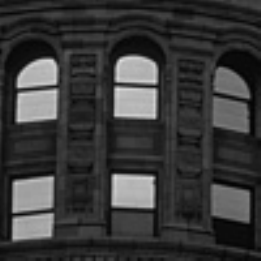}
                \caption{}
		\label{fig:regularRef1}
        \end{subfigure}        
        		\quad
        \begin{subfigure}[t]{0.13\textwidth}
                \centering
                \includegraphics[width=\textwidth]{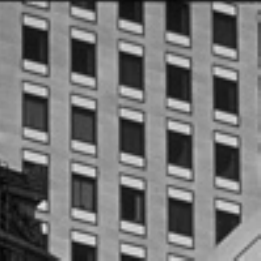}
                \caption{}
		\label{fig:regularRef2}
        \end{subfigure}        
	  \quad
        \begin{subfigure}[t]{0.13\textwidth}
                \centering
                \includegraphics[width=\textwidth]{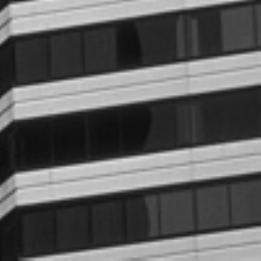}
                \caption{}
		\label{fig:regularRef3}
        \end{subfigure}        
\\
        \begin{subfigure}[t]{0.13\textwidth}
                \centering
                \includegraphics[width=\textwidth]{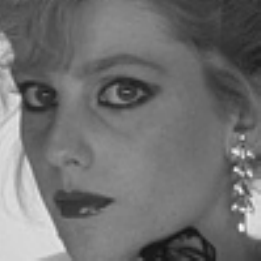}
                \caption{}
		\label{fig:hlRef1}
        \end{subfigure}
		\quad
        \begin{subfigure}[t]{0.13\textwidth}
                \centering
                \includegraphics[width=\textwidth]{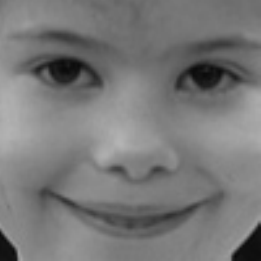}
                \caption{}
		\label{fig:hlRef2}
        \end{subfigure}        
        		\quad
        \begin{subfigure}[t]{0.13\textwidth}
                \centering
                \includegraphics[width=\textwidth]{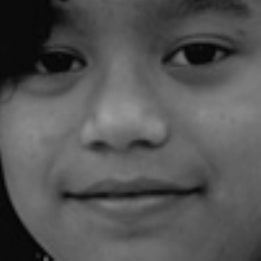}
                \caption{}
		\label{fig:hlRef3}
        \end{subfigure}        
\caption{The reference images of the generative image database: (a)-(i) subset of full-resolution 
images, (j)-(l) subset of random structured patches, (m)-(o) subset of regular structured patches, 
(p)-(r) subset of high-level structured patches. The images and patches are not shown at 
relative scale.}
\label{fig:generativeDatabase}
  \vspace{-3mm}
\end{figure*}
In~\cite{cit:Guo2017}, the authors proposed an one-to-many GAN to remove  
artifacts from JPEG-coded images. We adopted their network with some modifications
to build our generative image database. Fig.~\ref{fig:onetomany} shows the architecture 
of~\cite{cit:Guo2017}, where the image $Y$ is a JPEG compressed version of 
ground truth image $X$, and where random unit normal white Gaussian  image 
$Z\sim N(0,1)$ are used as the inputs to the CNNs. The outputs of the two branches 
are then concatenated into a matrix $\hat{X}$, and fed into the following network,
which is trained to choose the highest quality result, in the sense of both fidelity 
and naturalness. The objective function that is used to train the network has 
three terms:
\begin{equation} \label{eq:costFunc1}
\resizebox{ 0.9\hsize}{!}
{
$L({\hat{X}, X, Y}) = L_{percept}({\hat{X}, X}) + \lambda_1 \cdot L_{similar}({\hat{X}})
                                  + \lambda_2 \cdot L_{jpg}({\hat{X}, Y})$
}
\end{equation}
where $L_{percept}$ is the perceptual loss incurred when estimating the similarity in structure. 
$L_{similar}$ is an adversarial loss term, while $L_{jpg}$ measures color distortion. 
We depart from \cite{cit:Guo2017} by removing $L_{jpg}$ from (\ref{eq:costFunc1}),
hence our cost function becomes: 
\begin{equation} \label{eq:costFunc2}
\resizebox{ 0.8\hsize}{!}
{
$L({\hat{X}, X}) = L_{percept}({\hat{X}, X}) + \lambda \cdot L_{similar}({\hat{X}})$
}
\end{equation}
where $L_{percept} = \frac{1}{H_{\phi}}|\phi(\hat{X})-\phi(X)|^2$,
and $\phi$ are the activations of the last convolutional layer of VGG-16~\cite{cit:Simonyan2014}
while $H_{\phi}$ is the size of $\phi$. $L_{similar}$ is defined as $-log(D(\hat{X}))$,
where $D$ is an additional network that distinguishes whether an image is from the 
network or is like the original image. To train the network $D$, a binary entropy loss is used
as optimization cost: $L_D(X, \hat{X})=-(log(D(X)+log(1-D(\hat{X}))$. 
Unlike \cite{cit:Guo2017}, we employed the Microsoft COCO dataset~\cite{cit:COCO}
to train the network, and the truncated normal initializer was used to initialize the weights.\\
\begin{figure*}[t]
\centering
\includegraphics[width=0.89\textwidth]{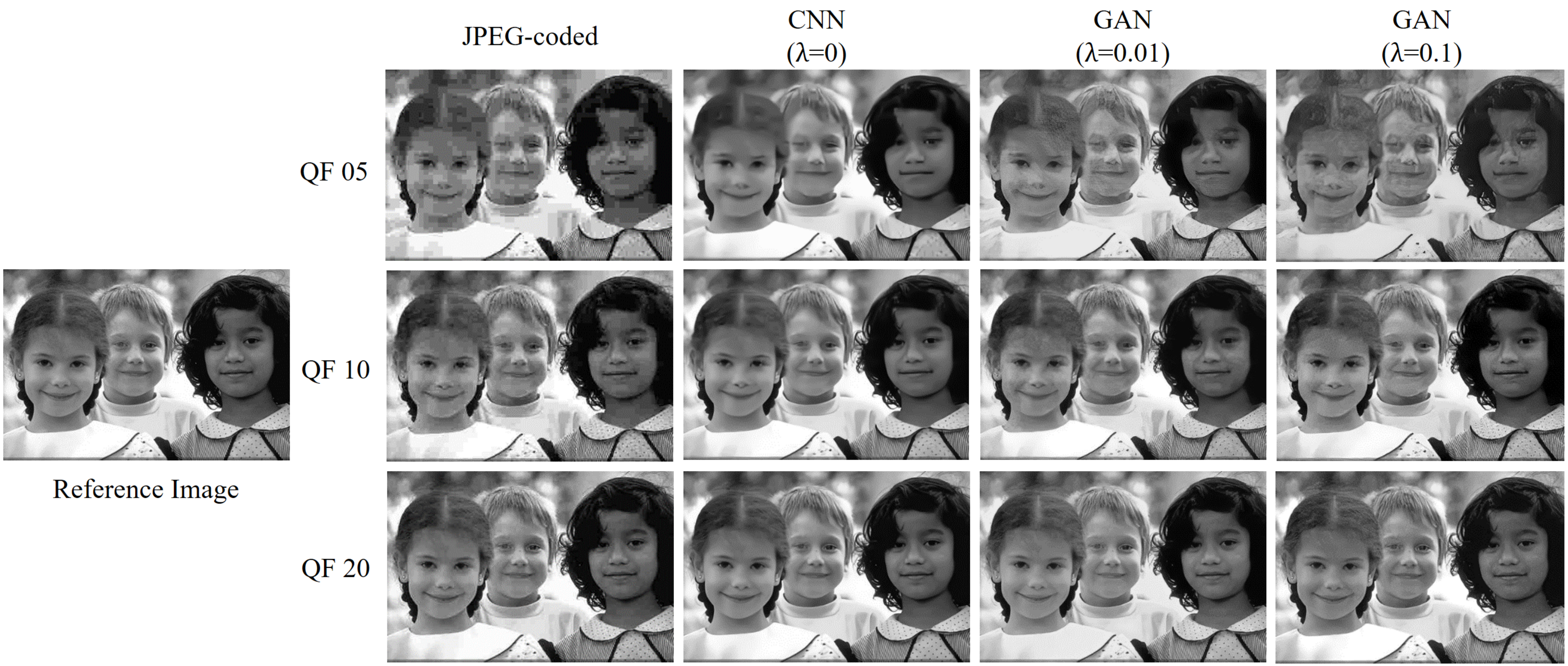}
\caption{Examples of generated test images (first column: JPEG-coded images, second column:
	CNN images, third column: GAN images with $\lambda$=0.01, fourth column: GAN images with 
	$\lambda$=0.1)}
\label{fig:testImgEx}
  \vspace{-3mm}
\end{figure*}
\indent To create the generative image database, we chose 9 reference images of resolutions 480x320 
or 320x480 from the BSDS500 dataset~\cite{cit:Arbelaez2011}, which contains a wide variety of
natural scene characteristics. We also created three sets of patch images of resolution 100x100 
cropped from the full-resolution reference images: 1) a randomly structured patch set including 
three textured patches of a lawn, a furry llama and a piece of cloth, 2) a regular, structured patch 
set containing three repetitive patterned patches from the reference image of buildings, and 
3) a high-level structured patch set containing human faces. Overall, the database contains 18 
reference images, all of which are shown in Fig.~\ref{fig:generativeDatabase}.\\
\indent Each reference image was subjected to (matlab) JPEG compression using three different quality 
factors: QF5 (low quality), QF10 (moderate quality) and QF20 (high quality). Next, for each of the three 
compressed images, we used the network in Fig.~\ref{fig:onetomany} to generate two generative images
using weighting factor $\lambda=0.01$ and $\lambda=0.1$, respectively, in Eq.~(\ref{eq:costFunc2}). 
We have observed that as the value of $\lambda$ is increased, the resulting generative image becomes 
shaper and naturalistic, but becomes less natural if $\lambda$ becomes too large. We also generated 
an image using the same network, but replacing $L_{percept}$ with the MSE (mean-squared error) 
and setting $\lambda=0$, which we designate as the CNN-generated image. In this case, since there is 
no adversarial loss term and the maintenance of pixel-wise fidelity is the only objective cost, 
the resulting generative images tend to be blurred. To sum up, there are 12 test images associated
with each of the 12 reference images: 3 JPEG quality factors $\times$ \{1 JPEG-coded images +
1 CNN image + 2 GAN generative images\}. An example of a reference image and the 12 images
generated from it are shown in Fig.~\ref{fig:testImgEx}.

\subsection{Subjective Study}\label{sec:3.2}
We conducted a human study to obtain MOS (mean opinion scores) on the generative 
image database. In general, there are two kinds of subjective evaluation methods that are
widely used in IQA studies. While the ACR (absolute category rating) recommended by the 
international telecommunication union (ITU)~\cite{cit:BT500} for image/video quality 
assessment is most widely used, we chose to instead use the pairwise comparison (PC) 
method. We decided this because GAN-generated distortions are a relatively new phenomenon, 
and we wanted to be sure that subtleties of texture and detailed distortion could be better 
detected. The obtained human judgments were converted into numerical scores to provide 
subjective ground truth. During each session, when a reference image was displayed, 
a test image A and a test image B were displayed below it in random left-right order
together per each viewing. Three questions were then asked: 
\begin{itemize}
  \item (N) Which test image looks more natural? (For this question, 
  it is not necessary to compare each test image with the reference image.)
  \item (S) Which test image better preserves structural fidelity with respect to 
  the reference image? 
  \item (C) Which test image better preserves the concept of the contents with respect to 
  the reference image? 
\end{itemize}
As a result, three different MOS values were acquired on each pair of test images. 
As mentioned in Section~\ref{sec:related}, GAN generative images may appear
very natural while numerically differing from an original image in a pixel-wise comparison,
especially in highly textured regions. Conversely, dominant structures such as 
the shapes and boundaries of objects might be well maintained in the GAN images. 
The final question was directed towards the preservation of recognizability in the image.\\
\indent The outcome of a series of pairwise comparisons by a single human is an ordered list 
of images. However, applying the PC method using a round-robin design is very 
time-consuming. For instance, if there are $N$ samples, then the total number of 
possible pairwise comparisons is $C_2^N$. To solve the complexity issue, we instead
adopt the more efficient Swiss-rule design as used in~\cite{cit:Lin2015}. 
Once an assessor finishes a session, a preference matrix is created. 
By aggregating the preference matrices of all the assessors, a group preference matrix
is obtained, which includes the count of preferred images over all the possible pairs. 
If $P(i,j)$ is the $(i,j)^{th}$ element of the group preference matrix, then $P(i,j)$ is the 
total number of times that the assessor preferred image $i$ to image $j$.
The averaged count data is then used as a form of MOS. There are also two well-known 
ways to convert pairwise comparison data into psychophysical scores: the Bradley-Terry 
(BT)~\cite{cit:Bradley1952} model, and the Thurstone-Mosteller (TM) model~\cite{cit:Handley2001}. 
We found that the averaged count data and the converted scores to be highly 
correlated with each other (Pearson correlation coefficient value = 0.984),
hence, we opted for simplicity and used the counting data. \\
\indent The experiment was conducted using an LG 65 inch UHDTV, and a graphical user 
interface (GUI) implemented in Matlab. In each presentation, a reference image was 
shown at the top of the screen and a randomly selected pair of test images was shown
below. Each assessor was given three choices: `the left one is better', `the right one is
better', or `no difference'. Further, the assessors were asked to answer the
abovementioned three questions. 
We divided each overall session into two sub-sessions, each of duration ranging 
from 20 to 30 min, to meet the recommendation of ITU~\cite{cit:BT500} that the 
duration of each session should not exceed 30 min to avoid subject fatigue. 
24 assessors participated in total, 
and the overall session duration and each decision made by every assessor was 
recorded. There were nineteen males and five females. Eight of them were 
in their 20s and the remining 16 were in their 30s. We also filtered abnormal results 
according to~\cite{cit:BT500}. Finally, we collected 20 
opinion scores on each test image. \\
\begin{figure}[t]
	\centering
	\begin{subfigure}[h!]{0.42\textwidth}
		\centering
		\includegraphics[width=1.8in]{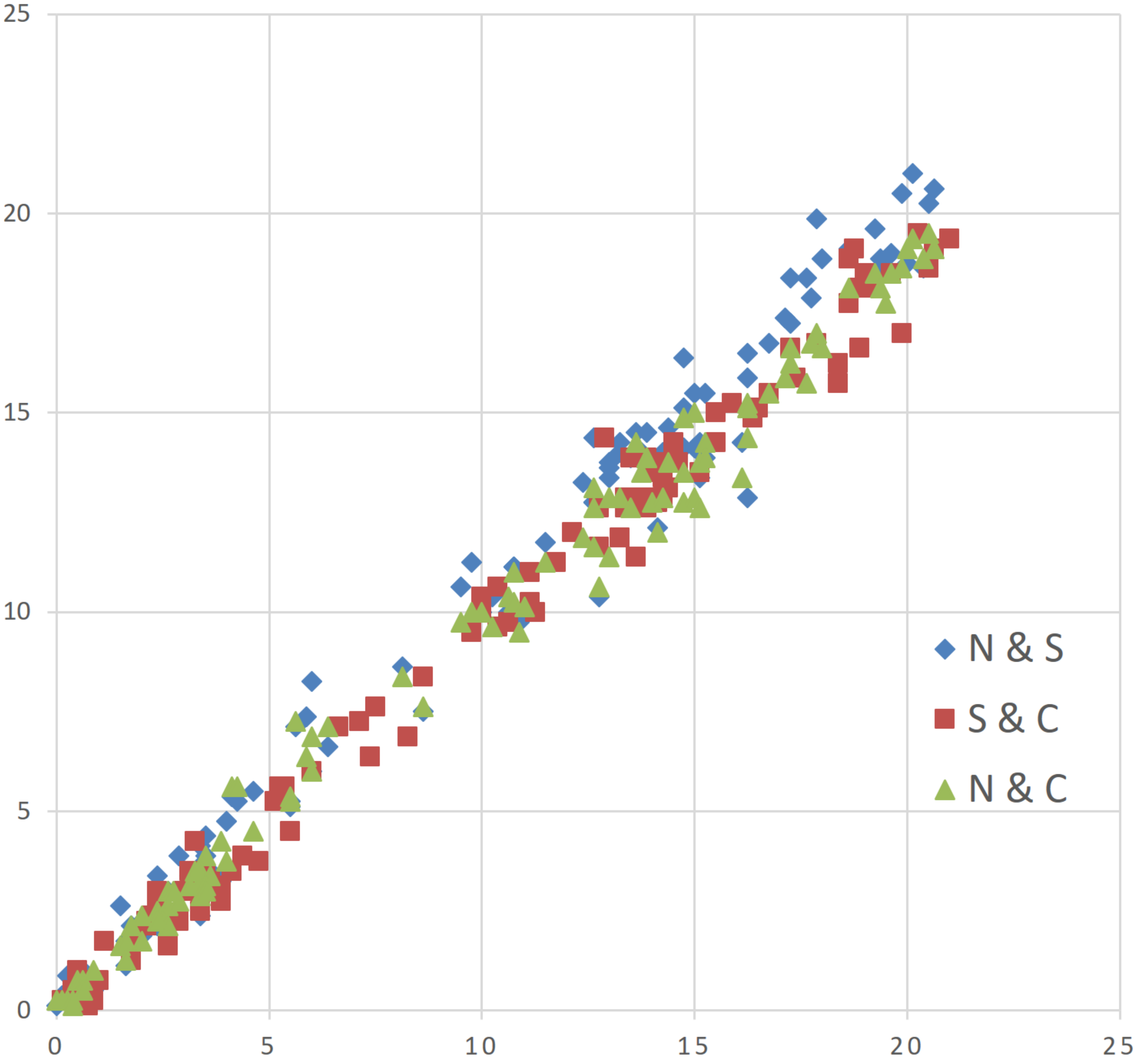} 
		\caption{Full-resolution images.}
		\label{fig:NSC1}
	\end{subfigure}
\\
	\vspace{0.1in}
	\begin{subfigure}[h!]{0.42\textwidth}
		\centering
		\includegraphics[width=1.8in]{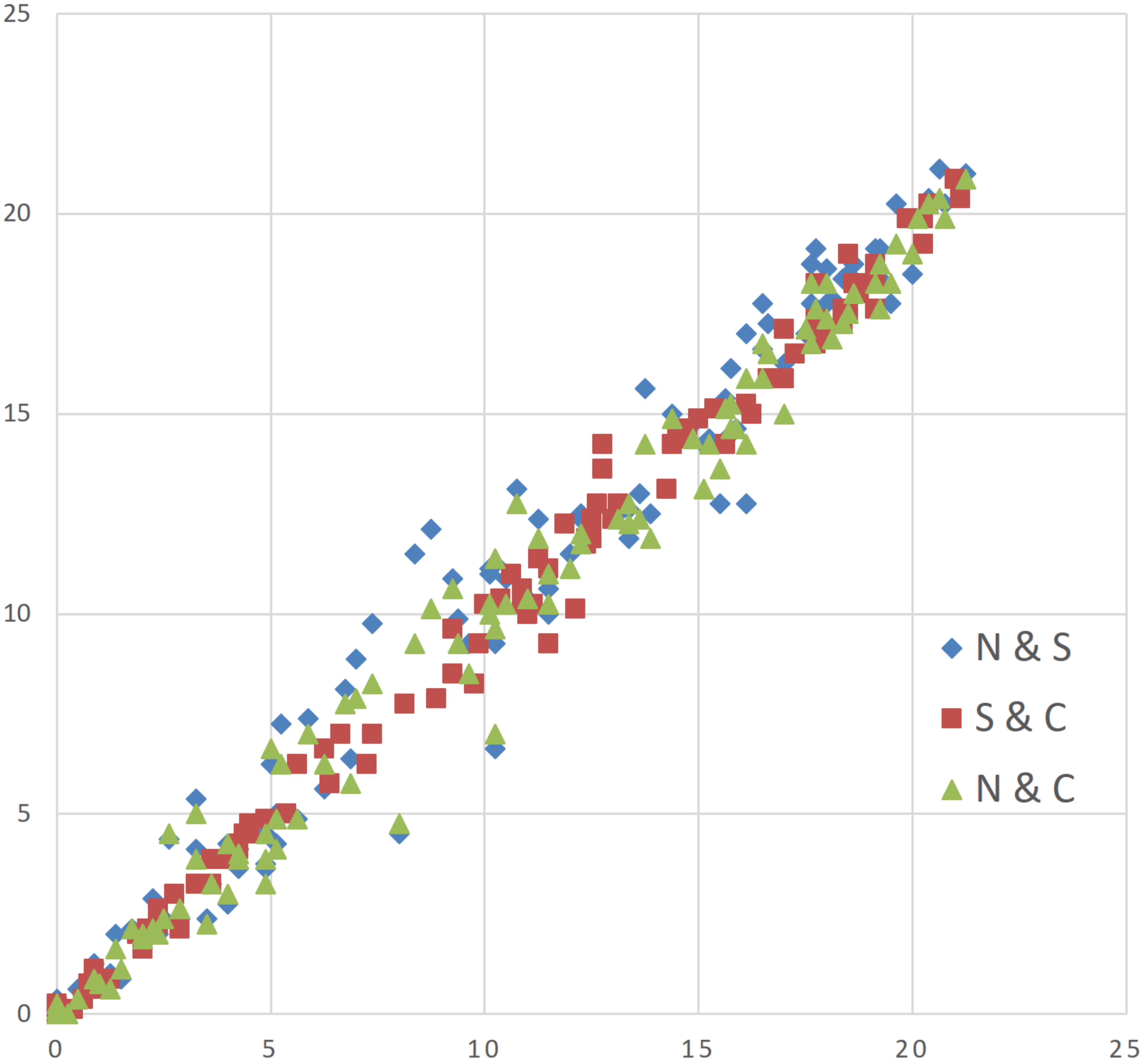} 
		\caption{Patch images.}
		\label{fig:NSC2}
	\end{subfigure}
	\caption{Scatter plots among MOS of naturalness, structural fidelity, and concept 
	preservation, where each dot represents one test image.}
	\label{fig:NSC}
 \vspace{-3mm}
\end{figure}
\begin{table*}[t]
  \centering
  \caption{Summary of the generative database and the subjective test.}
    \begin{tabular}{cc|llll}
    \toprule
          &       & \multicolumn{1}{c}{\multirow{2}[2]{*}{\makecell{Subset of \\Full-Resolution Images}}} & \multicolumn{3}{c}{Subsets of Image Patches} \\
          &       &       & \multicolumn{1}{c}{Random Structure} & \multicolumn{1}{c}{Regular Structure} & \multicolumn{1}{c}{High-level Structured} \\
    \midrule
    \multicolumn{1}{c}{\multirow{6}[2]{*}{\makecell{Generative \\Database}}} & \multicolumn{1}{p{10em}|}{\makecell{\# Reference images \\selected from BSDS500 \\(Resolution)}} & \multicolumn{1}{p{10em}}{\makecell{9 \\(480x320 or 320x480)}} & \multicolumn{1}{p{10em}}{\makecell{3 \\(100x100)}} & \multicolumn{1}{p{10em}}{\makecell{3 \\(100x100)}} & \multicolumn{1}{p{10em}}{\makecell{3 \\(100x100)}} \\
          & \multirow{5}[0]{*}{Test image design} & \multicolumn{4}{c}{Per each reference image:} \\
          &       & \multicolumn{4}{l}{[1] Three JPEG-coded images with QF5(low quality), QF10(moderate quality), and QF20(high quality)} \\
          &       & \multicolumn{4}{l}{[2] For each JPEG-coded image, two GAN images were generated using $\lambda=0.1$ and $\lambda=0.01$ in (\ref{eq:costFunc2}).} 
\\
          &       & \multicolumn{4}{l}{[3] For each JPEG-coded image, one CNN image was obtained using $\lambda=0$ and $L_{percept}$ = MSE} \\
          & \# of test images & \multicolumn{4}{c}{Total: 18 ref. images x {3 QF x (1 JPEG-coded + 1 CNN + 2 GAN img)} = 216 images} \\
    \midrule
    \multicolumn{1}{c}{\multirow{5}[2]{*}{\makecell{Subjective \\Test}}} & Study Methodology & \multicolumn{4}{c}{Pairwise Comparisons} \\
          & \# of participants & \multicolumn{4}{c}{20} \\
          & \multirow{3}[1]{*}{\makecell{Three independent \\mean subject scores}} & \multicolumn{4}{l}{[1] Naturalness (used in the experiments)} \\
          &       & \multicolumn{4}{l}{[2] Structural fidelity} \\
          &       & \multicolumn{4}{l}{[3] Concept preservation} \\
    \bottomrule
    \end{tabular}%
  \label{tab:databaseTest}%
\end{table*}%
\indent Figs.~\ref{fig:NSC1} and \ref{fig:NSC2} show the scatter plots among the subject 
responses with respect to naturalness (N), structural fidelity (S), and concept preservation 
(C), respectively. 
For example, the x- \& y-axes represent subjective scores, where each blue
dot represents one test image.
The PCC values between all six pairs of (N, S, C) exceeded 0.99, 
implying that N, S, and C are perceptually tightly coupled.
Hence, hereafter we use only utilize N as a mean subjective score in all the following
experiments. 
It is also noteworthy that 94\% of the assessors preferred the GAN-generative 
images to the CNN-generated images when we asked them about the  degree of naturalness 
of the test images, which supports our assumption that GANs are able to generate 
more naturalistic output images.
A summary of our generative database and the subjective test 
is given in Table~\ref{tab:databaseTest}

\section{Proposed Naturalness Assessment Metric}\label{sec:metric}
\begin{figure*}
        \centering
        \begin{subfigure}[t]{0.18\textwidth}
                \centering
                \includegraphics[width=\textwidth]{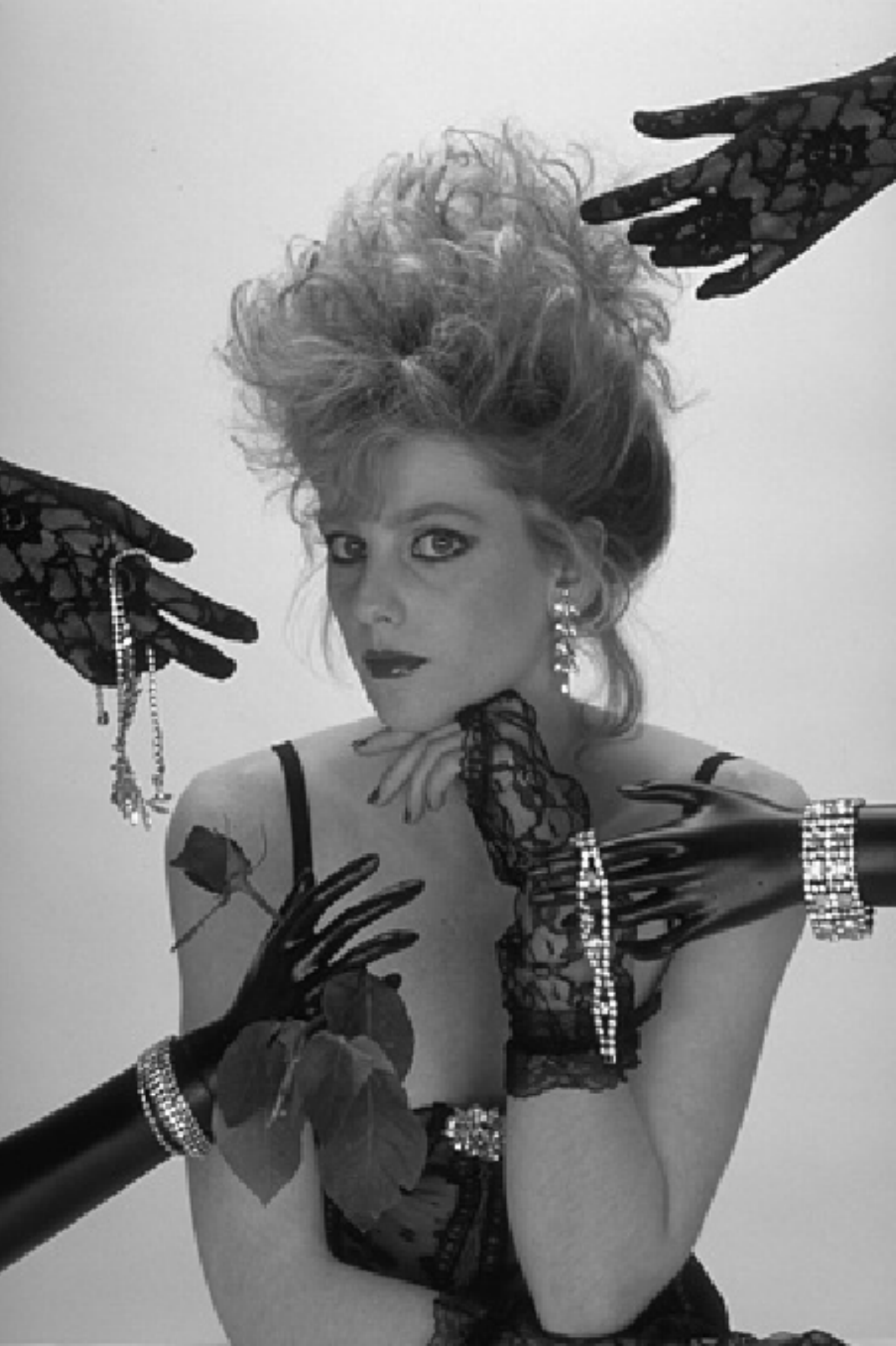}
                \caption{}
		\label{fig:EigenImg1}
        \end{subfigure}
		\quad \quad 
        \begin{subfigure}[t]{0.18\textwidth}
                \centering
                \includegraphics[width=\textwidth]{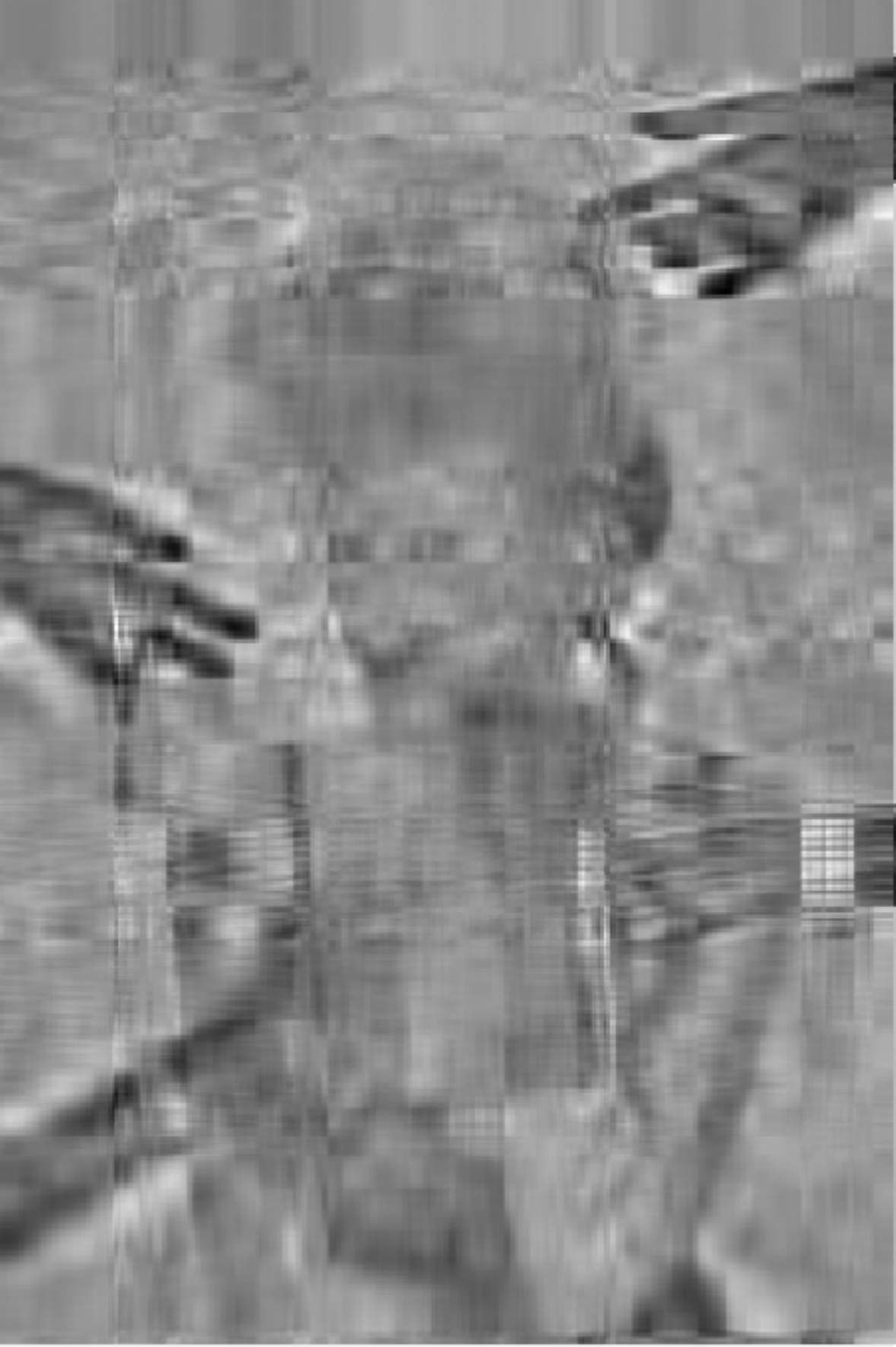}
                \caption{}
		\label{fig:EigenImg2}
        \end{subfigure}
		\quad \quad 
        \begin{subfigure}[t]{0.18\textwidth}
                \centering
                \includegraphics[width=\textwidth]{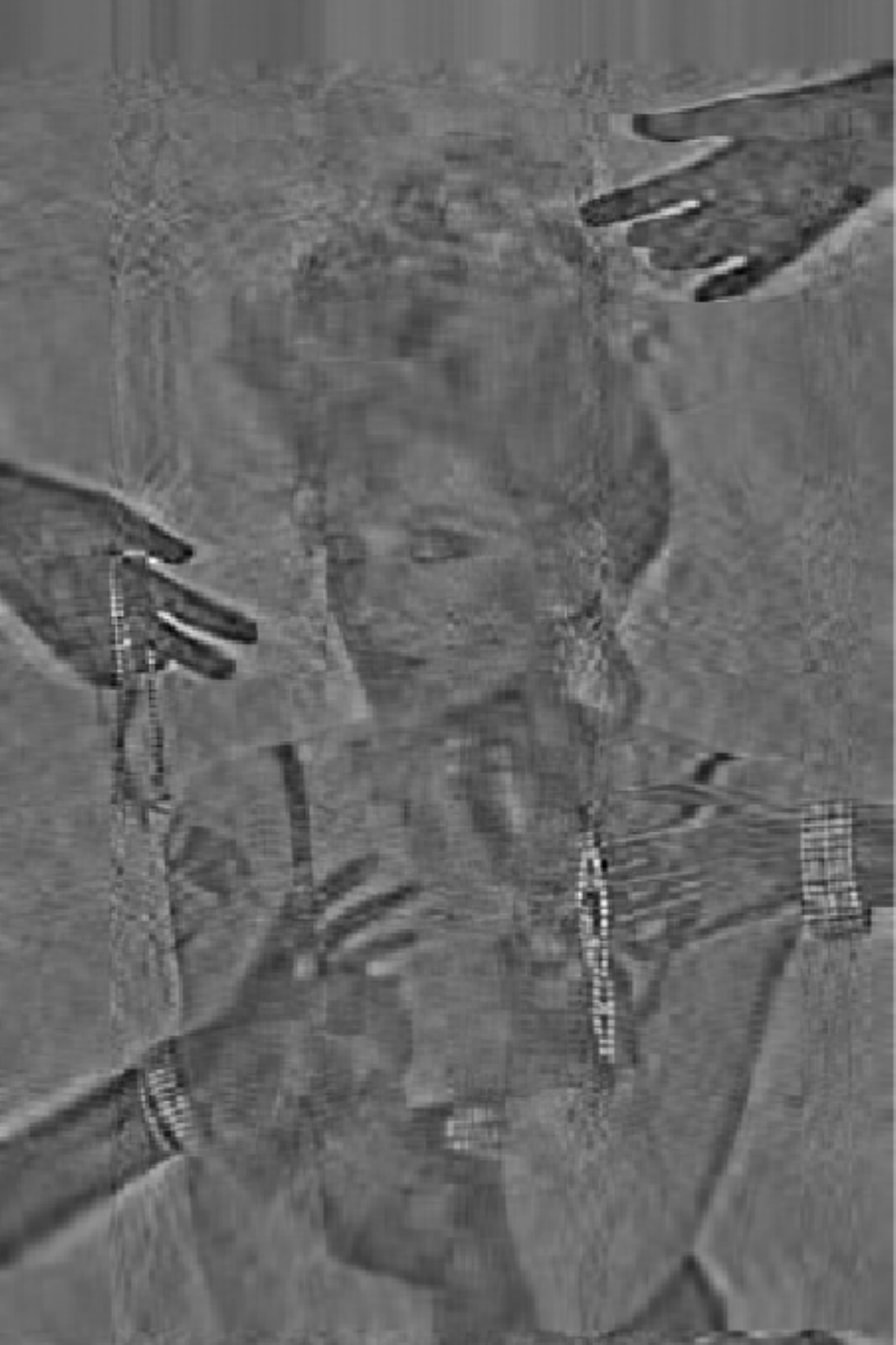}
                \caption{}
		\label{fig:EigenImg3}
        \end{subfigure}
        		\quad \quad 
        \begin{subfigure}[t]{0.18\textwidth}
                \centering
                \includegraphics[width=\textwidth]{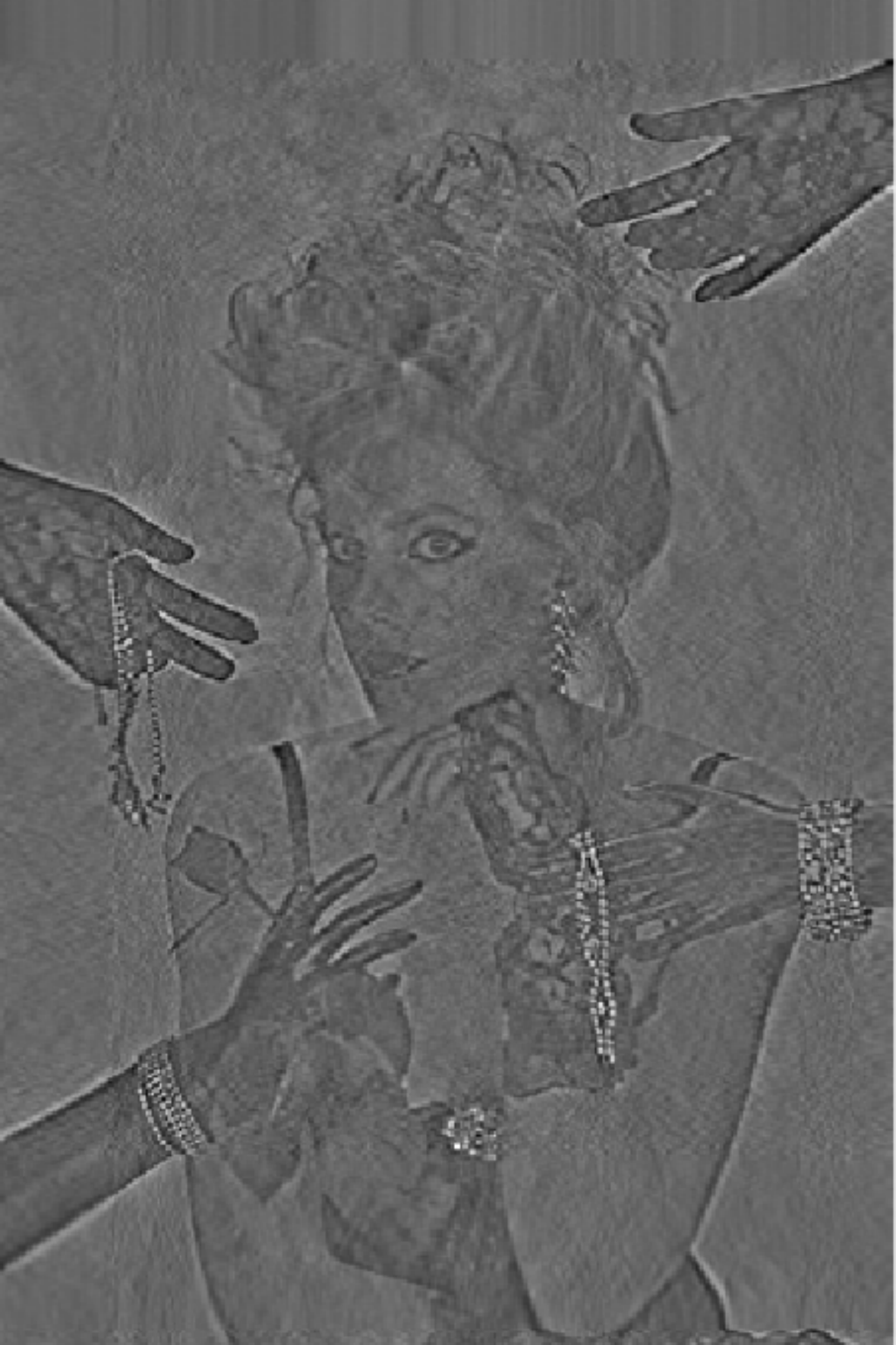}
                \caption{}
		\label{fig:EigenImg4}
        \end{subfigure}
\caption{Ensemble images of Eq.~\ref{eq:eigenImg} with different $k$ values: (a) Original image,
(b) $k$=15, (c) $k$=40, (d) $k$=100 }
\label{fig:EigenImg}
    \vspace{-3mm}
\end{figure*}
To develop an automatic predictor of the quality of generative images, we devised two 
different groups of features. We will refer to these as singular value decomposition (SVD) 
features and histogram-distance features. The former is designed to capture the structural 
similarity between an original and a test images while the latter is intended to measure the
statistical similarity of the images. 

\subsection{SVD based Features for the Structural Similarity}\label{sec:4.1}
Commonly used 2-D image transforms such as the DFT or DCT decompose an image using
a fixed basis set. In principle, any structural degradation of a test image with respect to a 
reference image can be measured from changes in the transform coefficients.
However, since the basis functions of SVD are unique to each image, changes 
in a test image can be measured both on an image's basis set as well as the transform 
coefficients. For an $r \times c$ input image $X$, the SVD is defined as: 
\[ X =U \cdot \sigma \cdot V^T\]
where $U=[u_1, u_2,...,u_r]$ is an $r \times r$ left singular vector matrix, 
$V=[v_1, v_2,...,v_c]$ is a $c\times c$ right singular vector matrix, and 
$\sigma= diag( \sigma_1, \sigma_2,...,\sigma_t)$ is a diagonal matrix 
of singular values in descending order: 
$\sigma_1>\sigma_2, ...>\sigma_t$   ($t$=min\{$r$, $c$\}).\\
\indent Singular vectors and singular values contain useful information related to image structure 
and frequencies~\cite{cit:Narwaria2012}. Given an SVD basis $u_i \cdot v_i^T$, then 
an ensemble image of accumulated basis images may be formed: 
\begin{equation} 
\label{eq:eigenImg}
X_k = \sum_{i=1}^{k} u_i \cdot v_i^T
\end{equation}
where $k \leq t$. Each basis implies a single layer of image structure while the sum 
of all layers yields the complete image structure. The first few layers contain the 
large-scale image structures, while the subsequent layers contain successive finer
details in the image. An example is depicted in Fig.~\ref{fig:EigenImg}, portraying different 
ensemble images obtained with different $k$ values. When just the first few basis 
images are used ($k$=15), the large structures in the image begin to appear, 
while finer structural details emerge as the number $k$ of basis images is increased. 
The singular vectors, $u_i$ and $v_j$, capture the structural elements images,
and may embody distortion-induced changes in them.\\
\indent The singular values function weight their corresponding basis, and thereby
represent the degree of luminance variation, strong textures versus smoothness 
or weak textures. For example, the ratio of the largest to the second largest singular 
value was used to estimate texture degree in~\cite{cit:Eskicioglu2006}. 
Since different distortions may modify the luminance patterns of original images
characteristically, it should be possible to likewise represent them in the singular values.
In short, the singular vectors and the singular values allow the possibility of 
separately analyzing changes in structures and in luminance variation. \\
\begin{figure}[t]
\centering
\includegraphics[width=0.50\textwidth]{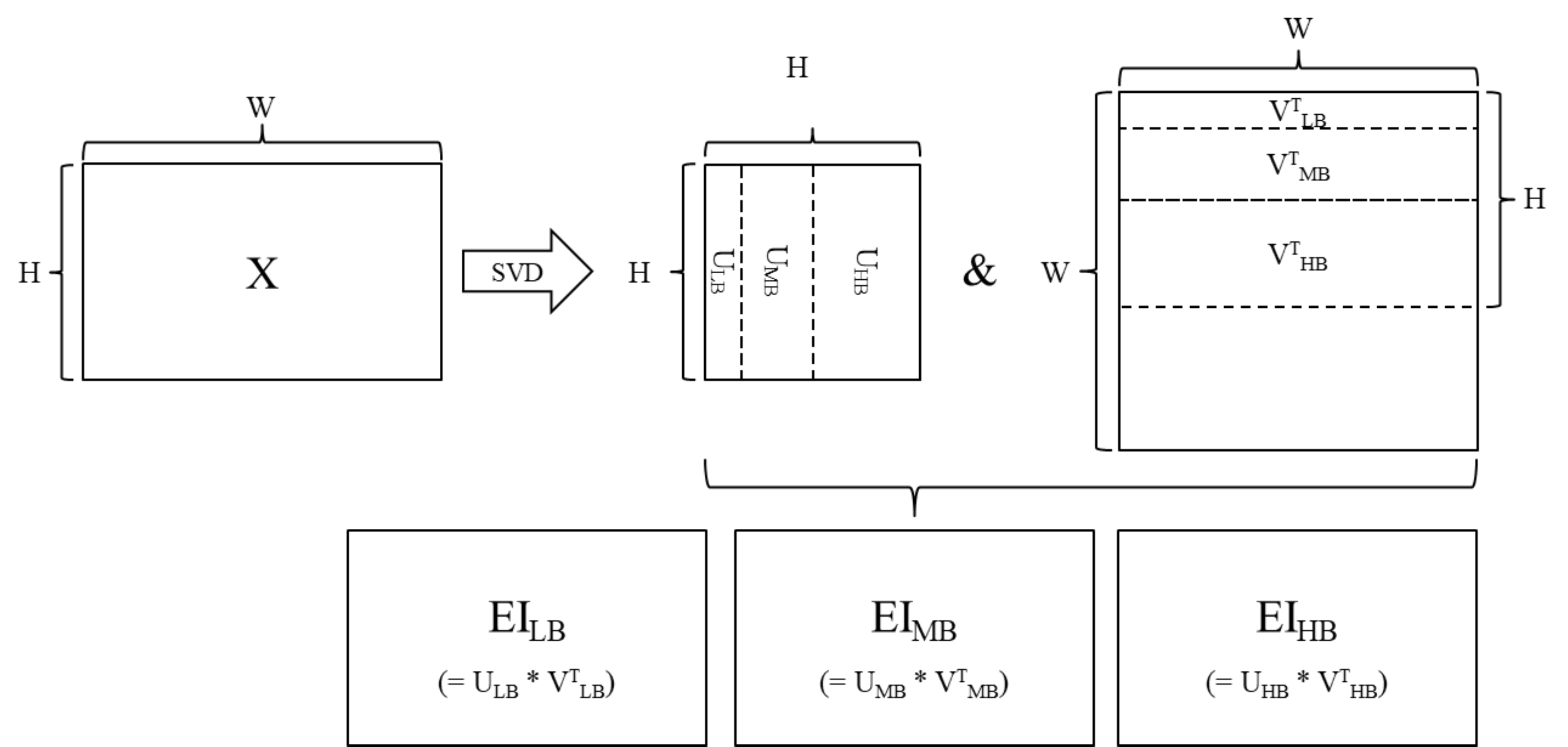}
\caption{SVD sub-band division used to create ensemble image bands ($H<W$).}
\label{fig:esbBand}
    \vspace{-3mm}
\end{figure}
\indent We utilize several SVD related features. Since each basis image contributes to an 
image's  structure/frequency content depending on the layer ordering, we divide the 
$U$ and $V$ matrices into three bands as shown in~Fig.~\ref{fig:esbBand}. 
Suppose that an input image of resolution $H$x$W$ is decomposed resulting in 
$U$ ($H$x$H$) and $V$ ($W$x$W$) matrices.
Given $k$=min\{$H$, $W$\}, we use the first $\frac{1}{6} \cdot k$ singular vectors in 
$U_{LB}$ and $V_{LB}$ to construct a low band ensemble image 
($EI_{LB}=U_{LB}\cdot V^T_{LB}$). Similarly, mid and high band ensembles images
are constructed using the subsequent  $\frac{2}{6} \cdot k$ 
and $\frac{3}{6} \cdot k$ singular vectors in $U_{MB}/V_{MB}$ and 
$U_{HB}/V_{HB}$, respectively. ($EI_{MB}=U_{MB}\cdot V^T_{MB}$ 
and $EI_{HB}=U_{HB}\cdot V^T_{HB}$).\\
\indent The first feature is the sum of absolute differences between ensemble images for 
each sub-band as:
\begin{equation} 
\label{eq:svdF1}
\resizebox{.91 \hsize}{!}
{
$F1^B_{svd} =\displaystyle\sum_{r=1}^{H}\sum_{c=1}^{W}abs\{ER_{B}(r,c) -ET_{B}(r,c)\}$
}
\end{equation}
where $ER_B$ and $ET_B$ are from ensemble images of the reference image and a test 
image for band $B$ ($\in \{LB, MB, HB\}$), respectively. 
The second SVD related feature utilizes the eigen images  
$X_k = \sum_{i=1}^{k} u_i \cdot \sigma_i \cdot v_i^T$, then form the sum of absolute 
differences between the eigen images from each sub-band as: 
\begin{equation} 
\label{eq:svdF2}
\resizebox{.91 \hsize}{!}
{
$F2^B_{svd} =\displaystyle\sum_{r=1}^{H}\sum_{c=1}^{W}abs\{R_{B}(r,c) -T_{B}(r,c)\}$
}
\end{equation}
where $R_B$ and $T_B$ are the eigen images of the reference and test images for band $B$. 
As discussed in~\cite{cit:Liu2008}, changes in an image's structure can significantly affect the 
singular vectors, hence our third feature is defined as:
\begin{equation} 
\label{eq:svdF3}
\resizebox{.91 \hsize}{!}
{
$F3^B_{svd} = \frac{1}{U_B\cdot V_B}\{UR_B\circ UT_B + VR_B\circ VT_B\}$
}
\end{equation}
where $XR_B$ and $XT_B$ ($X \in \{U, V\}$) are the singular vector matrices of band B 
for the reference and test images, respectively, and $U_B$ and $V_B$ are the number of 
singular vectors in $UR_B$ and $VR_B$. The operation $\circ$ denotes the 
matrix inner product. \\
\indent The last SVD feature is the sum of absolute differences of the singular values: 
\begin{equation} 
\label{eq:svdF5}
\resizebox{.91 \hsize}{!}
{
$F4^B_{svd} =\displaystyle\sum_{i=1}^{N_B}abs\{diagR_{B}(i) -diagT_{B}(i)\}$
}
\end{equation}
where $diagR_B$ and $diagT_B$ are the 1-D vectors of singular values in
band $B$, of the reference and test images, respectively, and where $N_B$ is 
the number of singular values in band $B$. \\
\indent The sub-bands for $F3^B_{svd} \sim F4^B_{svd}$ are divided into 
$\{LB, MB, HB\}$ in the same way as for $F1^B_{svd}$. Thus, the total 
number of SVD related features is 12 ($4$ features $\times 3$ bands).

\subsection{Histogram Features}\label{sec:4.2}

As discussed earlier, optimized GAN images may appear highly photorealistic,
since both the semantics/structural and statistical/spectral/textural characteristics
of the original image are well preserved. Although the local structure or detail may 
be slightly modified, preserving the statistical similarity yields a visually similar and
a natural viewing experiences. \\ 
\indent To quantify the degree of statistical similarity between reference and distorted 
images, we utilize a variety of histogram features. Begin with the coefficient of 
variation (CoV):
\begin{equation} 
\label{eq:rho}
\rho = \frac{\sigma}{\mu},
\end{equation}
where $\sigma$ and $\mu$ are the standard deviation and sample mean of a set of 
values, which will be drawn from both spatial and frequency domains. 
To derive the spatial features, first partition an image into 5x5 blocks. For example, 
6,144 values of $\rho$ would be computed on a 480x320 image. Likewise, compute 
the 2-D DCT of each 5x5 block and compute~(\ref{eq:rho}) on it. Then construct
histograms of the collected CoV values on both the reference image and the test
image. Denote these histograms as $\rho_{ab}$, where $a\in\{s,f\}$ indicates spatial 
and frequency CoV values, and $b\in\{r,t\}$ indicates whether measured on reference 
or test image. Then, measure the distances between histograms $\rho_{rs}$ and
$\rho_{ts}$ using the \textit{Kullback Leibler} (KL) distance measures:
\begin{equation} 
\label{eq:KL}
KL(\rho_{rs}, \rho_{ts}) = \sum_{i=1}^{N} \rho_{rs}(i)log\frac{\rho_{rs}(i)}{\rho_{ts}(i)}
\end{equation}
where $\rho_{rs}(i)$ and $\rho_{ts}(i)$ are the $i$-th bin values of the spatial CoV 
histograms of the reference and test images, respectively. Similarly, define 
$KL(\rho_{rf}, \rho_{tf})$ for the frequency CoV values. These distances become 
zero if the test image is the same as the reference image.\\
\begin{figure}[t]
\centering
\includegraphics[width=0.40\textwidth]{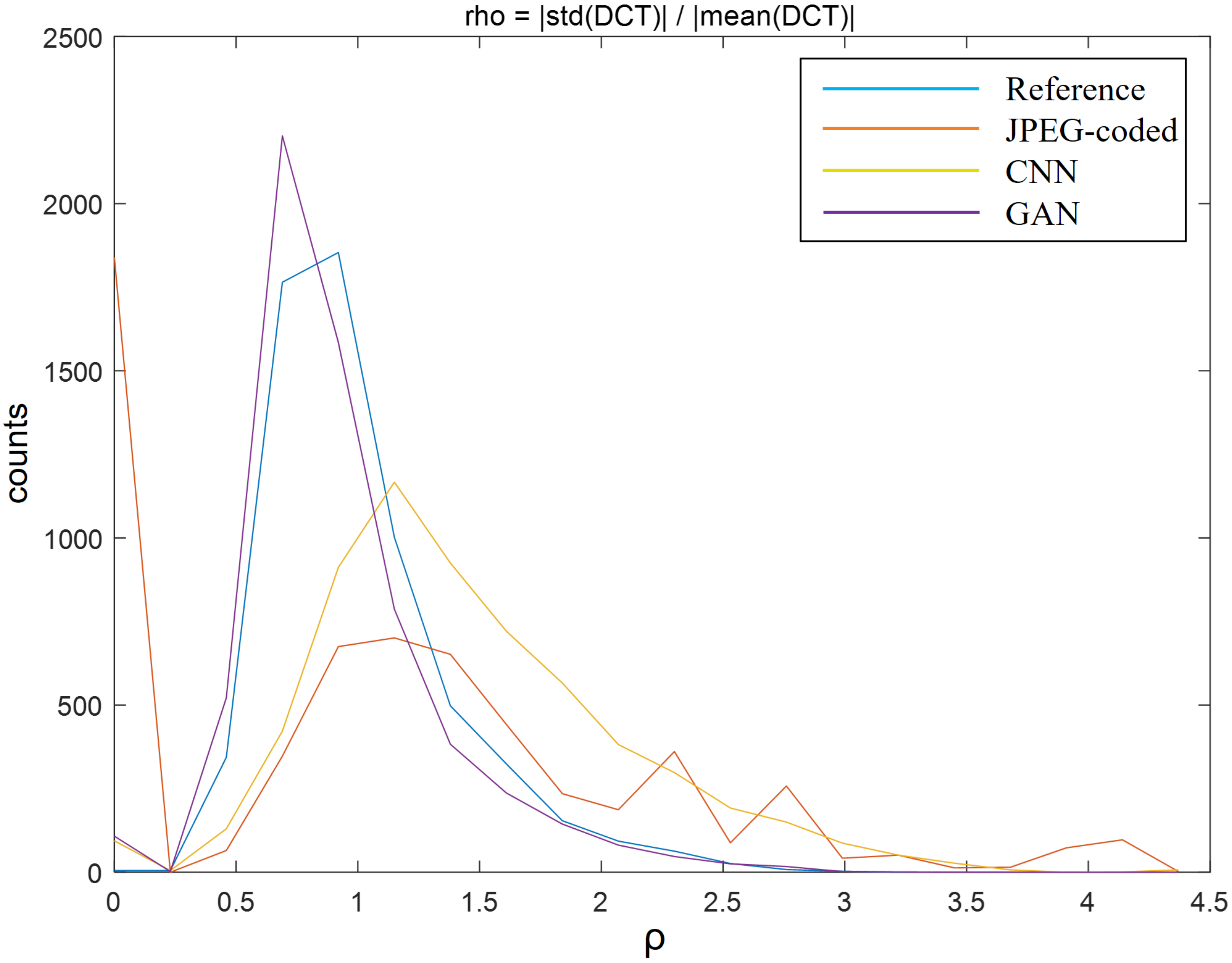}
\caption{Histograms $\rho_{rf}$ and $\rho_{tf}$ as a function of $\rho$ for the four images in Fig.~\ref{fig:ex1}. }
\label{fig:histFex}
\end{figure}
\begin{table}[t]
  \centering
  \caption{KL-distances between the distorted image histograms and the reference image histogram in Fig.~\ref{fig:histFex}}
    \begin{tabular}{cccc}
    \toprule
          & JPEG-coded & CNN & GAN \\
    \midrule
    \makecell{KL-distance (of $F2_{hist}$) \\ to reference} & 0.3710  & 0.2068  & 0.0182  \\
    \bottomrule
    \end{tabular}%
  \label{tab:KLex}%
    \vspace{-3mm}
\end{table}%
\begin{figure*}[t]
\centering
\includegraphics[width=0.74\textwidth]{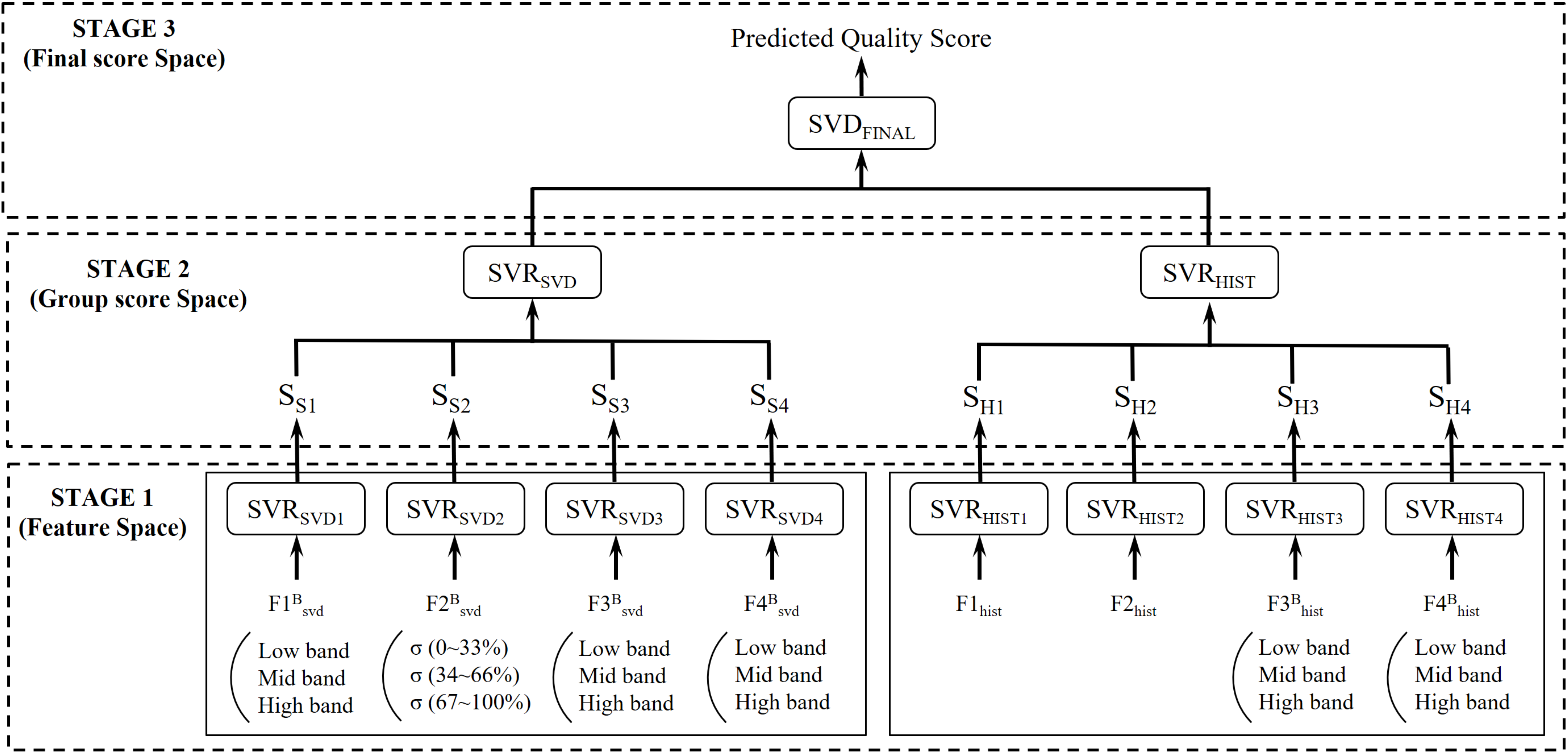}
\caption{Block diagram of the 3-stage parallel boosting system.}
\label{fig:boostSystem}
  \vspace{-3mm}
\end{figure*}
Given these measurements, define the first and second histogram features in the intensity 
and frequency domains as:
\begin{equation} 
\label{eq:histF1}
F1_{hist} = KL(\rho_{rs}, \rho_{ts})
\end{equation}
\begin{equation} 
\label{eq:histF2}
F2_{hist} = KL(\rho_{rf}, \rho_{tf})
\end{equation}
\indent Next, similar to the SVD based features, construct a set of ensemble images 
on which we define histogram features. Ensemble images for the three bands LB, MB and HB 
are constructed firstly, then the 2-D DCT is applied to them. The third and the fourth 
histogram features are expressed as KL distances in the space and frequency domains: 
\begin{equation} 
\label{eq:histF3}
F3^B_{hist} = KL(Hist^{ER_B}_{space}, Hist^{ET_B}_{space})
\end{equation}
\begin{equation} 
\label{eq:histF4}
F4^B_{hist} = KL(Hist^{ER_B}_{dct}, Hist^{ET_B}_{dct}).
\end{equation}
In sum, there are eight histogram features in total. \\
\indent Fig.~\ref{fig:histFex} plots four curves representing the histograms ($\rho_{rf}$ and 
$\rho_{tf}$ in Eq.~\ref{eq:histF2}) of the four images in Fig. \ref{fig:ex1}. 
The histogram curves show a good fit of the GAN image with the reference image, 
but poor fits of the CNN and JPEG-coded images. Table~\ref{tab:KLex} lists
the $KL$-distances of the distorted-image histograms to reference image histogram, 
reinforcing this result. 

\subsection{Block Feature Calculations}\label{sec:4.3}
In addition to calculating features on entire frames, as explained in 
Sections~\ref{sec:4.1} \& \ref{sec:4.2}, we also compute features on 
a block-wise basis, which are then pooled to also produce whole image features.
In this way, we are able to better capture local characteristics of distortions.
However, this benefit would get weakened if the block size becomes too large or too small.\\
\indent To analyze the relationship between block size and prediction accuracy, 
we performed some additional experiments. Table~\ref{tab:unitblk} shows the PCC/SRCC 
values obtained using 5-fold CV for different block sizes on the subset of 
full-frame images. For $F_{hist}$ features, the indicated block sizes were used for $\rho$ 
calculations, while the region over which the KL-distance calculation 
(in parentheses) increased accordingly. We found that block 
sizes of 10x10 and 5x5 for $F_{svd}$ and $F_{hist}$ provided
the best prediction accuracy. These block feature values were then pooled by averaging 
them to produce the final feature indices.
We provide assessments of the individual performance 
of the full-frame and block-based features in Section~\ref{sec:performance}.
\begin{table}[t]
  \centering
  \caption{Summary of results against block size on the subset of full-frame images.}
    \begin{tabular}{c|cc|c|cc}
    \toprule
    \multicolumn{3}{c|}{$F_{svd}$} & \multicolumn{3}{c}{$F_{hist}$} \\
    \midrule
    Block Size & PCC   & SROCC & Block Size & PCC   & SROCC \\
    \midrule
    5x5   & 0.837  & 0.841  & 5x5 (10x10) & \textbf{0.947}  & \textbf{0.903}  \\
    10x10 & \textbf{0.947}  & \textbf{0.903}  & 10x10 (20x20) & 0.871  & 0.822  \\
    20x20 & 0.852  & 0.827  & 25x25 (50x50) & 0.878  & 0.809  \\
    \bottomrule
    \end{tabular}%
  \label{tab:unitblk}%
  \vspace{-3mm}
\end{table}%

\subsection{Multi-stage Parallel Boosting System}\label{sec:4.4}
To learn a highly nonlinear model between the MOS and the proposed features, 
we employed a 3-stage parallel boosting system as demonstrated in 
Fig.~\ref{fig:boostSystem}. In the first stage, nine individual feature sets were each used
to train separately support vector regressor (SVR) to predict the MOS. In the second
stage, four $F_{svd}$ related scores ($S_{S1}\sim S_{S4}$) and four $F_{hist}$ related 
scores ($S_{H1}\sim S_{H4}$) were fed into two corresponding SVRs, respectively,
to further boost the prediction accuracy using the same group of feature scores.
The final predicted image quality score was obtained from the third stage, which fuses 
the two group scores. This hierarchical structure boosted the prediction of the
single feature by using the group scores. The boosted predictions were boosted 
further by using the across-group scores. \\
\indent To be more concrete, the SVRs in stage I take $(\boldsymbol{x}_n, y_n)$ as a set of
 training data, where $\boldsymbol{x}_n$ is a feature vector and $y_n$ is the target label, 
 \textit{e.g.,} the MOS of the $n$th image.
We deploy $\varepsilon$-SVR~\cite{cit:Scholkopf2002}, where the goal is to find a mapping 
function $f(\boldsymbol{x}_n)$ having a deviation of no more than $\varepsilon$ from the 
target label $y_n$ over all the training data. The mapping function has the  form:
\begin{equation} \label{eq:SVR1}
f(\boldsymbol{x}) = \boldsymbol{w_{f}}^T \phi(\boldsymbol{x})+b_{f},
\end{equation}
where $\boldsymbol{w_{f}}$ is a weighting vector, $\phi(\cdot)$ is a non-linear function, 
and $b_{f}$ is a bias term. The subscript $f$ implies that the SVRs in Stage I operate
in feature space, with feature vectors as input. It is desired to find $\boldsymbol{w}$ and 
$b$ satisfying the following condition: 
\begin{equation} \label{eq:SVR2}
|f(\boldsymbol{x}_n)-y_n| \le \varepsilon, \quad \forall n=1,2,\ldots ,N_t,
\end{equation}
where $N_t$ is the number of training data. We use the radial basis activation function 
(RBF), since it provides good performance in many image quality prediction applications
\cite{cit:Moorthy2011, cit:Saad2012, cit:Mittal2012}.
Since it is challenging to determine a proper value of $\varepsilon$ in (\ref{eq:SVR2}),
we used a modified version of the regression algorithm called $\nu$-SVR
\cite{cit:Basak2007}, where $\nu \in (0,1)$ is a control parameter to adjust
the number of support vectors and the accuracy level. Then,
$\varepsilon$ becomes a variable to be optimized, and we obtained $f(\boldsymbol{x})$ 
and $\boldsymbol{w}$ more easily.\\
\indent In Stage II and III, we fused all of the intermediate scores from the previous stage 
to determine a final predicted quality score. Suppose that there are $n$ SVRs 
fed by $m$ training images. On the $i$th image, compute the intermediate 
score $s_{i,j}$, where $i=1,2,...,m$ indexes the training images 
and $j=1,2,...,n$ is SVR index.  Let $\boldsymbol{s}_i= (s_{i,1},s_{i,2},\cdots,s_{i,n})$ 
be the intermediate score vector for the $i$th image. We trained the SVRs using 
$\boldsymbol{s_i}$ using all the images in the training set, and determined the 
weight vector $\boldsymbol{w_{s}}$ and bias parameter $b_{s}$ accordingly. 
The subscript $s$ indicates that the SVRs operate in score space with the 
intermediate score vectors as input. Finally, the ultimate designed image
quality model was found:
\begin{equation} \label{eq:ProposedIndex}
Q(\boldsymbol{s}) = \boldsymbol{w_{s}}^T \phi(\boldsymbol{s})+b_{s}.
\end{equation}
\begin{table*}[t]
  \centering
  \caption{Performances of single features on the proposed generative image database.}
\scalebox{0.55}{
    \begin{tabular}{ccc|cccccc|cccccc|cccccc|cccccc}
    \toprule
    \multirow{3}[2]{*}{Feature Group} & \multirow{3}[2]{*}{Feature} & \multirow{3}[2]{*}{Sub-band} & \multicolumn{6}{c|}{Subset of Full-Frame Images} & \multicolumn{6}{c|}{Subset of Randomly Structured Block Patches} & \multicolumn{6}{c|}{Subset of Regular Structured Block Patches} & \multicolumn{6}{c}{Subset of High-level Structured Block Patches} \\
          &       &       & \multicolumn{3}{c}{Full-frame Features} & \multicolumn{3}{c|}{Block Average Features} & \multicolumn{3}{c}{Full-frame Features} & \multicolumn{3}{c|}{Block Average Features} & \multicolumn{3}{c}{Full-frame Features} & \multicolumn{3}{c|}{Block Average Features} & \multicolumn{3}{c}{Full-frame Features} & \multicolumn{3}{c}{Block Average Features} \\
          &       &       & PCC   & SROCC & RMSE  & PCC   & SROCC & RMSE  & PCC   & SROCC & RMSE  & PCC   & SROCC & RMSE  & PCC   & SROCC & RMSE  & PCC   & SROCC & RMSE  & PCC   & SROCC & RMSE  & PCC   & SROCC & RMSE \\
    \midrule
    \multicolumn{1}{c}{\multirow{20}[10]{*}{\makecell{SVD \\ based \\ Feature}}} & \multirow{2}[2]{*}{$F1^B_{svd}$} & LB    & \textbf{0.61} & \textbf{0.60} & \textbf{5.11} & \textbf{0.66}  & \textbf{0.65}  & \textbf{4.84} & 0.65  & \textbf{0.70} & 5.07  & 0.64  & 0.62  & \textbf{5.14} & \textbf{0.90} & \textbf{0.89} & \textbf{2.65} & 0.72  & 0.70  & 4.28  & \textbf{0.91} & \textbf{0.92} & \textbf{2.97} & \textbf{0.92} & \textbf{0.90} & \textbf{3.08} \\
          &       & MB    & 0.55  & 0.52  & 5.42  & 0.46  & 0.48  & 6.48  & 0.55  & 0.59  & 5.56  & 0.42  & 0.31  & 6.05  & 0.83  & \textbf{0.83}  & 3.44  & 0.75  & 0.73  & 4.06  & \textbf{0.82}  & \textbf{0.80}  & \textbf{4.10}  & 0.78  & 0.80  & 7.22  \\
          &       & HB    & 0.19  & 0.41  & 6.36  & 0.17  & 0.33  & 6.39  & 0.26  & 0.21  & 6.44  & 0.54  & 0.52  & 5.62  & 0.49  & 0.47  & 5.37  & 0.28  & 0.05  & 5.92  & 0.30  & 0.31  & 6.92  & 0.25  & 0.05  & 6.99  \\
\cmidrule{2-27}          & \multirow{2}[2]{*}{$F2^B_{svd}$} & 0-33\% & 0.31  & 0.39  & 6.15  & 0.61  & 0.61  & 6.48  & 0.24  & 0.11  & 6.49  & 0.42  & 0.41  & 6.68  & 0.45  & 0.48  & 5.49  & 0.76  & 0.76  & \textbf{3.98}  & 0.27  & 0.52  & 6.95  & 0.64  & 0.62  & 7.22  \\
          &       & 34-66\% & 0.54  & \textbf{0.58}  & 5.45  & 0.30  & 0.28  & 6.48  & 0.55  & 0.48  & 5.57  & 0.37  & 0.47  & 6.23  & 0.74  & 0.71  & 4.17  & 0.38  & 0.40  & 6.16  & 0.69  & 0.64  & 5.25  & 0.36  & 0.32  & 7.22  \\
          &       & 67-100\% & 0.42  & 0.37  & 5.87  & 0.38  & 0.29  & 5.99  & 0.19  & 0.21  & 6.59  & 0.28  & 0.27  & 6.40  & 0.78  & 0.78  & 3.89  & 0.72  & 0.71  & 4.26  & 0.76  & 0.75  & 4.72  & 0.58  & 0.61  & 7.22  \\
\cmidrule{2-27}          & \multirow{2}[2]{*}{$F3^B_{svd}$} & LB    & \textbf{0.65} & \textbf{0.64} & \textbf{4.93} & \textbf{0.65}  & \textbf{0.64}  & \textbf{4.94} & 0.68  & 0.67  & \textbf{4.90}  & 0.63  & 0.61  & 5.18  & \textbf{0.86}  & 0.82  & \textbf{3.42}  & 0.71  & 0.71  & 4.33  & \textbf{0.89} & \textbf{0.87} & \textbf{3.25} & \textbf{0.84}  & \textbf{0.83}  & \textbf{3.86}  \\
          &       & MB    & \textbf{0.56}  & 0.54  & \textbf{5.36}  & 0.61  & 0.59  & 5.12  & 0.58  & 0.57  & 5.42  & 0.64  & \textbf{0.66}  & \textbf{5.15}  & \textbf{0.83} & \textbf{0.85} & \textbf{3.10} & \textbf{0.80} & \textbf{0.79} & \textbf{3.73} & 0.81  & 0.80  & 4.24  & \textbf{0.90} & \textbf{0.92} & \textbf{2.81} \\
          &       & HB    & 0.24  & 0.27  & 6.30  & 0.56  & 0.53  & 5.36  & 0.32  & 0.28  & 6.32  & \textbf{0.68} & 0.65  & \textbf{4.87} & 0.38  & 0.37  & 5.70  & \textbf{0.79} & \textbf{0.76}  & \textbf{3.79} & 0.29  & 0.32  & 6.91  & 0.76  & 0.74  & 4.72  \\
\cmidrule{2-27}          & \multirow{2}[2]{*}{$F4^B_{svd}$} & LB    & 0.50  & 0.45  & 5.62  & \textbf{0.69} & \textbf{0.68} & 6.48  & 0.11  & 0.14  & 6.64  & 0.25  & 0.26  & 6.68  & 0.35  & 0.32  & 5.77  & 0.54  & 0.57  & 6.16  & 0.35  & 0.30  & 6.75  & 0.51  & 0.52  & 7.22  \\
          &       & MB    & 0.51  & 0.44  & 5.59  & 0.60  & 0.62  & 6.48  & 0.55  & 0.51  & 5.58  & \textbf{0.64}  & \textbf{0.68} & 6.68  & 0.59  & 0.56  & 4.98  & 0.73  & 0.76  & 6.16  & 0.43  & 0.42  & 6.50  & 0.62  & 0.62  & 7.22  \\
          &       & HB    & 0.43  & 0.44  & 5.85  & 0.43  & 0.45  & 6.48  & \textbf{0.69}  & 0.68  & 4.91  & \textbf{0.65} & \textbf{0.69} & 6.68  & 0.65  & 0.62  & 4.70  & 0.71  & 0.71  & 6.16  & 0.45  & 0.44  & 6.45  & 0.41  & 0.42  & 7.22  \\
    \midrule
    \multicolumn{1}{c}{\multirow{10}[8]{*}{\makecell{Histogram- \\ distance \\ based \\ Feature}}} & \multicolumn{2}{c|}{$F1_{hist}$} & 0.30  & 0.39  & 6.48  & 0.55  & 0.55  & 6.48  & 0.44  & 0.40  & 6.68  & 0.56  & 0.56  & 6.68  & 0.32  & 0.42  & 6.16  & 0.61  & 0.61  & 4.90  & 0.37  & 0.37  & 7.22  & 0.65  & 0.67  & 7.22  \\
\cmidrule{2-27}          & \multicolumn{2}{c|}{$F2_{hist}$} & 0.44  & 0.47  & 5.81  & 0.33  & 0.43  & 6.48  & \textbf{0.71} & \textbf{0.70} & \textbf{4.81} & 0.32  & 0.32  & 6.68  & 0.58  & 0.59  & 6.16  & \textbf{0.79}  & \textbf{0.80} & 6.16  & 0.42  & 0.37  & 6.55  & 0.41  & 0.40  & 7.22  \\
\cmidrule{2-27}          & \multirow{2}[2]{*}{$F3^B_{hist}$} & LB    & 0.45  & 0.33  & 6.48  & 0.04  & 0.10  & 6.48  & 0.56  & 0.51  & 6.68  & 0.39  & 0.44  & 6.68  & 0.29  & 0.12  & 6.16  & 0.67  & 0.66  & 6.16  & 0.40  & 0.34  & 7.22  & 0.63  & 0.58  & 7.22  \\
          &       & MB    & 0.31  & 0.38  & 6.48  & 0.03  & 0.07  & 6.48  & 0.54  & 0.37  & 6.68  & 0.08  & 0.00  & 6.68  & 0.31  & 0.09  & 6.16  & 0.20  & 0.16  & 6.16  & 0.37  & 0.30  & 7.22  & 0.26  & 0.22  & 7.22  \\
          &       & HB    & 0.40  & 0.33  & 6.48  & 0.26  & 0.32  & 6.48  & 0.41  & 0.46  & 6.68  & 0.39  & 0.37  & 6.68  & 0.33  & 0.19  & 6.16  & 0.18  & 0.10  & 6.16  & 0.44  & 0.53  & 7.22  & 0.14  & 0.16  & 7.22  \\
\cmidrule{2-27}          & \multirow{2}[2]{*}{$F4^B_{hist}$} & LB    & 0.47  & 0.56  & 6.48  & 0.12  & 0.15  & 6.48  & \textbf{0.75} & \textbf{0.74} & \textbf{4.45} & 0.17  & 0.22  & 6.68  & 0.47  & 0.44  & 6.16  & 0.61  & 0.60  & 6.16  & 0.43  & 0.31  & 6.51  & 0.42  & 0.41  & 7.22  \\
          &       & MB    & 0.29  & 0.35  & 6.48  & 0.28  & 0.35  & 6.22  & 0.59  & 0.56  & 5.38  & 0.25  & 0.41  & 6.47  & 0.45  & 0.34  & 5.49  & 0.10  & 0.08  & 6.16  & 0.32  & 0.30  & 7.22  & 0.32  & 0.38  & 6.85  \\
          &       & HB    & 0.33  & 0.49  & 6.48  & 0.11  & 0.38  & 6.48  & 0.47  & 0.48  & 6.68  & 0.20  & 0.24  & 6.68  & 0.37  & 0.27  & 6.16  & 0.41  & 0.56  & 5.61  & 0.45  & 0.45  & 7.22  & 0.50  & 0.43  & 6.24  \\
    \bottomrule
    \end{tabular}
}
  \label{tab:singleFeature}%

\end{table*}%
\indent For the performance evaluation, we split the dataset into two training 
and testing subsets, consisting of 80\% and 20\% of the entire collection of images, respectively. 
The images in the training and testing subsets were drawn from non-overlapping content 
to avoid the SVRs learning the images.
The SVRs were trained on the training set, and the learned models were then tested on the 
testing set. To ensure that the proposed IQA model is robust across contents and was 
not dominated by the specific train-test split, we repeated this random split 1000 times 
on the dataset, and recorded the performances on each of the test sets. For all of the 
experiment results, we reported the median values across these 1000 train-test iterations 
as performance indices.
In addition, feature normalization was performed prior to the training and test
processes, to avoid features having larger numeric ranges dominating those having
smaller numeric ranges. We scaled the input of each SVR to the unit range [0,1]
using $(val – MIN) / (MAX-MIN)$.
During the training stage, the goal was to determine the optimal weighting
vector $\boldsymbol{w}$ and bias $b$ minimizing the error between the
MOS and the predicted scores:
\begin{equation} \label{eq:Training}
\sum_i |\mbox{MOS}_i- Q(\boldsymbol{s}_i)|^2.
\end{equation}
\indent Since we adopted the RBF kernel, the error penalty term
($C$) and the kernel parameter ($\gamma$) were optimized to achieve
the highest accuracy. We searched the optimal $C$ and $\gamma$ during
the training stage using the cross validation scheme in Section 3.2 
of \cite{cit:Hsu2003}. Specifically, we used the built-in training function in LIBSVM, 
which provides an option (-v) for running v-fold CV.
Various $(C, \gamma)$ pairs were tried, and the one yielding the highest
cross validation accuracy was selected. Finally, the entire training set was
used again to generate the final SVR predictor. At the test stage, we use 
the intermediate score vector $s_i$ in (\ref{eq:ProposedIndex}) to determine 
the predicted score. The score prediction was quite fast, since all of the model 
parameters were decided during the training stage. 

\section{Performance Evaluation and Analysis}\label{sec:performance}
Following the suggestions in ITU-T (p.1401)~\cite{cit:P1401}, we used three
measures to evaluate the performance of the proposed image quality predictor: 
(1) the Pearson correlation coefficient (PCC) measures the linear relationship 
between a model's score and the subjective data, (2) the Spearman rank-order 
correlation coefficient (SRCC), which measures the prediction monotonicity, 
and (3) the root mean squared error (RMSE), which quantifies the prediction 
accuracy. We apply the monotonic logistic function to the predicted scores to 
account for nonlinearity when fitting the subjective scores, but we did not apply
the nonlinearity when computing the rank order correlation. The logistic function
has the form:
\begin{equation} \label{eq:logistic}
L(s) = \frac{\beta_1-\beta_2}{1+exp(\frac{-s+\beta_3}{|\beta_4|})}+\beta_2
\end{equation}
where $s$ and $L(s)$ are the predicted scores and the adjusted predicted
scores, respectively, and $\beta_k$ ($k$ = 1,2,3,4) are the parameters that 
minimize the mean squared error between $L(s)$ and MOS. The choices
of the initial parameters are explained in~\cite{cit:VQEG2004}.\\

\subsection{Performance Analysis on Individual Features}\label{sec:5.1}
\begin{table*}[t]
  \centering
  \caption{Performance comparison on the proposed generative image database 
  (median PCC, SRCC and RMSE across 1,000 train-test trials for $SSQP_F$ and $SSQP_B$)}.
  \scalebox{0.9}{
    \begin{tabular}{c|ccc|ccc|ccc|ccc}
\cmidrule{2-13}    \multicolumn{1}{r}{} & \multicolumn{3}{c|}{Full-Frame Images} & \multicolumn{3}{c|}{Randomly Structured Patches} & \multicolumn{3}{c|}{Regular Structured Patches} & \multicolumn{3}{c}{High-level Structured Patches} \\
    \multicolumn{1}{r}{} & PCC   & SROCC & RMSE  & PCC   & SROCC & RMSE  & PCC   & SROCC & RMSE  & PCC   & SROCC & RMSE \\
    \midrule
    PSNR  & 0.58  & 0.56  & 5.29  & 0.43  & 0.41  & 6.04  & 0.78  & 0.79  & 3.87  & 0.72  & 0.75  & 4.99  \\
    SSIM  & 0.60  & 0.59  & 5.18  & 0.62  & 0.60  & 5.26  & 0.85  & 0.85  & 3.21  & 0.77  & 0.79  & 4.57  \\
    MSIM  & 0.74  & 0.74  & 4.32  & 0.68  & 0.70  & 4.90  & 0.76  & 0.77  & 4.03  & 0.84  & 0.80  & 3.96  \\
    VSNR  & 0.67  & 0.67  & 4.82  & 0.66  & 0.60  & 5.00  & 0.64  & 0.65  & 4.71  & 0.77  & 0.75  & 4.63  \\
    VIF   & 0.81  & 0.79  & 3.77  & 0.69  & 0.60  & 4.85  & 0.84  & 0.85  & 3.34  & 0.90  & 0.88  & 3.20  \\
    VIFP  & 0.67  & 0.66  & 4.80  & 0.68  & 0.56  & 4.90  & 0.77  & 0.78  & 3.91  & 0.84  & 0.82  & 3.95  \\
    UQI   & 0.71  & 0.69  & 4.58  & 0.61  & 0.60  & 5.27  & 0.71  & 0.69  & 4.32  & 0.81  & 0.80  & 4.28  \\
    IFC   & 0.79  & 0.76  & 4.00  & 0.66  & 0.66  & 4.99  & 0.82  & 0.83  & 3.50  & 0.90  & 0.90  & 3.10  \\
    NQM   & 0.57  & 0.55  & 5.32  & 0.57  & 0.54  & 5.49  & 0.71  & 0.68  & 4.32  & 0.88  & 0.85  & 3.42  \\
    WSNR  & 0.53  & 0.51  & 5.48  & 0.53  & 0.59  & 5.66  & 0.52  & 0.51  & 5.25  & 0.59  & 0.57  & 5.85  \\
    SNR   & 0.54  & 0.51  & 5.47  & 0.28  & 0.25  & 6.42  & 0.72  & 0.72  & 4.25  & 0.72  & 0.68  & 5.00  \\
    FSIM  & 0.84  & 0.82  & 3.54  & 0.70  & 0.65  & 4.78  & 0.93  & \textbf{0.91}  & 2.33  & 0.88  & 0.88  & 3.41  \\
    GMSD  & 0.89  & 0.88  & 6.48  & 0.71  & 0.65  & 6.68  & 0.80  & 0.80  & 6.16  & 0.87  & 0.88  & 7.22  \\
    \textit{$SSQP_F$} & 0.93  & 0.88  & 2.32  & \textbf{0.91}  & \textbf{0.86}  & \textbf{2.44}  & \textbf{0.96} & 0.88 & \textbf{1.54} & \textbf{0.96} & \textbf{0.90}  & \textbf{1.87}  \\
    \textit{$SSQP_B$} & \textbf{0.95} & \textbf{0.89} & \textbf{2.03} & 0.87 & 0.81 & 2.84 & 0.95  & 0.88  & 1.70  & 0.94 & 0.86 & 2.13 \\
    \bottomrule
    \end{tabular}%
    }
  \label{tab:PerfComp1}%
  \vspace{-3mm}
\end{table*}%
\begin{figure*}
        \centering
        \begin{subfigure}[t]{0.25\textwidth}
                \centering
                \includegraphics[width=\textwidth]{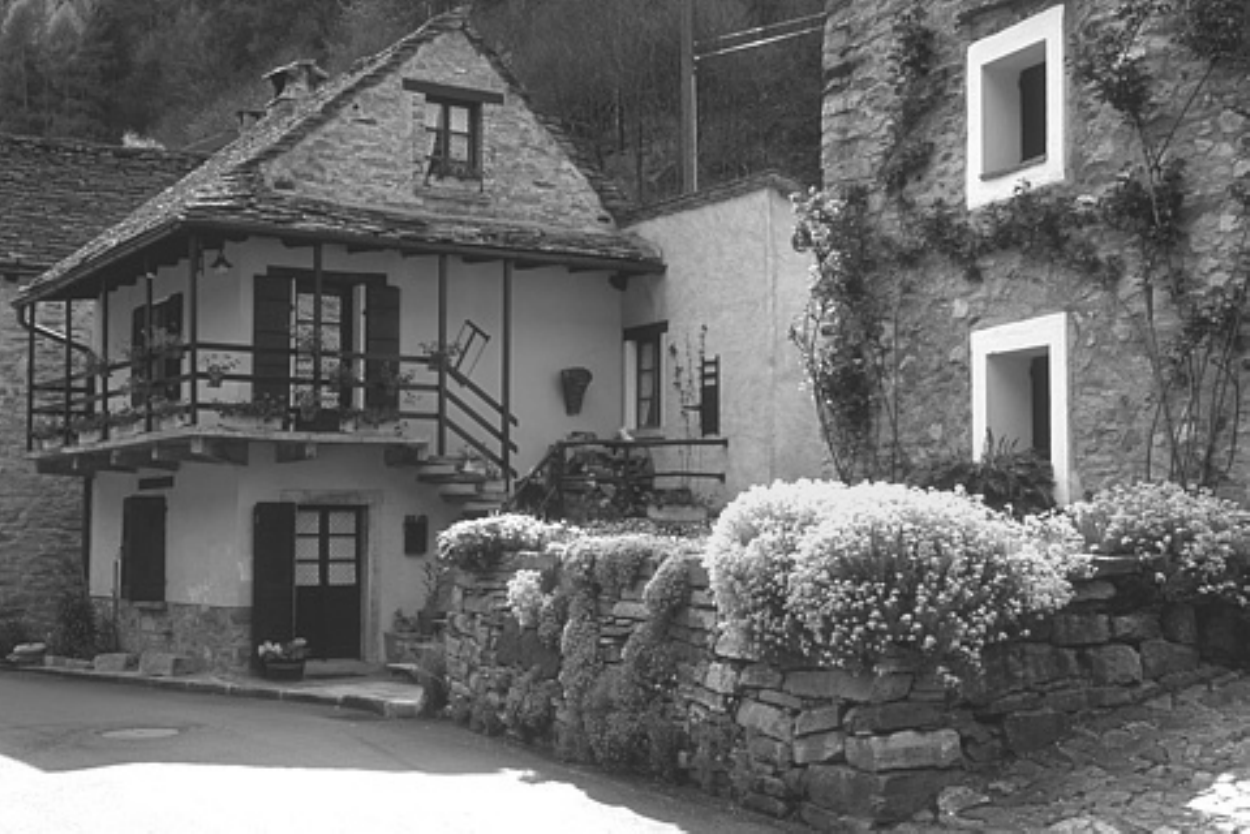}
                \caption{}
		\label{fig:EigenImg1}
        \end{subfigure}
		\quad
        \begin{subfigure}[t]{0.166\textwidth}
                \centering
                \includegraphics[width=\textwidth]{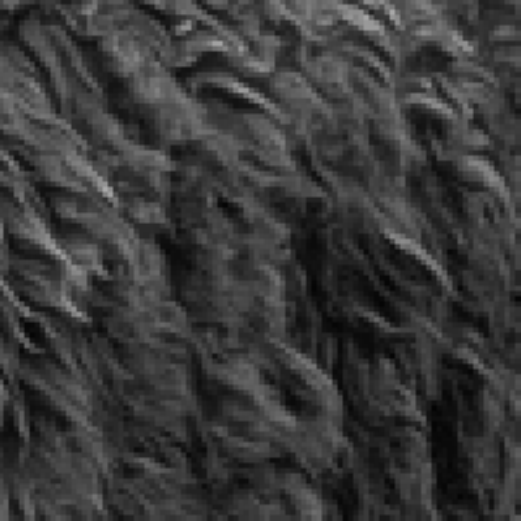}
                \caption{}
		\label{fig:EigenImg2}
        \end{subfigure}
		\quad
        \begin{subfigure}[t]{0.166\textwidth}
                \centering
                \includegraphics[width=\textwidth]{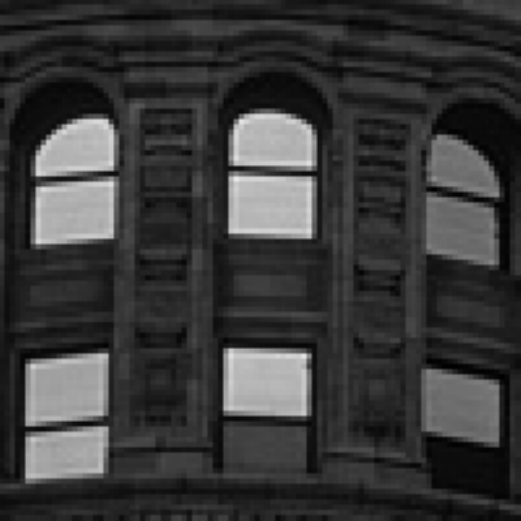}
                \caption{}
		\label{fig:EigenImg3}
        \end{subfigure}
        		\quad
        \begin{subfigure}[t]{0.166\textwidth}
                \centering
                \includegraphics[width=\textwidth]{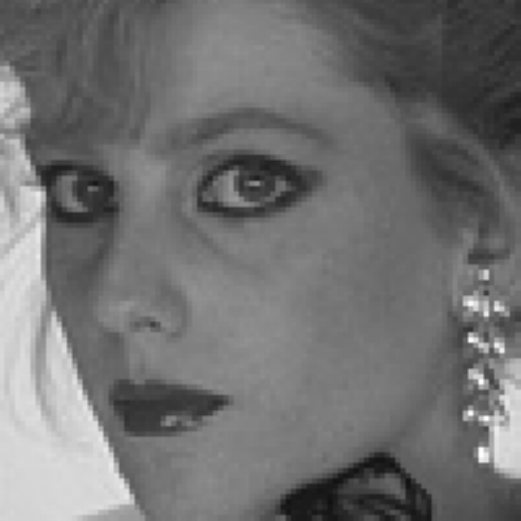}
                \caption{}
		\label{fig:EigenImg4}
        \end{subfigure}
\\
        \begin{subfigure}[t]{0.25\textwidth}
                \centering
                \includegraphics[width=\textwidth]{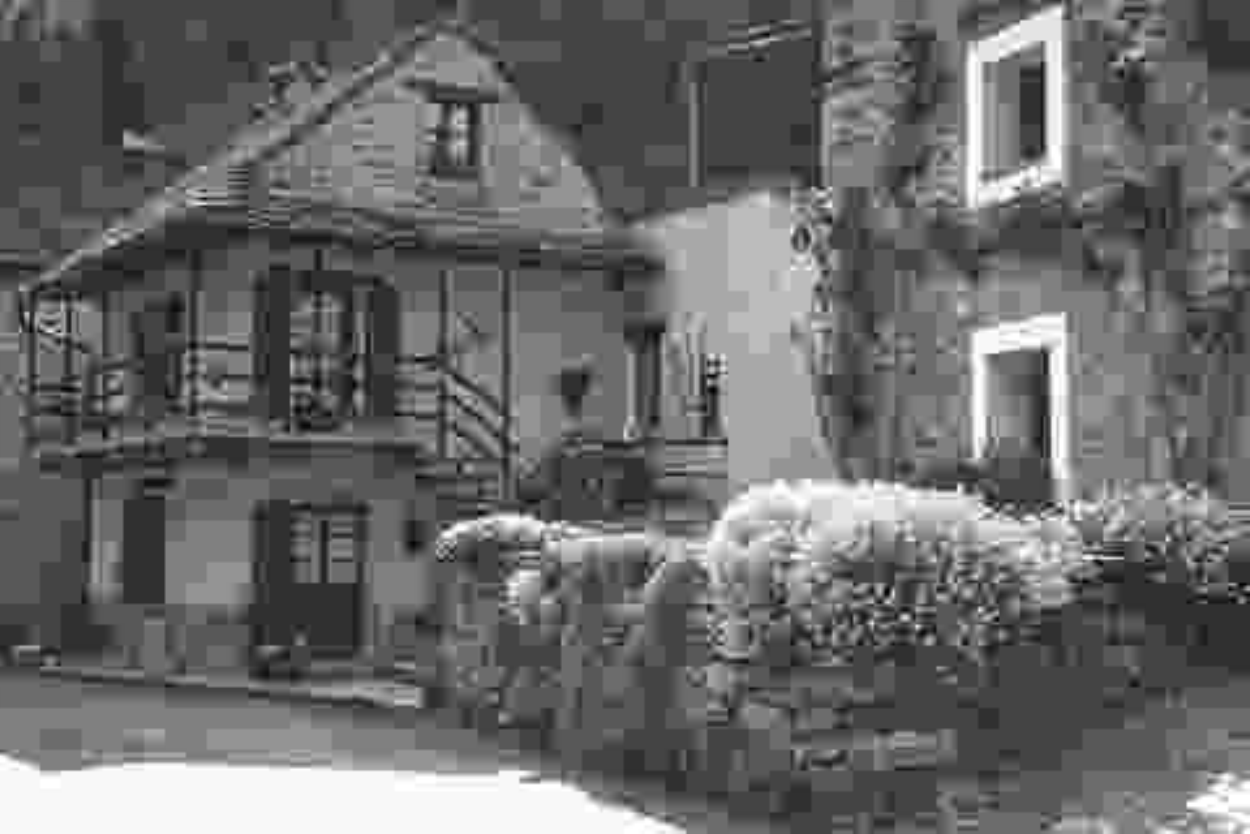}
                \caption{}
		\label{fig:EigenImg1}
        \end{subfigure}
		\quad
        \begin{subfigure}[t]{0.166\textwidth}
                \centering
                \includegraphics[width=\textwidth]{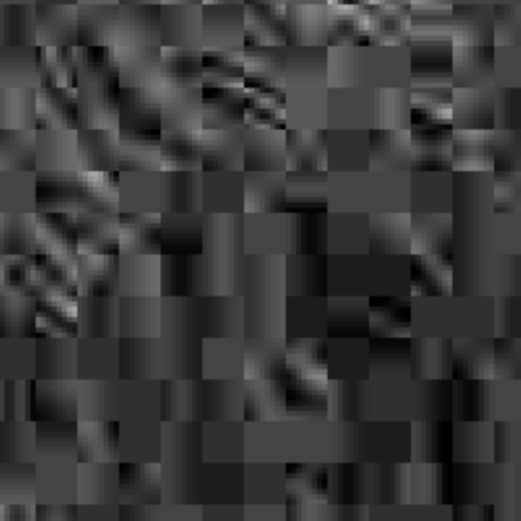}
                \caption{}
		\label{fig:EigenImg2}
        \end{subfigure}
		\quad
        \begin{subfigure}[t]{0.166\textwidth}
                \centering
                \includegraphics[width=\textwidth]{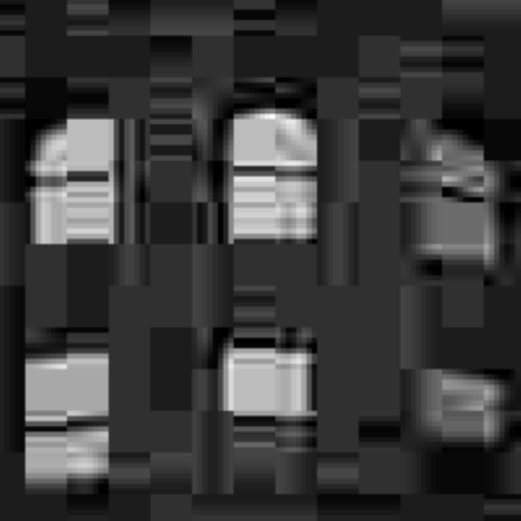}
                \caption{}
		\label{fig:EigenImg3}
        \end{subfigure}
        		\quad
        \begin{subfigure}[t]{0.166\textwidth}
                \centering
                \includegraphics[width=\textwidth]{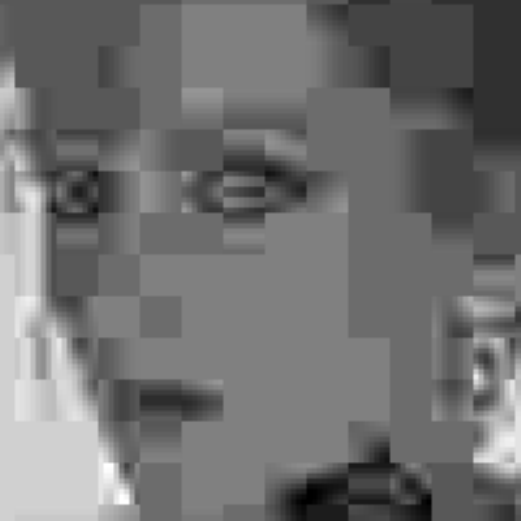}
                \caption{}
		\label{fig:EigenImg4}
        \end{subfigure}
\\
        \begin{subfigure}[t]{0.25\textwidth}
                \centering
                \includegraphics[width=\textwidth]{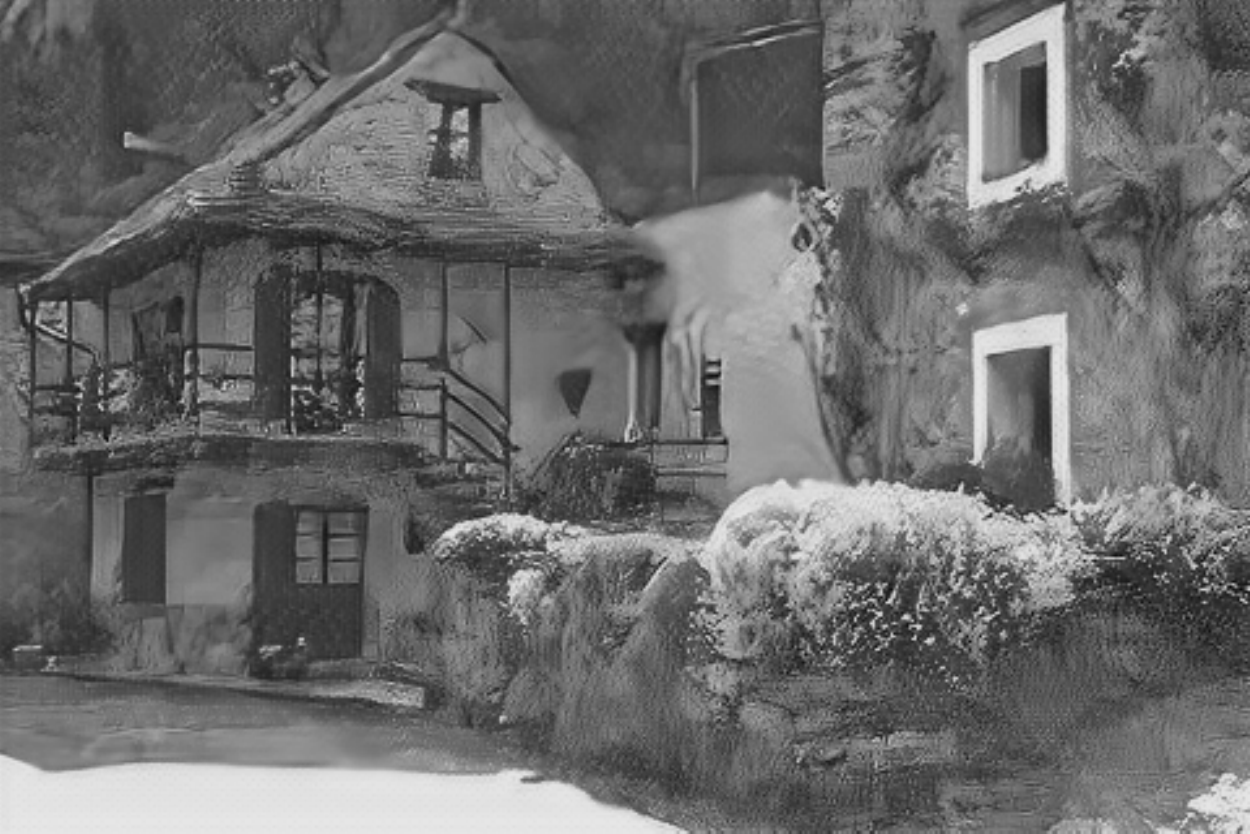}
                \caption{}
		\label{fig:EigenImg1}
        \end{subfigure}
		\quad
        \begin{subfigure}[t]{0.166\textwidth}
                \centering
                \includegraphics[width=\textwidth]{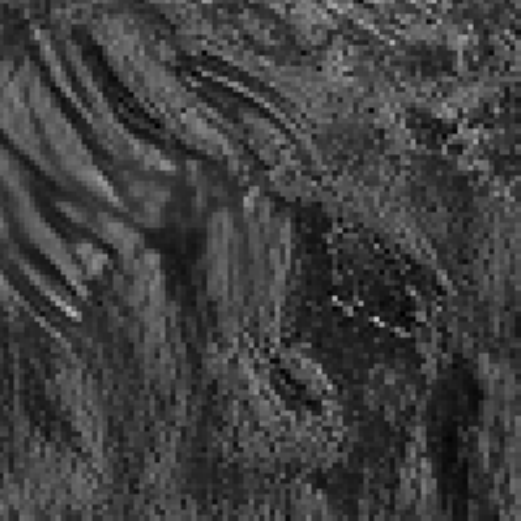}
                \caption{}
		\label{fig:EigenImg2}
        \end{subfigure}
		\quad
        \begin{subfigure}[t]{0.166\textwidth}
                \centering
                \includegraphics[width=\textwidth]{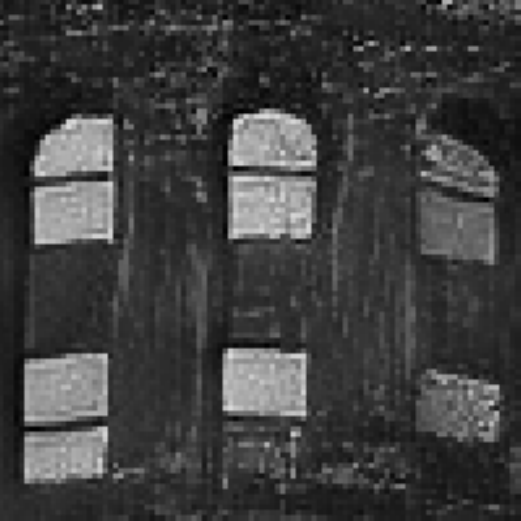}
                \caption{}
		\label{fig:EigenImg3}
        \end{subfigure}
        		\quad
        \begin{subfigure}[t]{0.166\textwidth}
                \centering
                \includegraphics[width=\textwidth]{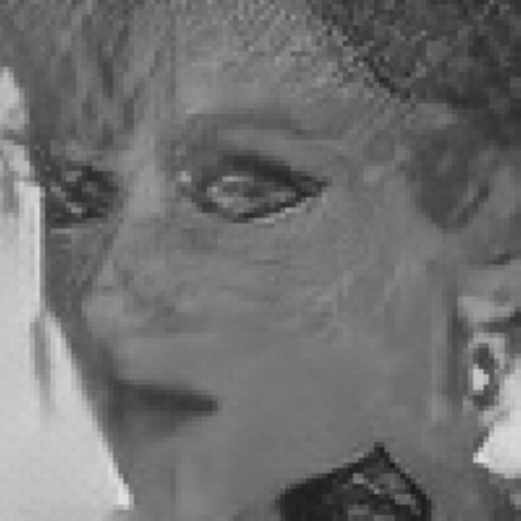}
                \caption{}
		\label{fig:EigenImg4}
        \end{subfigure}

\caption{The subjective quality comparison: (a) House image from subset of full-frame images, 
(b) Llama fur patch from subset of random structured patches, (c) building patch from subset of regular structured patches, (d) face patch from subset of high-level structured patches, (e)-(h): the corresponding JPEG QF05 images, 
(i)-(l):  the corresponding GAN images ($\lambda$=0.1 with input of JPEG QF05).}
\label{fig:comparison}
\vspace{-3mm}
\end{figure*}
Next, we analyze the performance of each individual feature on the proposed generative 
image database, which consists of four sub-datasets. Table~\ref{tab:singleFeature} 
lists the experimental results on all the sub-datasets, where the three top-performing 
features in each index are marked in bold. In the following analysis, the designation 
$F$ or $B$ means that the feature is calculated on full frames or is a block average, 
respectively.\\
\indent For the first subset of full-frame images, the SVD related features, such as 
$F3^{LB}_{svd_F}$ (PCC=0.65) and $F1^{LB}_{svd_F}$ (PCC=0.61) (full-frame
features) and $F5^{LB}_{svd_B}$ (PCC=0.69) and $F1^{LB}_{svd_B}$ (PCC=0.66)
(block average features) yielded good performance. In general, the block average based 
features provided better performance than did the full-frame based features. Also, 
the low band features yielded better predictors than those from the mid and high bands,
probably because preservation of the overall image structure is more important than 
retaining fine details on full-frame generative images. The histogram-distance based 
features yielded relatively low prediction accuracy on this subset. \\
\indent For the second subset of randomly structured patches, meaningful differences
as compared to the previous subset were observed. The histogram-distance based 
features delivered improved performance, possibly due to the reasons explained in 
Section~\ref{sec:4.2}. For example, $F4^{LB}_{hist_F}$ gave the best PCC value (0.75).  
The block-average features also provided performance increases, but not as good as the
full-frame based features, possibly because the fixed block sizes (10x10/20x20) restricted 
performance as compared to full-frame calculation. Among the SVD related features, 
$F3^{HB}_{svd_B}$ (0.68) and $F5^{HB}_{svd_F}$ (0.69) also delivered high prediction 
accuracy, perhaps because the high band captures high frequency image and distortion 
details, which strongly characterize the second subset. 
The results on the subset of regular structured patches strongly suggest that 
structure-representing features are well correlated with MOS. For example, 
$F1^{LB}_{svd_F}$ (PCC=0.90) and $F3^{LB}_{svd_F}$ (PCC=0.86) yielded very good 
predictions. Lastly, on the subset of high-level structured patches, the same tendency 
was observed, and $F1^{LB}_{svd_B}$ (PCC=0.92) and $F1^{LB}_{svd_F}$ (PCC=0.91) 
yielded very good prediction performance.\\
\indent In sum, the single feature analysis showed that the SVD-related features 
effectively captured the similarities in structure, while the histogram-distance related 
features were more useful for representing statistical similarities.

\subsection{Algorithm Comparison on the Generative IQA Database}\label{sec:5.2}
\begin{table*}[t]
  \centering
  \caption{Experimental results on four selected test image sets: In each row, two test images are 
highlighted which obtained the highest and the lowest scores in green and red, respectively.}
\scalebox{0.57}{
    \begin{tabular}{c|rrr|rrr|rrr|rrr|c|rrr|rrr|rrr|rrr}
    \toprule
    \multicolumn{13}{c|}{House Image from the Subset of Full-frame Images}                                    & \multicolumn{13}{c}{Llama Fur patch from the Subset of Randomly Structured Patches} \\
    \midrule
    \multirow{2}[2]{*}{Metric} & \multicolumn{3}{c|}{JPEG} & \multicolumn{3}{c|}{CNN} & \multicolumn{3}{c|}{GAN (lambda=0.1)} & \multicolumn{3}{c|}{GAN (lambda=0.01)} & \multirow{2}[2]{*}{Metric} & \multicolumn{3}{c|}{JPEG} & \multicolumn{3}{c|}{CNN} & \multicolumn{3}{c|}{GAN (lambda=0.1)} & \multicolumn{3}{c}{GAN (lambda=0.01)} \\
          & \multicolumn{1}{c}{QF05} & \multicolumn{1}{c}{QF10} & \multicolumn{1}{c|}{QF20} & \multicolumn{1}{c}{QF05} & \multicolumn{1}{c}{QF10} & \multicolumn{1}{c|}{QF20} & \multicolumn{1}{c}{QF05} & \multicolumn{1}{c}{QF10} & \multicolumn{1}{c|}{QF20} & \multicolumn{1}{c}{QF05} & \multicolumn{1}{c}{QF10} & \multicolumn{1}{c|}{QF20} &       & \multicolumn{1}{c}{QF05} & \multicolumn{1}{c}{QF10} & \multicolumn{1}{c|}{QF20} & \multicolumn{1}{c}{QF05} & \multicolumn{1}{c}{QF10} & \multicolumn{1}{c|}{QF20} & \multicolumn{1}{c}{QF05} & \multicolumn{1}{c}{QF10} & \multicolumn{1}{c|}{QF20} & \multicolumn{1}{c}{QF05} & \multicolumn{1}{c}{QF10} & \multicolumn{1}{c}{QF20} \\
    \midrule
    \midrule
    PSNR  & 22.68  & 24.61  & 26.46  & 23.82  & 25.71  & \cellcolor[rgb]{ .776,  .937,  .808}\textcolor[rgb]{ 0,  .38,  0}{27.64 } & \cellcolor[rgb]{ 1,  .78,  .808}\textcolor[rgb]{ .612,  0,  .024}{20.40 } & 23.06  & 24.69  & 20.94  & 23.62  & 25.07  & PSNR  & 22.862  & 25.100  & 27.265  & 23.935  & 26.268  & \cellcolor[rgb]{ .776,  .937,  .808}\textcolor[rgb]{ 0,  .38,  0}{28.572 } & \cellcolor[rgb]{ 1,  .78,  .808}\textcolor[rgb]{ .612,  0,  .024}{20.176 } & 24.895  & 25.466  & 21.563  & 23.762  & 26.922  \\
    SSIM  & 0.57  & 0.70  & 0.80  & 0.62  & 0.73  & \cellcolor[rgb]{ .776,  .937,  .808}\textcolor[rgb]{ 0,  .38,  0}{0.83 } & \cellcolor[rgb]{ 1,  .78,  .808}\textcolor[rgb]{ .612,  0,  .024}{0.47 } & 0.64  & 0.73  & 0.47  & 0.65  & 0.77  & SSIM  & 0.521  & 0.684  & 0.809  & 0.555  & 0.716  & \cellcolor[rgb]{ .776,  .937,  .808}\textcolor[rgb]{ 0,  .38,  0}{0.838 } & \cellcolor[rgb]{ 1,  .78,  .808}\textcolor[rgb]{ .612,  0,  .024}{0.369 } & 0.682  & 0.741  & 0.415  & 0.626  & 0.782  \\
    MSIM  & 0.87  & 0.94  & 0.97  & 0.89  & 0.95  & \cellcolor[rgb]{ .776,  .937,  .808}\textcolor[rgb]{ 0,  .38,  0}{0.97 } & \cellcolor[rgb]{ 1,  .78,  .808}\textcolor[rgb]{ .612,  0,  .024}{0.83 } & 0.92  & 0.96  & 0.84  & 0.92  & 0.96  & MSIM  & 0.827  & 0.923  & 0.966  & 0.843  & 0.931  & \cellcolor[rgb]{ .776,  .937,  .808}\textcolor[rgb]{ 0,  .38,  0}{0.971 } & 0.787  & 0.919  & 0.959  & \cellcolor[rgb]{ 1,  .78,  .808}\textcolor[rgb]{ .612,  0,  .024}{0.776 } & 0.895  & 0.960  \\
    VSNR  & 18.89  & 22.87  & 27.46  & 20.38  & 24.15  & \cellcolor[rgb]{ .776,  .937,  .808}\textcolor[rgb]{ 0,  .38,  0}{28.70 } & \cellcolor[rgb]{ 1,  .78,  .808}\textcolor[rgb]{ .612,  0,  .024}{16.97 } & 21.54  & 25.85  & 17.25  & 21.76  & 25.06  & VSNR  & 9.477  & 13.647  & 18.055  & 10.569  & 14.430  & \cellcolor[rgb]{ .776,  .937,  .808}\textcolor[rgb]{ 0,  .38,  0}{19.242 } & \cellcolor[rgb]{ 1,  .78,  .808}\textcolor[rgb]{ .612,  0,  .024}{8.073 } & 14.114  & 17.252  & 8.772  & 13.872  & 17.143  \\
    VIF   & 0.17  & 0.30  & 0.47  & 0.21  & 0.34  & \cellcolor[rgb]{ .776,  .937,  .808}\textcolor[rgb]{ 0,  .38,  0}{0.50 } & \cellcolor[rgb]{ 1,  .78,  .808}\textcolor[rgb]{ .612,  0,  .024}{0.15 } & 0.29  & 0.43  & 0.16  & 0.28  & 0.44  & VIF   & 0.128  & 0.260  & 0.427  & 0.145  & 0.286  & \cellcolor[rgb]{ .776,  .937,  .808}\textcolor[rgb]{ 0,  .38,  0}{0.452 } & \cellcolor[rgb]{ 1,  .78,  .808}\textcolor[rgb]{ .612,  0,  .024}{0.101 } & 0.257  & 0.402  & 0.103  & 0.237  & 0.406  \\
    VIFP  & 0.21  & 0.30  & 0.39  & 0.26  & 0.35  & \cellcolor[rgb]{ .776,  .937,  .808}\textcolor[rgb]{ 0,  .38,  0}{0.44 } & \cellcolor[rgb]{ 1,  .78,  .808}\textcolor[rgb]{ .612,  0,  .024}{0.17 } & 0.28  & 0.35  & 0.18  & 0.28  & 0.38  & VIFP  & 0.151  & 0.256  & 0.352  & 0.180  & 0.289  & \cellcolor[rgb]{ .776,  .937,  .808}\textcolor[rgb]{ 0,  .38,  0}{0.389 } & \cellcolor[rgb]{ 1,  .78,  .808}\textcolor[rgb]{ .612,  0,  .024}{0.110 } & 0.257  & 0.301  & 0.115  & 0.225  & 0.330  \\
    UQI   & 0.47  & 0.63  & 0.75  & 0.51  & 0.65  & \cellcolor[rgb]{ .776,  .937,  .808}\textcolor[rgb]{ 0,  .38,  0}{0.77 } & \cellcolor[rgb]{ 1,  .78,  .808}\textcolor[rgb]{ .612,  0,  .024}{0.38 } & 0.57  & 0.68  & 0.40  & 0.58  & 0.71  & UQI   & 0.523  & 0.707  & 0.828  & 0.545  & 0.732  & \cellcolor[rgb]{ .776,  .937,  .808}\textcolor[rgb]{ 0,  .38,  0}{0.855 } & \cellcolor[rgb]{ 1,  .78,  .808}\textcolor[rgb]{ .612,  0,  .024}{0.386 } & 0.701  & 0.769  & 0.416  & 0.636  & 0.802  \\
    IFC   & 1.53  & 2.71  & 4.28  & 1.87  & 3.10  & \cellcolor[rgb]{ .776,  .937,  .808}\textcolor[rgb]{ 0,  .38,  0}{4.69 } & \cellcolor[rgb]{ 1,  .78,  .808}\textcolor[rgb]{ .612,  0,  .024}{1.36 } & 2.57  & 3.93  & 1.41  & 2.55  & 4.07  & IFC   & 0.904  & 1.861  & 3.116  & 1.055  & 2.081  & \cellcolor[rgb]{ .776,  .937,  .808}\textcolor[rgb]{ 0,  .38,  0}{3.345 } & \cellcolor[rgb]{ 1,  .78,  .808}\textcolor[rgb]{ .612,  0,  .024}{0.710 } & 1.852  & 2.917  & 0.739  & 1.718  & 2.994  \\
    NQM   & 17.39  & 23.48  & 28.96  & 19.14  & 24.46  & \cellcolor[rgb]{ .776,  .937,  .808}\textcolor[rgb]{ 0,  .38,  0}{29.54 } & \cellcolor[rgb]{ 1,  .78,  .808}\textcolor[rgb]{ .612,  0,  .024}{12.94 } & 20.63  & 23.13  & 15.07  & 21.04  & 20.41  & NQM   & 3.232  & 5.827  & 8.620  & 4.358  & 7.173  & \cellcolor[rgb]{ .776,  .937,  .808}\textcolor[rgb]{ 0,  .38,  0}{10.171 } & \cellcolor[rgb]{ 1,  .78,  .808}\textcolor[rgb]{ .612,  0,  .024}{1.272 } & 6.181  & 7.367  & 2.381  & 5.723  & 8.648  \\
    WSNR  & 25.16  & 30.82  & 36.42  & 26.67  & 31.68  & \cellcolor[rgb]{ .776,  .937,  .808}\textcolor[rgb]{ 0,  .38,  0}{37.07 } & \cellcolor[rgb]{ 1,  .78,  .808}\textcolor[rgb]{ .612,  0,  .024}{19.86 } & 27.59  & 28.88  & 21.34  & 27.20  & 25.56  & WSNR  & 18.157  & 23.618  & 29.022  & 19.090  & 24.177  & \cellcolor[rgb]{ .776,  .937,  .808}\textcolor[rgb]{ 0,  .38,  0}{29.402 } & \cellcolor[rgb]{ 1,  .78,  .808}\textcolor[rgb]{ .612,  0,  .024}{16.009 } & 20.130  & 25.522  & 16.518  & 16.698  & 25.584  \\
    SNR   & 16.02  & 17.95  & 19.80  & 17.16  & 19.05  & \cellcolor[rgb]{ .776,  .937,  .808}\textcolor[rgb]{ 0,  .38,  0}{20.98 } & \cellcolor[rgb]{ 1,  .78,  .808}\textcolor[rgb]{ .612,  0,  .024}{13.74 } & 16.40  & 18.03  & 14.28  & 16.96  & 18.41  & SNR   & 10.883  & 13.121  & 15.285  & 11.956  & 14.289  & \cellcolor[rgb]{ .776,  .937,  .808}\textcolor[rgb]{ 0,  .38,  0}{16.593 } & \cellcolor[rgb]{ 1,  .78,  .808}\textcolor[rgb]{ .612,  0,  .024}{8.197 } & 12.916  & 13.487  & 9.583  & 11.783  & 14.943  \\
    FSIM  & \cellcolor[rgb]{ 1,  .78,  .808}\textcolor[rgb]{ .612,  0,  .024}{0.77 } & 0.85  & 0.90  & 0.79  & 0.87  & \cellcolor[rgb]{ .776,  .937,  .808}\textcolor[rgb]{ 0,  .38,  0}{0.92 } & 0.80  & 0.87  & 0.90  & 0.80  & 0.86  & 0.91  & FSIM  & 0.780  & 0.860  & 0.914  & \cellcolor[rgb]{ 1,  .78,  .808}\textcolor[rgb]{ .612,  0,  .024}{0.761 } & 0.860  & \cellcolor[rgb]{ .776,  .937,  .808}\textcolor[rgb]{ 0,  .38,  0}{0.917 } & 0.780  & 0.856  & 0.899  & 0.761  & 0.840  & 0.891  \\
    GMSD  & \cellcolor[rgb]{ 1,  .78,  .808}\textcolor[rgb]{ .612,  0,  .024}{0.18 } & 0.09  & 0.04  & 0.13  & 0.08  & \cellcolor[rgb]{ .776,  .937,  .808}\textcolor[rgb]{ 0,  .38,  0}{0.04 } & 0.14  & 0.09  & 0.05  & 0.13  & 0.09  & 0.04  & GMSD  & \cellcolor[rgb]{ 1,  .78,  .808}\textcolor[rgb]{ .612,  0,  .024}{0.148 } & 0.076  & 0.040  & 0.139  & 0.070  & \cellcolor[rgb]{ .776,  .937,  .808}\textcolor[rgb]{ 0,  .38,  0}{0.034 } & 0.137  & 0.080  & 0.045  & 0.142  & 0.090  & 0.044  \\
    \midrule
    \textit{$SSQP_F$} & \cellcolor[rgb]{ 1,  .78,  .808}\textcolor[rgb]{ .612,  0,  .024}{1.45 } & 5.65  & 15.78  & 2.84  & 8.98  & 15.24  & 1.65  & 17.69  & \cellcolor[rgb]{ .776,  .937,  .808}\textcolor[rgb]{ 0,  .38,  0}{18.98 } & 2.40  & 16.62  & 18.34  & \textit{$SSQP_F$} & \cellcolor[rgb]{ 1,  .78,  .808}\textcolor[rgb]{ .612,  0,  .024}{0.79 } & 11.45  & 15.78  & 7.10  & 8.40  & 17.75  & 2.12  & 4.53  & 15.68  & 4.01  & 13.23  & \cellcolor[rgb]{ .776,  .937,  .808}\textcolor[rgb]{ 0,  .38,  0}{17.89 } \\
    \textit{$SSQP_B$} & \cellcolor[rgb]{ 1,  .78,  .808}\textcolor[rgb]{ .612,  0,  .024}{1.37 } & 5.01  & 18.47  & 2.10  & 11.06  & 17.81  & 2.63  & 15.04  & \cellcolor[rgb]{ .776,  .937,  .808}\textcolor[rgb]{ 0,  .38,  0}{18.73 } & 2.22  & 12.30  & 17.64  & \textit{$SSQP_B$} & \cellcolor[rgb]{ 1,  .78,  .808}\textcolor[rgb]{ .612,  0,  .024}{1.86 } & 12.23  & 18.66  & 3.87  & 14.32  & 18.81  & 2.51  & 12.53  & 12.50  & 2.19  & 4.47  & \cellcolor[rgb]{ .776,  .937,  .808}\textcolor[rgb]{ 0,  .38,  0}{19.61 } \\
    \midrule
    MOS   & \cellcolor[rgb]{ 1,  .78,  .808}\textcolor[rgb]{ .612,  0,  .024}{0.00 } & 5.88  & 12.38  & 2.63  & 11.00  & 14.75  & 3.25  & 15.13  & \cellcolor[rgb]{ .776,  .937,  .808}\textcolor[rgb]{ 0,  .38,  0}{18.00 } & 3.50  & 13.00  & 16.25  & MOS   & \cellcolor[rgb]{ 1,  .78,  .808}\textcolor[rgb]{ .612,  0,  .024}{0.00 } & 0.50  & 8.38  & 5.13  & 9.38  & 18.00  & 4.88  & 13.13  & 17.63  & 8.00  & 15.25  & \cellcolor[rgb]{ .776,  .937,  .808}\textcolor[rgb]{ 0,  .38,  0}{18.63 } \\
    \midrule
    \multicolumn{13}{c|}{Building patch from the Subset of Regular Structured Patches}                        & \multicolumn{13}{c}{Face patch from the Subset of High-level Structured Patches} \\
    \midrule
    \multirow{2}[2]{*}{Metric} & \multicolumn{3}{c|}{JPEG} & \multicolumn{3}{c|}{CNN} & \multicolumn{3}{c|}{GAN (lambda=0.1)} & \multicolumn{3}{c|}{GAN (lambda=0.01)} & \multirow{2}[2]{*}{Metric} & \multicolumn{3}{c|}{JPEG} & \multicolumn{3}{c|}{CNN} & \multicolumn{3}{c|}{GAN (lambda=0.1)} & \multicolumn{3}{c}{GAN (lambda=0.01)} \\
          & \multicolumn{1}{c}{QF05} & \multicolumn{1}{c}{QF10} & \multicolumn{1}{c|}{QF20} & \multicolumn{1}{c}{QF05} & \multicolumn{1}{c}{QF10} & \multicolumn{1}{c|}{QF20} & \multicolumn{1}{c}{QF05} & \multicolumn{1}{c}{QF10} & \multicolumn{1}{c|}{QF20} & \multicolumn{1}{c}{QF05} & \multicolumn{1}{c}{QF10} & \multicolumn{1}{c|}{QF20} &       & \multicolumn{1}{c}{QF05} & \multicolumn{1}{c}{QF10} & \multicolumn{1}{c|}{QF20} & \multicolumn{1}{c}{QF05} & \multicolumn{1}{c}{QF10} & \multicolumn{1}{c|}{QF20} & \multicolumn{1}{c}{QF05} & \multicolumn{1}{c}{QF10} & \multicolumn{1}{c|}{QF20} & \multicolumn{1}{c}{QF05} & \multicolumn{1}{c}{QF10} & \multicolumn{1}{c}{QF20} \\
    \midrule
    \midrule
    PSNR  & 23.183  & 25.958  & 28.455  & 26.299  & 29.424  & \cellcolor[rgb]{ .776,  .937,  .808}\textcolor[rgb]{ 0,  .38,  0}{31.954 } & \cellcolor[rgb]{ 1,  .78,  .808}\textcolor[rgb]{ .612,  0,  .024}{21.087 } & 26.364  & 27.776  & 23.912  & 26.218  & 29.805  & PSNR  & 25.414  & 27.890  & 30.193  & 26.995  & 29.857  & \cellcolor[rgb]{ .776,  .937,  .808}\textcolor[rgb]{ 0,  .38,  0}{32.168 } & \cellcolor[rgb]{ 1,  .78,  .808}\textcolor[rgb]{ .612,  0,  .024}{23.907 } & 26.703  & 29.299  & 24.385  & 27.874  & 29.267  \\
    SSIM  & 0.598  & 0.764  & 0.857  & 0.692  & 0.838  & \cellcolor[rgb]{ .776,  .937,  .808}\textcolor[rgb]{ 0,  .38,  0}{0.909 } & \cellcolor[rgb]{ 1,  .78,  .808}\textcolor[rgb]{ .612,  0,  .024}{0.482 } & 0.766  & 0.849  & 0.586  & 0.749  & 0.871  & SSIM  & 0.704  & 0.795  & 0.868  & 0.791  & 0.870  & \cellcolor[rgb]{ .776,  .937,  .808}\textcolor[rgb]{ 0,  .38,  0}{0.910 } & 0.627  & 0.711  & 0.824  & \cellcolor[rgb]{ 1,  .78,  .808}\textcolor[rgb]{ .612,  0,  .024}{0.619 } & 0.782  & 0.872  \\
    MSIM  & 0.902  & 0.956  & 0.981  & 0.925  & 0.969  & \cellcolor[rgb]{ .776,  .937,  .808}\textcolor[rgb]{ 0,  .38,  0}{0.987 } & \cellcolor[rgb]{ 1,  .78,  .808}\textcolor[rgb]{ .612,  0,  .024}{0.860 } & 0.956  & 0.979  & 0.898  & 0.948  & 0.981  & MSIM  & 0.882  & 0.944  & 0.976  & 0.907  & 0.965  & \cellcolor[rgb]{ .776,  .937,  .808}\textcolor[rgb]{ 0,  .38,  0}{0.982 } & \cellcolor[rgb]{ 1,  .78,  .808}\textcolor[rgb]{ .612,  0,  .024}{0.817 } & 0.932  & 0.958  & 0.838  & 0.938  & 0.970  \\
    VSNR  & 21.122  & 26.609  & 32.235  & 24.829  & 30.120  & \cellcolor[rgb]{ .776,  .937,  .808}\textcolor[rgb]{ 0,  .38,  0}{35.915 } & \cellcolor[rgb]{ 1,  .78,  .808}\textcolor[rgb]{ .612,  0,  .024}{18.117 } & 27.438  & 30.820  & 21.986  & 26.756  & 32.506  & VSNR  & 18.325  & 22.985  & 27.589  & 20.081  & 25.222  & \cellcolor[rgb]{ .776,  .937,  .808}\textcolor[rgb]{ 0,  .38,  0}{29.897 } & \cellcolor[rgb]{ 1,  .78,  .808}\textcolor[rgb]{ .612,  0,  .024}{16.721 } & 22.330  & 25.900  & 17.618  & 22.721  & 27.157  \\
    VIF   & 0.223  & 0.375  & 0.533  & 0.304  & 0.466  & \cellcolor[rgb]{ .776,  .937,  .808}\textcolor[rgb]{ 0,  .38,  0}{0.609 } & \cellcolor[rgb]{ 1,  .78,  .808}\textcolor[rgb]{ .612,  0,  .024}{0.215 } & 0.385  & 0.542  & 0.250  & 0.379  & 0.544  & VIF   & 0.183  & 0.316  & 0.477  & 0.239  & 0.386  & \cellcolor[rgb]{ .776,  .937,  .808}\textcolor[rgb]{ 0,  .38,  0}{0.532 } & \cellcolor[rgb]{ 1,  .78,  .808}\textcolor[rgb]{ .612,  0,  .024}{0.174 } & 0.328  & 0.460  & 0.184  & 0.320  & 0.475  \\
    VIFP  & 0.249  & 0.359  & 0.455  & 0.348  & 0.468  & \cellcolor[rgb]{ .776,  .937,  .808}\textcolor[rgb]{ 0,  .38,  0}{0.565 } & \cellcolor[rgb]{ 1,  .78,  .808}\textcolor[rgb]{ .612,  0,  .024}{0.213 } & 0.382  & 0.474  & 0.275  & 0.371  & 0.488  & VIFP  & 0.223  & 0.346  & 0.460  & 0.301  & 0.438  & \cellcolor[rgb]{ .776,  .937,  .808}\textcolor[rgb]{ 0,  .38,  0}{0.534 } & \cellcolor[rgb]{ 1,  .78,  .808}\textcolor[rgb]{ .612,  0,  .024}{0.186 } & 0.327  & 0.408  & 0.187  & 0.343  & 0.460  \\
    UQI   & 0.545  & 0.745  & 0.852  & 0.607  & 0.804  & \cellcolor[rgb]{ .776,  .937,  .808}\textcolor[rgb]{ 0,  .38,  0}{0.898 } & \cellcolor[rgb]{ 1,  .78,  .808}\textcolor[rgb]{ .612,  0,  .024}{0.436 } & 0.733  & 0.840  & 0.508  & 0.695  & 0.855  & UQI   & 0.443  & 0.624  & 0.755  & 0.550  & 0.710  & \cellcolor[rgb]{ .776,  .937,  .808}\textcolor[rgb]{ 0,  .38,  0}{0.801 } & \cellcolor[rgb]{ 1,  .78,  .808}\textcolor[rgb]{ .612,  0,  .024}{0.378 } & 0.574  & 0.686  & 0.400  & 0.604  & 0.730  \\
    IFC   & 1.556  & 2.644  & 3.865  & 2.168  & 3.367  & \cellcolor[rgb]{ .776,  .937,  .808}\textcolor[rgb]{ 0,  .38,  0}{4.547 } & \cellcolor[rgb]{ 1,  .78,  .808}\textcolor[rgb]{ .612,  0,  .024}{1.500 } & 2.738  & 3.941  & 1.749  & 2.728  & 3.986  & IFC   & 0.899  & 1.568  & 2.470  & 1.288  & 2.061  & \cellcolor[rgb]{ .776,  .937,  .808}\textcolor[rgb]{ 0,  .38,  0}{2.942 } & \cellcolor[rgb]{ 1,  .78,  .808}\textcolor[rgb]{ .612,  0,  .024}{0.859 } & 1.630  & 2.372  & 0.910  & 1.606  & 2.517  \\
    NQM   & 10.265  & 13.606  & 16.828  & 13.582  & 17.215  & \cellcolor[rgb]{ .776,  .937,  .808}\textcolor[rgb]{ 0,  .38,  0}{20.296 } & \cellcolor[rgb]{ 1,  .78,  .808}\textcolor[rgb]{ .612,  0,  .024}{8.901 } & 14.591  & 17.416  & 11.361  & 14.378  & 18.235  & NQM   & 9.296  & 12.192  & 15.191  & 11.171  & 14.454  & \cellcolor[rgb]{ .776,  .937,  .808}\textcolor[rgb]{ 0,  .38,  0}{17.289 } & \cellcolor[rgb]{ 1,  .78,  .808}\textcolor[rgb]{ .612,  0,  .024}{8.289 } & 11.657  & 14.301  & 8.948  & 12.491  & 15.305  \\
    WSNR  & 19.850  & 25.593  & 30.892  & 22.418  & 27.117  & \cellcolor[rgb]{ .776,  .937,  .808}\textcolor[rgb]{ 0,  .38,  0}{32.003 } & \cellcolor[rgb]{ 1,  .78,  .808}\textcolor[rgb]{ .612,  0,  .024}{15.938 } & 21.888  & 22.089  & 19.493  & 20.919  & 28.291  & WSNR  & 24.920  & 30.330  & 35.278  & 25.937  & 31.936  & \cellcolor[rgb]{ .776,  .937,  .808}\textcolor[rgb]{ 0,  .38,  0}{36.152 } & \cellcolor[rgb]{ 1,  .78,  .808}\textcolor[rgb]{ .612,  0,  .024}{21.387 } & 27.221  & 29.293  & 22.530  & 28.165  & 27.267  \\
    SNR   & 12.706  & 15.481  & 17.977  & 15.821  & 18.947  & \cellcolor[rgb]{ .776,  .937,  .808}\textcolor[rgb]{ 0,  .38,  0}{21.476 } & \cellcolor[rgb]{ 1,  .78,  .808}\textcolor[rgb]{ .612,  0,  .024}{10.609 } & 15.886  & 17.299  & 13.434  & 15.741  & 19.327  & SNR   & 18.950  & 21.426  & 23.729  & 20.531  & 23.394  & \cellcolor[rgb]{ .776,  .937,  .808}\textcolor[rgb]{ 0,  .38,  0}{25.704 } & \cellcolor[rgb]{ 1,  .78,  .808}\textcolor[rgb]{ .612,  0,  .024}{17.443 } & 20.240  & 22.836  & 17.921  & 21.410  & 22.803  \\
    FSIM  & 0.795  & 0.846  & 0.899  & 0.825  & 0.908  & \cellcolor[rgb]{ .776,  .937,  .808}\textcolor[rgb]{ 0,  .38,  0}{0.947 } & \cellcolor[rgb]{ 1,  .78,  .808}\textcolor[rgb]{ .612,  0,  .024}{0.786 } & 0.888  & 0.930  & 0.819  & 0.882  & 0.932  & FSIM  & \cellcolor[rgb]{ 1,  .78,  .808}\textcolor[rgb]{ .612,  0,  .024}{0.800 } & 0.876  & 0.921  & 0.867  & 0.913  & \cellcolor[rgb]{ .776,  .937,  .808}\textcolor[rgb]{ 0,  .38,  0}{0.940 } & 0.826  & 0.882  & 0.922  & 0.844  & 0.897  & 0.930  \\
    GMSD  & \cellcolor[rgb]{ 1,  .78,  .808}\textcolor[rgb]{ .612,  0,  .024}{0.170 } & 0.098  & 0.051  & 0.148  & 0.082  & \cellcolor[rgb]{ .776,  .937,  .808}\textcolor[rgb]{ 0,  .38,  0}{0.038 } & 0.146  & 0.096  & 0.051  & 0.161  & 0.104  & 0.050  & GMSD  & \cellcolor[rgb]{ 1,  .78,  .808}\textcolor[rgb]{ .612,  0,  .024}{0.136 } & 0.070  & 0.035  & 0.109  & 0.053  & \cellcolor[rgb]{ .776,  .937,  .808}\textcolor[rgb]{ 0,  .38,  0}{0.027 } & 0.118  & 0.072  & 0.042  & 0.116  & 0.070  & 0.037  \\
    \midrule
    \textit{$SSQP_F$} & \cellcolor[rgb]{ 1,  .78,  .808}\textcolor[rgb]{ .612,  0,  .024}{0.51 } & 3.10  & 12.67  & 4.27  & 15.08  & \cellcolor[rgb]{ .776,  .937,  .808}\textcolor[rgb]{ 0,  .38,  0}{17.90 } & 0.70  & 9.01  & 16.12  & 2.70  & 6.34  & 16.75  & \textit{$SSQP_F$} & \cellcolor[rgb]{ 1,  .78,  .808}\textcolor[rgb]{ .612,  0,  .024}{1.15 } & 7.49  & 16.13  & 1.45  & 14.86  & 14.86  & 4.84  & 2.30  & \cellcolor[rgb]{ .776,  .937,  .808}\textcolor[rgb]{ 0,  .38,  0}{16.36 } & 1.67  & 8.53  & 13.98  \\
    \textit{$SSQP_B$} & \cellcolor[rgb]{ 1,  .78,  .808}\textcolor[rgb]{ .612,  0,  .024}{1.32 } & 6.81  & 14.19  & 10.85  & 15.84  & 17.04  & 1.54  & 8.02  & 16.54  & 5.39  & 6.86  & \cellcolor[rgb]{ .776,  .937,  .808}\textcolor[rgb]{ 0,  .38,  0}{17.27 } & \textit{$SSQP_B$} & \cellcolor[rgb]{ 1,  .78,  .808}\textcolor[rgb]{ .612,  0,  .024}{1.32 } & 6.81  & 14.19  & 10.85  & 15.84  & 17.27  & 1.43  & 8.02  & 16.54  & 5.39  & 6.86  & \cellcolor[rgb]{ .776,  .937,  .808}\textcolor[rgb]{ 0,  .38,  0}{17.73 } \\
    \midrule
    MOS   & \cellcolor[rgb]{ 1,  .78,  .808}\textcolor[rgb]{ .612,  0,  .024}{0.00 } & 5.25  & 8.75  & 5.63  & 11.50  & 16.50  & 2.00  & 13.88  & 17.75  & 4.88  & 13.38  & \cellcolor[rgb]{ .776,  .937,  .808}\textcolor[rgb]{ 0,  .38,  0}{18.38 } & MOS   & \cellcolor[rgb]{ 1,  .78,  .808}\textcolor[rgb]{ .612,  0,  .024}{0.25 } & 5.00  & 11.25  & 1.75  & 16.13  & 19.50  & 2.00  & 10.13  & 16.63  & 1.38  & 13.63  & \cellcolor[rgb]{ .776,  .937,  .808}\textcolor[rgb]{ 0,  .38,  0}{20.75 } \\
    \bottomrule
    \end{tabular}%
}
  \label{tab:comparison}%
\end{table*}%
We call the proposed IQA model the $SSQP$ (Structural and Statistical Quality Predictor), and 
compared its performance against many leading 2D FR IQA models on the new 
generative image database. The experimental results are summarized in 
Table~\ref{tab:PerfComp1}. Note that  
$SSQP_F$ and $SSQP_B$ are two versions of $SSQP$ depending on the feature calculation 
method. The thirteen existing models that were used for performance benchmarking
are PSNR, SSIM~\cite{cit:Wang2004}, multi-scale SSIM index (MSSIM)
\cite{cit:Wang2003}, visual signal-to-noise ratio (VSNR)~\cite{cit:Chandler2007},
visual information fidelity (VIF)~\cite{cit:Bovik2005}, pixel-based VIF (VIFP),
universal quality index (UQI)~\cite{cit:Wang2002}, information fidelity criterion (IFC)
\cite{cit:Sheikh2005}, noise quality measure (NQM)~\cite{cit:Damera2000}, weighted 
signal-to-noise ratio (WSNR), signal-to-noise ratio (SNR), feature similarity index  (FSIM)
\cite{cit:Zhang2011}, and gradient magnitude similarity deviation (GMSD)~\cite{cit:Xue2014}. 
The parameters used in each were the default settings mentioned in their original papers.\\
\indent Table~\ref{tab:PerfComp1} shows that $SSQP$ significantly outperformed all of the 
compared FR on the subset of full-frame images, $SSQP_B$ and $SSQP_F$ achieved 
PCC=0.95 and PCC=0.93 while the best performance of a previously existing FR metric 
was GMSD with PCC=0.89. Note that as compared to performance on well-known image 
quality databases such as LIVE~\cite{cit:Sheikh2006} and TID~\cite{cit:Ponomarenko2015}, 
where several state-of-the-art 2D FR metrics have already achieved excellent performance 
($>0.95$ in PCC), the best PCC value achieved by any of the existing models on the new
database was below 0.90, reflecting the more challenging aspects of the new data resource. 
On the subset of random structured blocks, the performance gap between $SSQP$ and 
the best existing models was even larger. Specifically, the PCC values attained by $SSQP_B$ 
and $SSQP_F$ were 0.91 and 0.87, respectively, whereas that of GMSD was 0.71. 
This could be because that most existing benchmark FR IQA models lack the ability to 
capture the requisite types of statistical similarity.
The prediction accuracy of $SSQP$ improves even further (PCC=0.96 for $SSQP_F$) 
on the subset of regular structured patches and PCC=0.96 for $SSQP_F$ 
on the subset of high-level structured patches, while the best existing models were 
FSIM (PCC=0.93) and IFC (PCC=0.90), respectively. Overall, $SSQP_B$ was a better 
predictor than $SSQP_F$ because it was better able to account for the diversity of local 
characteristics in images.\\
\indent In order to better understand the superiority of $SSQP$, we analyzed the limitations
of the existing models. Fig.~\ref{fig:comparison} show the four reference images from all 
the subsets and their corresponding JPEG QF05 images and $GAN$ images 
($\lambda=0.1$ with input of JPEG QF05). Table~\ref{tab:comparison} shows the 
prediction performances of the benchmark models as well as their MOS. 
For each model, we highlighted two test images that caused the best and the worst scores 
in green and red, respectively. 
As shown in the results, most of the existing models selected CNN (with input of 
JPEG Q20) as the highest quality image and $GAN$ ($\lambda$=0.1 with input of 
JPEG QF05) as the worst. However, this was not always the case. In Table
\ref{tab:comparison}, the subjective scores indicate that JPEG QF05 was the worst 
image while $GAN$ ($\lambda$=0.1 with input of JPEG QF20) was the best image 
among the given examples. One likely reason for the inaccuracies of the existing 
metrics is that they aim to assess preservation of pixel-wise fidelity, rather than 
innate quality. This could explain why they tolerate severe blocking artifacts, but not 
moderate structural changes, even though the former are more annoying. 
This might explain why the existing models choose CNN-generated images rather than 
$GAN$ images as higher quality. However, CNN images introduce blur
(as in Fig.~\ref{fig:exbase}), which deteriorates the viewing experience. 
$SSQP$ evaluates the natural quality based on both structural and statistical similarities, 
which are not as strongly affected by pixel-wise differences. Moreover, the parallel 
boosting system is able to optimize the relative weights by considering the significance 
of the structural degradations against statistical degradations. This was experimentally 
verified in Table~\ref{tab:comparison}, where the $SSQP$ objective scores closely 
fit the MOS. \\
\begin{table}[t]
  \centering
  \caption{Percentages of the worst and best case matches bewteen $SSQP_B$ and MOS.}
    \begin{tabular}{c|c}
    \toprule
    \makecell{Percentage of matches \\ on lowest MOS images} & \makecell{Percentage of matches \\ on highest MOS images} \\
    \midrule
    83.3\% (15/18) & 27.8\% (5/18) \\
    \bottomrule
    \end{tabular}%
  \label{tab:PctTable}%
    \vspace{-3mm}
\end{table}%
\begin{table}[t]
  \centering
  \caption{Distributions of highest MOS and $SSQP_B$ predictions 
  against method of image generation.}
    \begin{tabular}{c|ccc}
    \toprule
    \multicolumn{1}{c|}{} & \multicolumn{3}{c}{\# of images having the highest score} \\
    \makecell{Input \\ image} & \multicolumn{1}{c}{CNN w/ QF20} & \multicolumn{1}{c}{\makecell{GAN w/ QF20 \\ ($\lambda=0.1$)}} & \multicolumn{1}{c}{\makecell{GAN w/ QF20 \\ ($\lambda=0.01$)}} \\
    \midrule
    MOS   & 1     & 13    & 4 \\
    $SSQP_B$ & 3     & 2     & 13 \\
    \bottomrule
    \end{tabular}%
  \label{tab:highest}%
      \vspace{-3mm}
\end{table}%
\begin{figure*}
        \centering
        \begin{subfigure}[t]{0.15\textwidth}
                \centering
                \includegraphics[width=\textwidth]{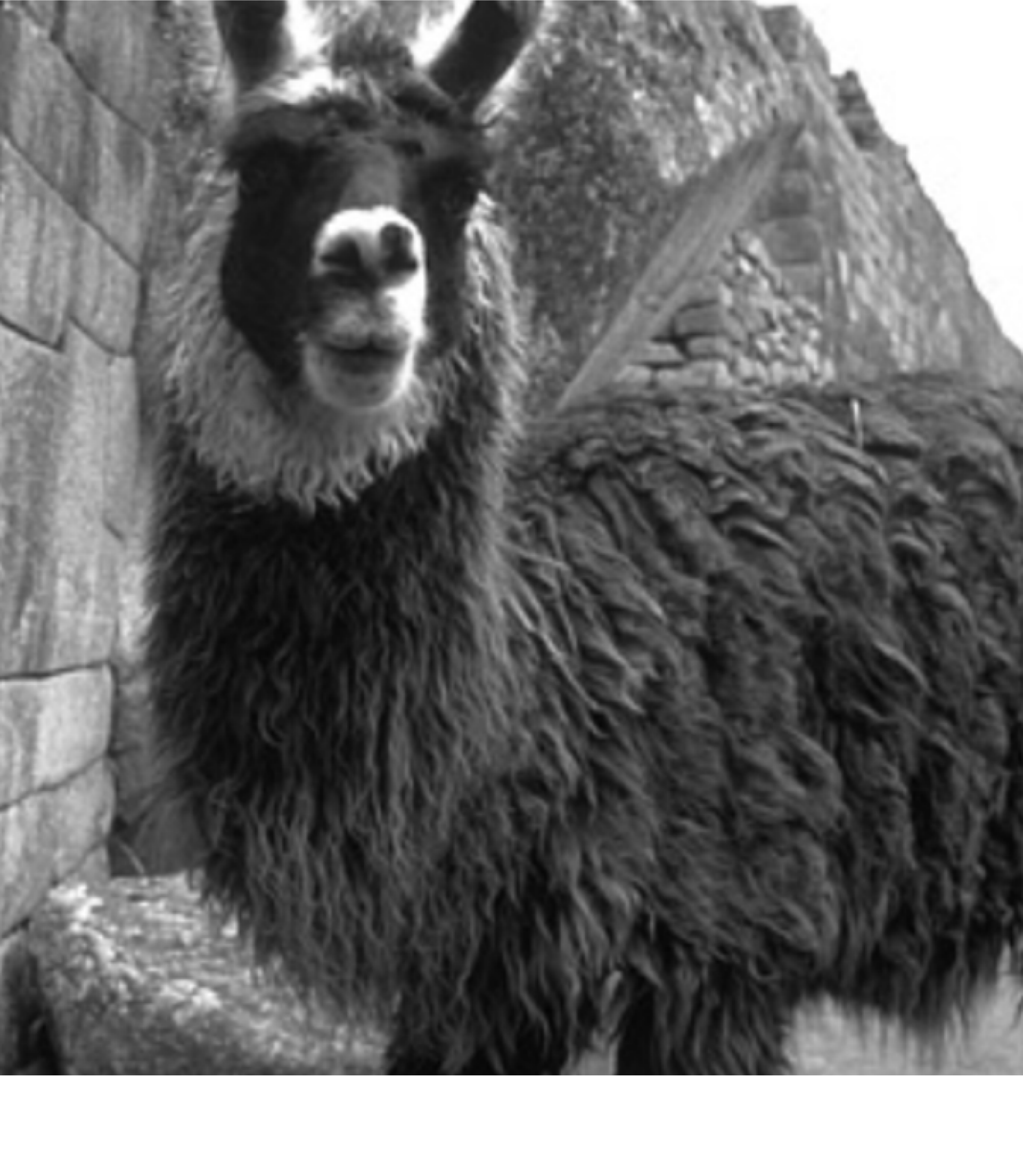}
                \caption{}
		\label{fig:Llama1}
        \end{subfigure}
        \begin{subfigure}[t]{0.15\textwidth}
                \centering
                \includegraphics[width=\textwidth]{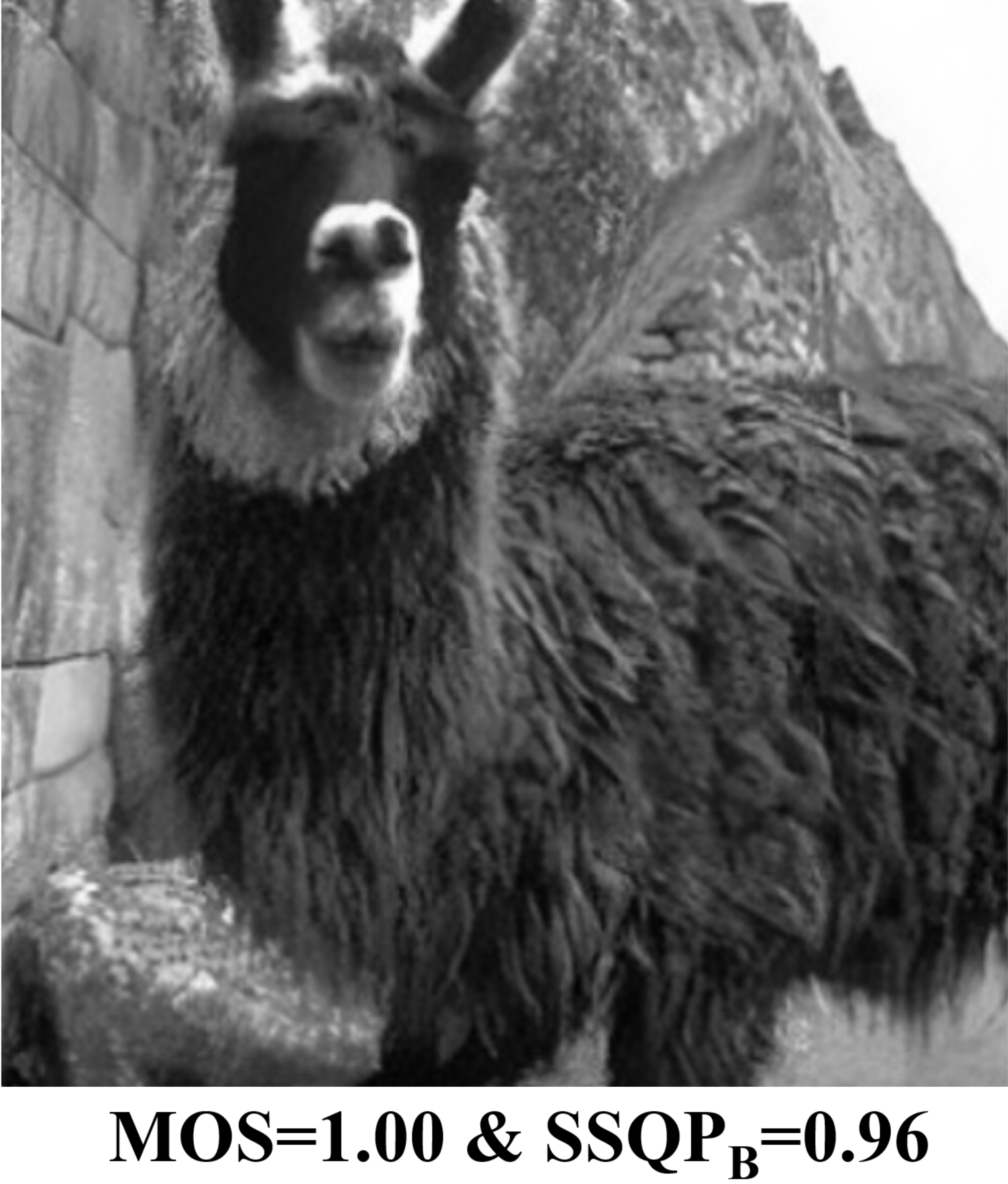}
                \caption{}
		\label{fig:Llama2}
        \end{subfigure}
        \begin{subfigure}[t]{0.15\textwidth}
                \centering
                \includegraphics[width=\textwidth]{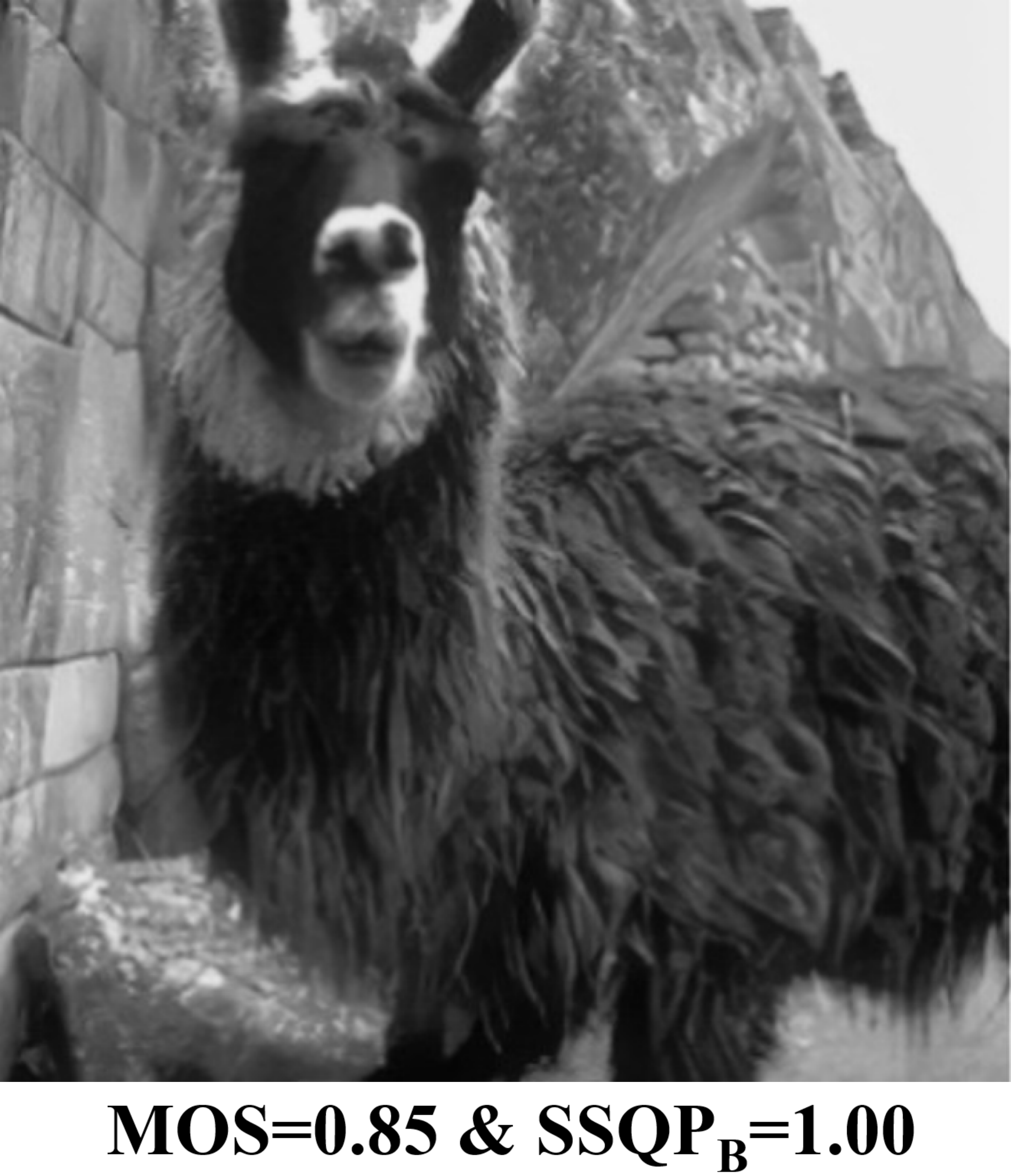}
                \caption{}
		\label{fig:Llama3}
        \end{subfigure}
        \begin{subfigure}[t]{0.16\textwidth}
                \centering
                \includegraphics[width=\textwidth]{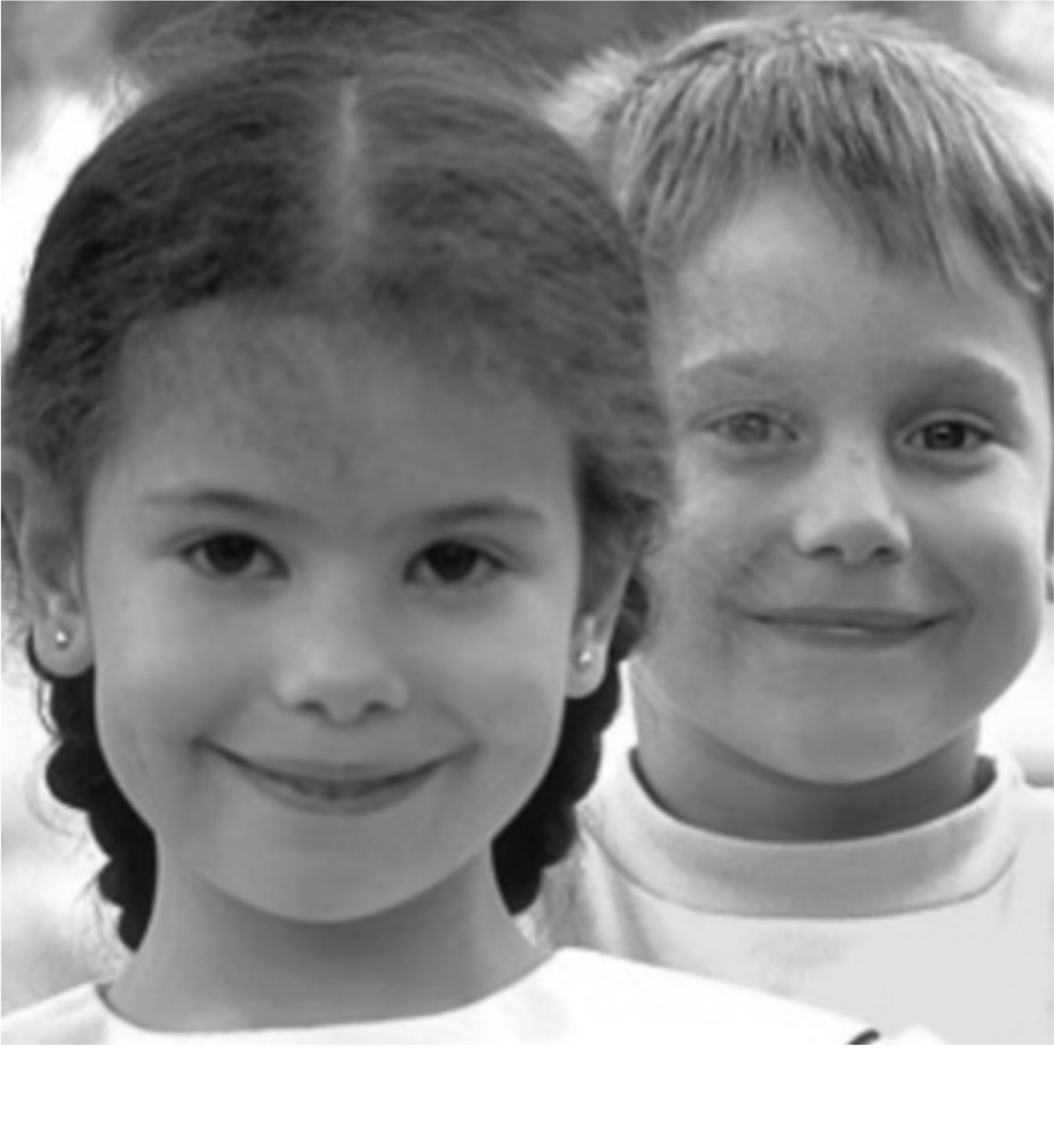}
                \caption{}
		\label{fig:EigenImg4}
        \end{subfigure}
        \begin{subfigure}[t]{0.16\textwidth}
                \centering
                \includegraphics[width=\textwidth]{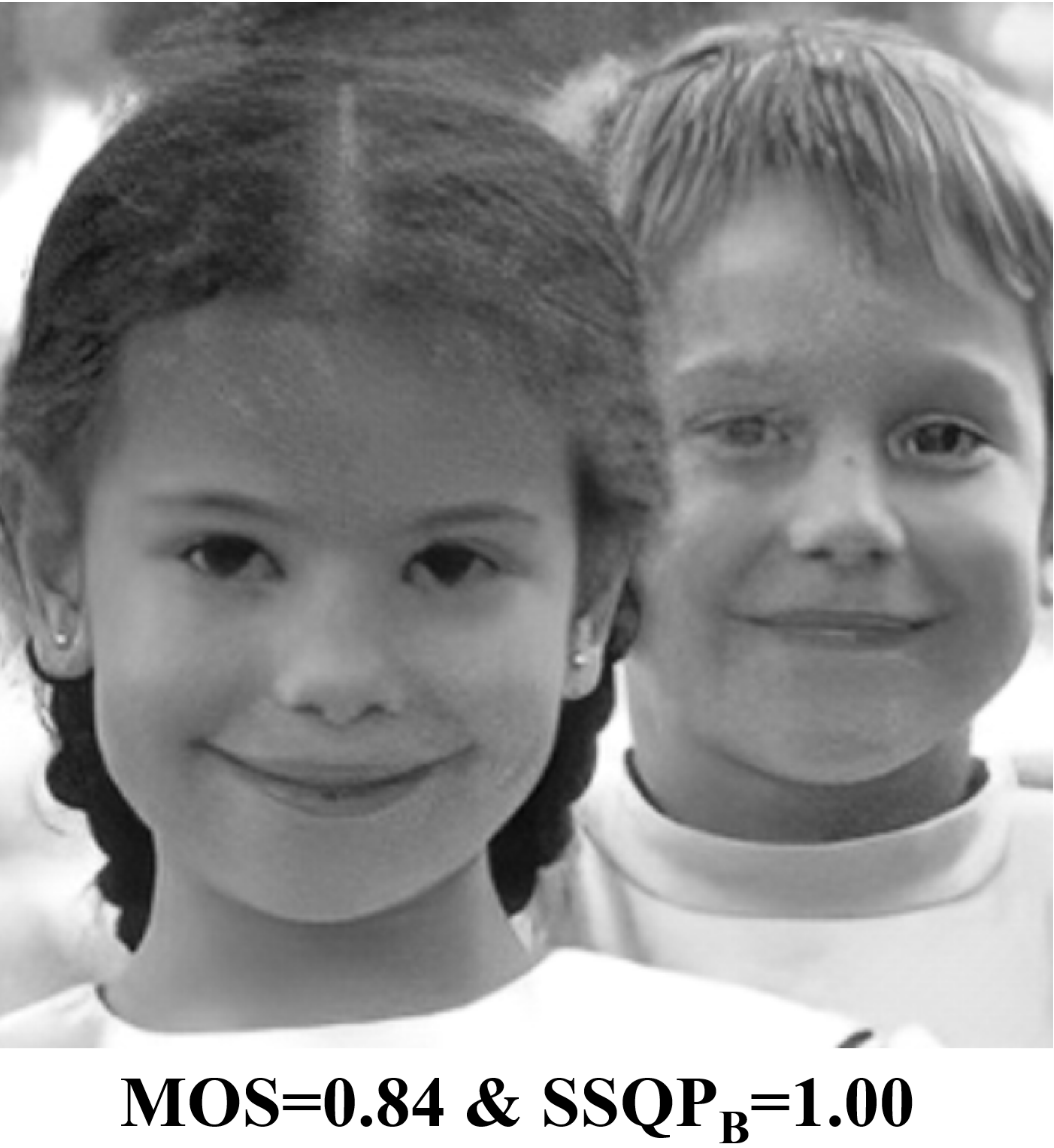}
                \caption{}
		\label{fig:EigenImg4}
        \end{subfigure}
        \begin{subfigure}[t]{0.16\textwidth}
                \centering
                \includegraphics[width=\textwidth]{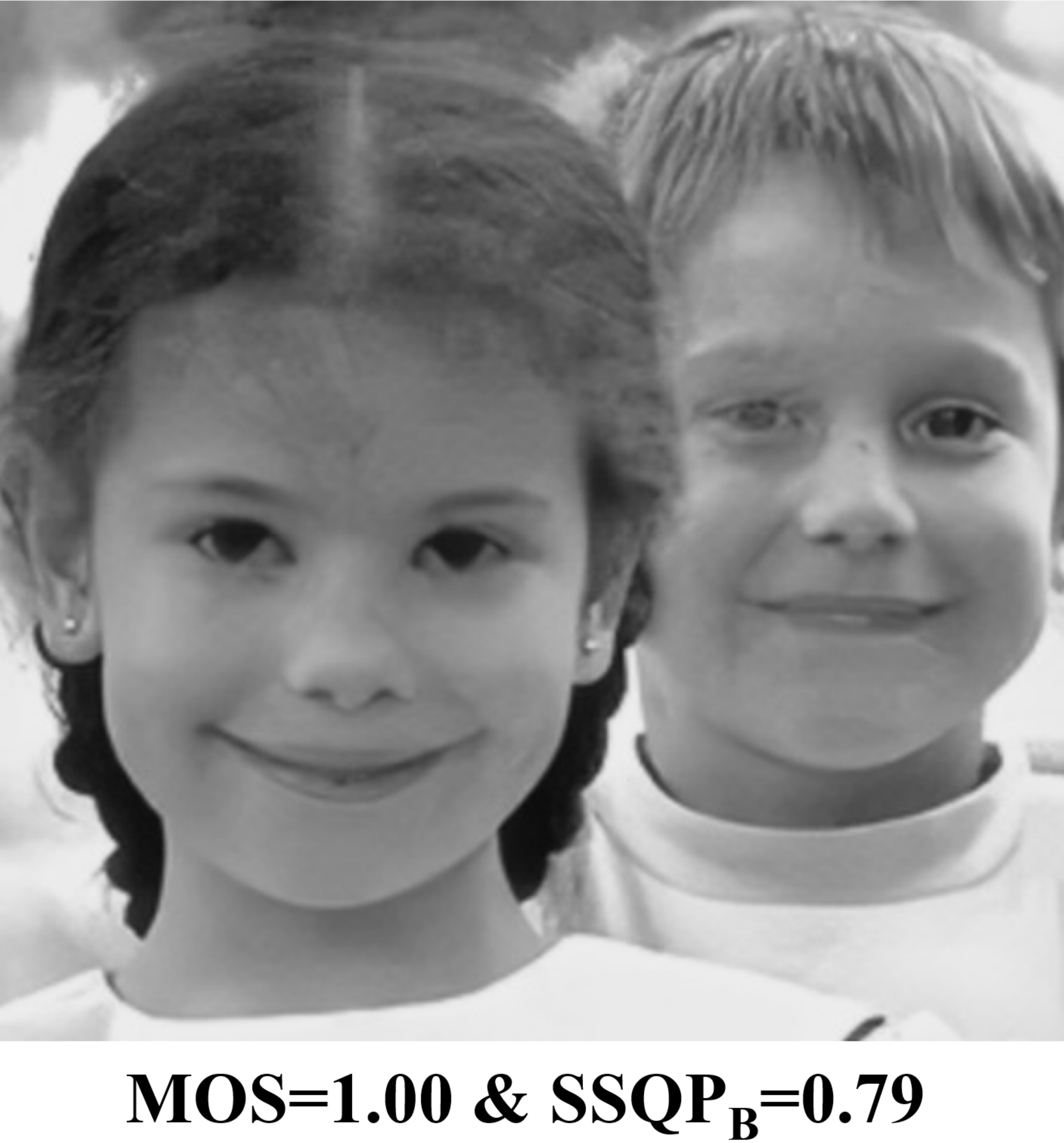}
                \caption{}
		\label{fig:EigenImg4}
        \end{subfigure}
\caption{The examples of the mismatch between MOS and $SSQP_B$ for the highest subjective
score: (a), (d) the original images, (b), (e) the output images w/ QF20 ($\lambda=0.1$), (c), (f) the output
images w/ QF20 ($\lambda=0.01$), where the values of MOS and $SSPQ_B$ are scaled
from 0 to 1.}
\label{fig:failure}
\vspace{-3mm}
\end{figure*}
\indent  In addition, we analyzed those cases where the proposed model fails. 
Specifically, we calculated the percentages of cases where the worst and the best case
MOS and $SSQP_B$ matched, as summarized in Table~\ref{tab:PctTable}.
The lowest MOS were mostly observed on JPEG-QF05 images (94\%), which 
$SSQP_B$ predicted easily. The highest MOS were 
distributed among three types of images, but the distributions diverged between MOS 
and $SSQP_B$, as shown in Table~\ref{tab:highest}, where each entry 
represents the number of times (of the 18 possible) that the highest MOS or $SSQP_B$ 
prediction occurred. The observed discrepancy is that the majority of highest MOS occurred 
on images generated using the GAN with $\lambda=0.1$,
while the highest $SSQP_B$ predictions tended to occur when $\lambda=0.01$. 
As the value of $\lambda$ increased, the resulting generative images become sharper 
and more naturalistic, but they also become less
natural if $\lambda$ becomes too large. Two examples are given in Fig.~\ref{fig:failure}
(MOS and $SSQP_B$ are scaled from 0 to 1). The human subjects preferred the GAN
to generate an abundant texture on the llama’s furry parts, while they felt it was unnatural 
to have excessive texture on people’s faces. Although $SSQP_B$ controls the weights 
between structural/statistical aspects of predicted quality via a multi-stage system, 
it sometimes fails to precisely determine the proper weights.

\subsection{Traditional Image Quality Prediction}\label{sec:5.3}
\begin{table*}[t]
  \centering
  \caption{Performance comparison on the LIVE image quality database 
  (median PCC, SRCC and RMSE across 1,000 train-test trials for $SSQP_F$ and $SSQP_B$).}
\scalebox{0.8}{
    \begin{tabular}{c|ccc|ccc|ccc|ccc|ccc|ccc}
\cmidrule{2-19}    \multicolumn{1}{r}{} & \multicolumn{3}{c|}{JP2K} & \multicolumn{3}{c|}{JPEG} & \multicolumn{3}{c|}{WN} & \multicolumn{3}{c|}{GBLUE} & \multicolumn{3}{c|}{FF} & \multicolumn{3}{c}{ALL} \\
    \multicolumn{1}{r}{} & PCC   & SROCC & RMSE  & PCC   & SROCC & RMSE  & PCC   & SROCC & RMSE  & PCC   & SROCC & RMSE  & PCC   & SROCC & RMSE  & PCC   & SROCC & \multicolumn{1}{c}{RMSE} \\
    \midrule
    PSNR  & 0.90  & 0.89  & 7.19  & 0.86  & 0.84  & 8.17  & 0.99  & 0.99  & 2.68  & 0.78  & 0.78  & 9.77  & 0.89  & 0.89  & 7.52  & 0.82  & 0.82  & 9.12  \\
    SSIM  & 0.90  & 0.93  & 16.20  & 0.85  & 0.90  & 15.99  & 0.96  & 0.96  & 15.97  & 0.85  & 0.89  & 15.72  & 0.90  & 0.94  & 16.45  & 0.74  & 0.85  & 16.10  \\
    MSIM  & 0.83  & 0.95  & 16.20  & 0.77  & 0.91  & 15.99  & 0.93  & 0.97  & 15.97  & 0.85  & 0.96  & 15.72  & 0.80  & 0.93  & 16.45  & 0.69  & 0.90  & 16.10  \\
    VSNR  & 0.95  & 0.94  & 4.94  & 0.94  & 0.91  & 5.35  & 0.98  & 0.98  & 3.34  & 0.93  & 0.94  & 5.64  & 0.90  & 0.90  & 7.10  & 0.89  & 0.89  & 7.36  \\
    VIF   & 0.94  & 0.95  & 16.20  & 0.93  & 0.91  & 15.99  & 0.96  & 0.98  & 15.97  & 0.96  & \textbf{0.97} & 15.72  & 0.96  & \textbf{0.96} & 16.45  & 0.94  & 0.95  & 16.10  \\
    VIFP  & 0.93  & 0.95  & 16.20  & 0.91  & 0.90  & 15.99  & 0.96  & 0.99  & 15.97  & 0.94  & 0.96  & 15.72  & 0.95  & 0.96  & 16.45  & 0.92  & 0.93  & 16.10  \\
    UQI   & 0.84  & 0.85  & 16.20  & 0.80  & 0.83  & 15.99  & 0.93  & 0.91  & 15.97  & 0.95  & 0.94  & 15.72  & 0.94  & 0.94  & 16.45  & 0.85  & 0.86  & 16.10  \\
    IFC   & 0.90  & 0.89  & 7.11  & 0.90  & 0.86  & 6.86  & 0.96  & 0.94  & 4.64  & 0.96  & 0.96  & 4.39  & 0.96  & 0.96  & 4.52  & 0.91  & 0.91  & 6.70  \\
    NQM   & 0.94  & 0.93  & 5.69  & 0.93  & 0.90  & 5.77  & 0.99  & \textbf{0.99} & 2.62  & 0.88  & 0.84  & 7.42  & 0.84  & 0.82  & 9.02  & 0.87  & 0.87  & 7.89  \\
    WSNR  & 0.92  & 0.91  & 6.48  & 0.93  & 0.89  & 5.80  & 0.98  & 0.97  & 3.50  & 0.92  & 0.91  & 6.26  & 0.72  & 0.76  & 12.08  & 0.88  & 0.88  & 7.79  \\
    SNR   & 0.87  & 0.86  & 8.09  & 0.85  & 0.83  & 8.50  & 0.97  & 0.97  & 3.80  & 0.76  & 0.75  & 10.19  & 0.89  & 0.90  & 7.36  & 0.81  & 0.81  & 9.41  \\
    FSIM  & 0.87  & 0.96  & 16.20  & 0.73  & 0.91  & 15.99  & 0.91  & 0.97  & 15.97  & 0.91  & 0.97  & 15.72  & 0.85  & 0.95  & 16.45  & 0.78  & 0.92  & 16.10  \\
    GMSD  & 0.96  & \textbf{0.96} & 4.36  & 0.94  & 0.91  & 5.25  & 0.97  & 0.97  & 4.16  & 0.96  & 0.96  & 4.34  & 0.94  & 0.94  & 5.63  & 0.91  & 0.91  & 6.73  \\
    \textit{$SSQP_F$} & 0.95  & 0.94  & 4.86  & 0.94  & 0.89  & 5.53  & 0.99  & 0.97  & 2.62  & \textbf{0.96} & 0.96  & \textbf{4.08} & 0.96  & 0.95  & 4.39  & 0.95  & 0.95  & 4.88  \\
    \textit{$SSQP_B$} & \textbf{0.96} & 0.95  & \textbf{4.20} & \textbf{0.95} & \textbf{0.91} & \textbf{5.03} & \textbf{0.99} & 0.98  & \textbf{2.51} & 0.96  & 0.95  & 4.49  & \textbf{0.97} & 0.95  & \textbf{3.98} & \textbf{0.96} & \textbf{0.95} & \textbf{4.68} \\
    \bottomrule
    \end{tabular}
}
  \label{tab:PerfComp2}%
  \vspace{-3mm}
\end{table*}%
We also found that our $SSQP$ IQA model is also capable of predicting traditional
perceptual image quality. We compared $SSQP$ against the same benchmark algorithms 
on the well-known LIVE IQA database~\cite{cit:Sheikh2006},
It consists of 29 reference images and 982 distorted image with five distortion 
types: (1) JPEG2000 (JP2K), (2) JPEG, (3) white noise (WN), (4) Gaussian blur (GBLUR), 
and (5) Fast Fading (FF). For each distorted image, difference mean opinion scores
(DMOS) are provided, which is scaled and shifted to the range of [0, 100] where smaller
values mean better perceptual quality. \\
\indent The experimental results are given in Table~\ref{tab:PerfComp2}, where the best 
performing model is boldfaced. For all distortion types, $SSQP$ provided the highest 
prediction accuracy. These results suggest that GAN quality assessment subsumes 
some elements of more traditional image quality assessment, and the proposed 
$SSQP$ model provides more universally robust performance than existing FR metrics. 

\subsection{Perfomance Comparison against DNN-based IQA Models}\label{sec:5.4}
\begin{table}[t]
  \centering
  \caption{Performance comparison with two deep learning-based 
IQA models on the proposed generative image database.}
    \begin{tabular}{c|cc|cc}
    \toprule
    \multirow{2}[2]{*}{Model} & \multicolumn{2}{c|}{$Subset_{FI}$} & \multicolumn{2}{c}{$Subset_{PI}$} \\
          & PCC   & SROCC & PCC   & SROCC \\
    \midrule
    DeepQA & 0.948  & \textbf{0.918} & 0.739  & 0.712  \\
    BIECON & 0.676  & 0.865  & 0.124  & 0.155  \\
    \textit{SNPF} & 0.931  & 0.877  & \textbf{0.877} & \textbf{0.863} \\
    \textit{SNPB} & \textbf{0.953} & 0.891  & 0.831  & 0.832  \\
    \bottomrule
    \end{tabular}%
  \label{tab:CompDeepIQA}%
\vspace{-3mm}
\end{table}%
We compared the $SSQP$ against two deep learning based IQA 
models in~\cite{cit:Zeng2017}. One is $DeepQA$~\cite{cit:Jong2017}, which is a full-reference 
model, and the other is $BIECON$~\cite{cit:Kim2017} which is a no-reference model.
In~\cite{cit:Zeng2017}, $DeepQA$ achieved state-of-the-art performance in an FR
benchmark comparison, while $BIECON$ also showed top prediction accuracy in the 
category of no-reference models. The authors provide source codes of both models in 
the Theano framework, so that we could reproduce the results. \\
\indent We tested both models on two 
subsets of our generative image dataset: one is the subset of full-frame images ($subset_{FI}$)
and the other is the subset of patch images ($subset_{PI}$) which includes three different type 
of patches (random/regular/high-level structures). For $BIECON$, following the authors’ instruction, 
we used SSIM~\cite{cit:Wang2004} to generate local quality score maps as intermediate targets (see more 
details in~\cite{cit:Kim2017}). For $DeepQA$, we trained the network using the patch-based approach
adopted in~\cite{cit:Jong2017} on the $subset_{FI}$, since it consists of two different resolutions 
and the size of the input images needs to be fixed for training. We used the same patch size of 
$112$ x $112$ as in~\cite{cit:Jong2017}. On the single resolution $subset_{PI}$, 
$DeepQA$ network was trained image-wise. During the experiment, we randomly divided
the reference images into two subsets, 80\% for training and 20\% for testing.
The correlation coefficients were averaged after the procedure was repeated 10 times 
on randomly divided training and testing sets. \\
\indent The experimental results are given in Table~\ref{tab:CompDeepIQA}.
On $subset_{FI}$, $SSQP_B$ and $DeepQA$ gave comparable 
performances while $BIECON$’s prediction accuracy was relatively low. On $subset_{PI}$, 
the performance of $DeepQA$ dropped significantly, while $SSQP_B$ 
still delivered good prediction accuracy. Fig.~\ref{fig:comparison} helps 
explain this phenomenon. In the case of a full-frame image, even if introduced distortions get 
stronger, the main structure of the original image could be still maintained (the first column). 
However, patch images tend to lose their structure as generative 
distortions get stronger (the second through the fourth columns). $SSQP_B$ 
may cope with this structural collapse by alternatively measuring statistical 
similarity using the proposed histogram-distance features, as demonstrated in 
the results of Table~\ref{tab:singleFeature}. CNNs like those used by $DeepQA$ and 
$BIECON$ have far better abstraction ability than shallow regression methods 
when representing structures from low-level to high-level. However, they 
may not capture statistical similarity as well, which plays an important role in assessing 
the quality of generative images. $BIECON$ failed to provide reliable 
prediction accuracy. 

\subsection{Cross Database Test}\label{sec:5.5}
\begin{table*}[t]
  \caption{The database and cross-database tests: performance comparison 
between $SSQP_B$ and $DeepQA$, where the higher value in each pair of corresponding 
coefficients is marked in boldface.}
\begin{subtable}{\textwidth}
  \centering
  \caption{PCC comparison}
    \begin{tabular}{c|c|cccc}
    \toprule
    \multicolumn{2}{c|}{\multirow{2}[4]{*}{$SSQP_B$ / $DeepQA$}} & \multicolumn{4}{c}{Dataset for training} \\
\cmidrule{3-6}    \multicolumn{2}{c|}{} & \multicolumn{1}{c}{Generative} & LIVE  & TID2013 & CSIQ \\
    \midrule
    \multicolumn{1}{c|}{\multirow{4}[2]{*}{\makecell{Dataset \\ for\\ testing}}} & \multicolumn{1}{c|}{Generative} & \textbf{0.953} / 0.948 & 0.671 / \textbf{0.829} & 0.805 / \textbf{0.956} & 0.414 / \textbf{0.841}\\
          & LIVE  & \textbf{0.900} / 0.898 & \textbf{0.963} / 0.962 & \textbf{0.819} / 0.667 & 0.881/ \textbf{0.890} \\
          & TID2013 & \textbf{0.656} / 0.509 & \textbf{0.651} / 0.495 & 0.872 / \textbf{0.884} & \textbf{0.734} / 0.679 \\
          & CSIQ  & \textbf{0.828} / 0.819 & 0.830 / \textbf{0.841} & 0.815 / \textbf{0.878} & 0.888 / \textbf{0.962} \\
    \bottomrule
    \end{tabular}%
  \label{tab:crossPCC}%
\end{subtable}%
\\
\begin{subtable}{\textwidth}
  \centering
  \caption{SRCC comparison}
    \begin{tabular}{c|c|cccc}
    \toprule
    \multicolumn{2}{c|}{\multirow{2}[4]{*}{$SSQP_B$ / $DeepQA$}} & \multicolumn{4}{c}{Dataset for training} \\
\cmidrule{3-6}    \multicolumn{2}{c|}{} & \multicolumn{1}{c}{Generative} & LIVE  & TID2013 & CSIQ \\
    \midrule
    \multicolumn{1}{c|}{\multirow{4}[2]{*}{\makecell{Dataset \\ for\\ testing}}} & \multicolumn{1}{c|}{Generative} & 0.891 / \textbf{0.918} & 0.666 / \textbf{0.851} & 0.789 / \textbf{0.952} & 0.417 / \textbf{0.876} \\
          & LIVE  & 0.896 / \textbf{0.918} & 0.953 / \textbf{0.964} & \textbf{0.810} / 0.556 & 0.885/ \textbf{0.900} \\
          & TID2013 & \textbf{0.519} / 0.427 & \textbf{0.532} / 0.430 & 0.841 / \textbf{0.865} & \textbf{0.624} / 0.571 \\
          & CSIQ  & \textbf{0.841} / 0.833 & 0.782 / \textbf{0.868} & 0.749 / \textbf{0.889} & 0.806 / \textbf{0.958} \\
    \bottomrule
    \end{tabular}%
  \label{tab:crossSRCC}%
\end{subtable}%
  \label{tab:crossdata}%
\vspace{-5mm}
\end{table*}%
To demonstrate the generalization ability of $SSQP$, we conducted a  comprehensive set of
database and cross-database experiments. First, in addition to the proposed Generative IQA dataset 
and the LIVE IQA dataset, we added two existing databases: the TID2013 
database~\cite{cit:Ponomarenko2015} and the CSIQ database~\cite{cit:Larson2010}. 
The TID2013 database consists of 25 reference images and 3,000 distorted images with 24 
different distortion types at five levels of degradation, and the MOS of the distorted images 
is provided. The CSIQ database includes 30 reference images and 866 distorted images of 
six types: JPEG, JPEG2000, global contrast decrements, AWN, pink gaussian noise and 
gaussian blur, and it provides DMOS. We compared $SSQP_B$ against the 
DNN-based $DeepQA$ model. On the Generative IQA database, the subset of full-frame 
images ($subset_{FI}$) was used. \\
\indent Table~\ref{tab:crossdata} shows both the database and the cross-database 
test results. In each pair of corresponding correlation coefficients, we marked the one having 
higher value in boldface.
For the results where training and testing were done on the same databases (the diagonal),
$SSQP_B$ and $DeepQA $ attained comparable performance. 
$SSQP_B$ provided slightly better performance on the Generative and the LIVE datasets 
in terms of PCC, whilst $DeepQA$ was  
advantageous on TID2013 and CSIQ. It is noteworthy that $SSQP_B$ was able to 
provide comparable prediction accuracy as DeepQA, although it has a
much smaller number of parameters and a simpler system architecture. \\ 
\indent Next, for the cross-database experiments, the model trained on one database was 
tested on the other database, where the DMOS of the LIVE/CSIQ databases were converted 
to MOS. For example, when the model is trained on the Generative database and tested 
on the others, the PCC values were still quite reasonable: 0.900 and 0.828 on LIVE 
and CSIQ, respectively. The performance drop on TID2013 was due to the fact that it contains 
many distortion types that do not exist in the Generative database (or arguably, anywhere!) 
while the number of test images is far larger than in the train database. For the same reason, 
when we use TID2013 as a train dataset, the trained model delivers PCC values for 
cross-database tests, close to the results attained when the same database was used for 
testing as training. It is also noteworthy that although the three existing datasets consist 
of images of larger resolutions than  the Generative dataset, $SSQP$ still achieved 
reasonable prediction accuracy. It was able to cope with diverse distortion types,
which suggests that the structural/statistical features extracted by it reflect general aspects 
of distortions.

\section{Conclusion}\label{sec:conclusions}
We proposed a GAN image quality assessment model called $SSQP$ that
was devised using two groups of features representing structural and statistical 
similarities. We also used a multi-stage parallel boosting system to uncover the 
nonlinear relationship between the subjective scores and the proposed features.
We built a generative image quality database consisting of GAN generative 
images, and conducted a subjective study on it. The experimental results 
demonstrate the superiority of $SSQP$ on the new database, outperforming
existing FR models by significant margins. 
Furthermore, it also attained  comparable prediction accuracies as recent
DNN-based IQA models on three traditional image quality databases.

\end{document}